\documentclass[12pt]{article}
\usepackage[utf8x]{inputenc}
\usepackage{amsmath,amsthm,amssymb}
\usepackage{mathtools} % for prescript
\usepackage{mathrsfs}
\usepackage{verbatim}
\usepackage{hyperref}
\usepackage{xcolor}
\usepackage{enumitem}

\usepackage{color}
\usepackage{authblk}
\usepackage{graphicx,tikz,xcolor}
\usepackage{xcolor}
\usepackage{setspace}

\usepackage{floatrow}

\newcommand{\R}{\mathbb{R}}  % real nunmbers
\newcommand{\C}{\mathbb{C}}  % complex numbers
\newcommand{\Z}{\mathbb{Z}}  % integers
\newcommand{\N}{\mathbb{N}}  % natural numbers

\newcommand{\bfa}{{\bf a}}

\newcommand{\bfr}{{\bf r}}

\newcommand{\bfz}{{\bfr}}
\newcommand{\Frho}{G_{\rho}}

\newcommand{\bfR}{{\bf R}}

\newcommand{\SPL}{\mathrm{SPL}}

\def\Vee{{V}_{ee}}

\def\WN{\mathcal{W}^{N}}

\def\Te{T}
\def\Vne{{V}_{{\rm ne}}}
\def\vne{v_{{\rm ne}}}
\def\Sym{\mathfrak{S}}

\def\VSCE {\operatorname{V}_{ee}^{\operatorname{SCE}}}

\def\Vee{V_{ee}}
\def\Te{T}
\def\Vne{V_{ne}}

\def\Veered {\operatorname{V}_{ee}^{\operatorname{rad}}}

\def\Vtau{{\rm V}_{\rm ee}^{\tau}}

\def\ep{\varepsilon}
\newcommand{\pitau}{\pi^{\tau}}

\newtheorem{theorem}{Theorem}[section]

\newtheorem{corollary}[theorem]{Corollary}
\newtheorem{lemma}[theorem]{Lemma}

\newtheorem{example}[theorem]{Example}

\newtheorem{remark}[theorem]{Remark}

\def\mycolor(#1){%
   \ifnum#1=2
     green%
   \else
   \ifnum#1=3
     blue%
   \else
   \ifnum#1=4
     red%
   \else
   \ifnum#1=5
     violet%
   \else
   \ifnum#1=6
     yellow%
   \else
   \ifnum#1=7
     purple%
   \else
     black%
   \fi
   \fi
   \fi
   \fi
   \fi
   \fi
   }

\title{The strong-interaction limit of \\ density functional theory\footnote{Chapter in the book `Density Functional Theory' edited by Eric Canc\`{e}s and Gero Friesecke, Springer}}

\author[1]{Gero Friesecke}
\author[2,3]{Augusto Gerolin}
\author[4]{Paola Gori-Giorgi}
\affil[1]{Department of Mathematics, Technische Universit\"at M\"unchen}
\affil[2]{Department of Chemistry and Biomolecular Sciences, University of Ottawa}
\affil[3]{Department of Mathematics and Statistics, University of Ottawa}
\affil[4]{Department of Theoretical Chemistry, Vrije Universiteit Amsterdam}

\date{February 17, 2022}

\begin{document}
\maketitle

\begin{abstract}
\noindent
This is a comprehensive review of the strong-interaction limit of density functional theory. It covers the derivation of the limiting strictly correlated electrons (SCE) functional from exact Hohenberg-Kohn DFT, basic aspects of SCE physics such as the nonlocal dependence of the SCE potential on the density, equivalent formulations and the mathematical interpretation as optimal transport with Coulomb cost, rigorous results (including exactly soluble cases), approximations, numerical methods, integration into Kohn-Sham DFT (KS SCE), and applications to molecular systems, an example being that KS SCE, unlike the local density approximation or generalized gradient approximations, dissociates H$_2$ correctly. We have made an effort to make this review accessible to a broad audience of physicists, chemists, and mathematicians. 
\end{abstract}

%\noindent
{\small Keywords: Density functional theory, strongly correlated electrons, strictly correlated electrons, optimal transport}

\vspace*{4mm}

{\bf Introduction.} The strong-interaction limit of DFT is the inhomogeneous low-density limit associated with the uniform coordinate scaling
$$
  \rho_\gamma(\bfr) = \gamma^3 \rho(\gamma\bfr)
$$
of the single-particle density at fixed particle number, with $\gamma\to 0$. In this limit, the Levy-Lieb functional which gives the minimum kinetic and interaction energy subject to the given density has the leading order asymptotics
$$
   F_{\rm LL}[\rho_\gamma] \sim \gamma \VSCE[\rho],
$$
and the corresponding optimal wavefunction $ \Psi_\gamma$ has the asymptotics
$$
   \sum_{s_1,...,s_N\in\Z_2}|\Psi_\gamma(\bfr_1,s_1,...,\bfr_N,s_N)|^2 \sim \gamma^{3N} \rho_N(\gamma\bfr_1,...,\gamma\bfr_N)
$$
where $\rho_N$ solves the variational principle of having minimal Coulomb energy subject to the given density $\rho$  and  $\VSCE[\rho]$ denotes the resulting minimal energy.

This appears to be the only case in which one can obtain insight into how to extract information about the interaction energy directly from the density. As turns out, in this limit
%The limit clearly shows that 
none of the ingredients from the traditional ``Jacob's ladder'' of DFT approximations (local density, local density gradients, Kohn-Sham kinetic energy density, Hartree-Fock exchange, virtual orbitals) play any role. Instead, maps based on integrals not derivatives of the density appear. These maps are mathematically related to the field of optimal transport, and physically describe strictly correlated electrons (SCE). The SCE functional $\VSCE$ appearing above is the limiting Hartree-exchange-correlation functional.

While the strong-interaction limit is, of course, not reached in nature, it points the way towards the real physics happening in molecular systems containing strong correlations, without having to leave the realm of Kohn-Sham DFT. Two important examples whose physics is missed by Kohn-Sham DFT with semilocal or hybrid exchange-correlation functionals but captured correctly by integrating the SCE functional into Kohn-Sham DFT (KS SCE) are weakly charged nanosystems, see Figure \ref{fig:KSSCEwire}, and H$_2$ near the dissociation limit, see Figure \ref{F:H2binding}.

This chapter provides a self-contained introduction to this limit and its fascinating physics and mathematics which has been unearthed in the past two decades, and reviews the current state of the art.

%It is clear that the strong correlation problem remains in DFT...

%- Standard KS DFT re-models electron correlations in a semi-empirical manner 
% (LDA, B3LYP, PBE, ...)
 
%- But the accuracy of these approximations worsens as the molecules become more complex and long-range interactions become dominant. In particular, DFT approximations fail to accurately predict the physics of systems in which electronic correlation plays a prominent role, e.g., stretched bonds, transition metals, dispersion (van der Waals) interactions, ... [other regimes? hydrogen-bonding interaction in DNA? ...]

%- traditional KS ansatz is based on the weakly interacting limit; much insight into how to capture electron correlations within DFT can be gained from the strongly interacting limit. 

\begin{small}
\setstretch{0.8}{
\tableofcontents

}

\end{small}

\section{Many-electron Schr\"odinger equation and universal density functional}

In this section we quickly introduce the time-independent electronic Schr\"odinger equation and the exact reformulation of the ground state problem via a universal density functional.

\subsection{Many-electron Schr\"odinger equation}

\quad We consider a quantum mechanical system of $N$ non-relativistic electrons (of mass $m_e$ and charge $-e$), moving around classical nuclei with positions $\bfR_1,\dots,\bfR_M \in \R^d$ and charges $Z_1e,\dots,Z_Me$ (Born-Oppenheimer approximation). Our main interest is in the physical space $\R^3$, but we consider the general space dimension $d\ge 1$ since it will be instructive to illustrate key properties of the strong interaction limit with lower dimensional examples. The electrons are described by a wave function $\Psi:(\R^d\times \Z_2)^N\to\mathbb{C}$ of $N$ positions $\bfr_i \in \R^d$ and spin coordinates $s_i\in\lbrace\uparrow,\downarrow\rbrace=\Z_2$. 

\quad The Pauli exclusion principle states that the electronic wave function must be antisymmetric with respect to permutations of the electron coordinates,
\begin{equation} \label{eq:anti}
\Psi(\bfr_{\sigma(1)},s_{\sigma(1)},\dots,\bfr_{\sigma(N)},s_{\sigma(N)}) =
\operatorname{sign}(\sigma)\Psi(\bfr_1,s_1,\dots,\bfr_N,s_N), \quad \sigma
\in \Sym_N, 
\end{equation}
where $\Sym_N$ denotes the group of permutations of the indices $1,...,N$. The set of square-integrable $N$-electron wave functions, $\{\Psi\in L^2((\R^d\times\Z_2)^N;\C) \, : \, \eqref{eq:anti}\}$, will be denoted $\bigwedge^N_{i=1}L^2(\R^d\times\Z_2;\C)$. The one-body density of an electronic wave function $\Psi\in\bigwedge^N_{i=1}L^2(\R^d\times\Z_2;\C)$ is defined by 
\[
\rho_{\Psi}(\bfr_j) = N \sum_{s_1,\dots,s_N\in\Sym_N}\int_{\R^{d(N-1)}}   \vert \Psi(\bfr_1,s_1,\dots,\bfr_N,s_N)\vert^2 \prod_{i\neq j}d\bfr_i, \;\;\; \forall j\in\{1,...,N\}.
\]

\quad The energy $E[\Psi,v]$ of a fermionic state $\Psi$ with external potential $v:\R^d\to\R$ is given, in atomic units, by
\begin{equation}\label{intro:energyE}
\quad E[\Psi,v] =  \Te[\Psi] + \Vee[\Psi] + \Vne[\Psi,v],
\end{equation}

where $\Te[\Psi]$ is the \textit{kinetic energy},
\[
\Te[\Psi] = \dfrac{1}{2}\sum_{s_1 \in \Z_2} \int_{\R^3}\dots\sum_{s_N \in \Z_2} \int_{\R^3}\sum^N_{i=1}\vert
\nabla_{\bfr_i}\Psi(\bfr_1,s_1\dots,\bfr_N,s_N)\vert^2d\bfr_1\dots d\bfr_N;
\]
$\Vee[\Psi]$ is the electron-electron interaction energy 
\[
\Vee[\Psi] = \sum_{s_1 \in \Z_2} \int_{\R^d}\dots\sum_{s_N \in \Z_2}
\int_{\R^d} \sum^N_{1\leq i<j<N} w(
\bfr_i - \bfr_j) \,  \Psi(\bfr_1,s_1\dots,\bfr_N,s_N)\vert^2d\bfr_1\dots d\bfr_N,
\medskip
\]
and $\Vne[\Psi,v]$ is the electron-nuclei interaction energy,
\[
\Vne[\Psi,v] = \sum_{s_1 \in \Z_2} \int_{\R^d}\dots\sum_{s_N \in \Z_2} \int_{\R^d}\sum^N_{i=1}v(\bfr_i)\vert
\Psi(\bfr_1,s_1\dots,\bfr_N,s_N)\vert^2d\bfr_1\dots d\bfr_N,
\]
where $w \, : \, \R^d\to\R$ is an interaction potential satisfying $w(\bfr)=w(-\bfr)$, so that the total interaction potential 
\begin{equation} \label{eq:2body}
   V_{ee}(\bfr_1,...,\bfr_N) = \sum_{1\le i<j\le N}w(\bfr_i-\bfr_j)
\end{equation}
is symmetric.\footnote{We follow the usual convention to use the same letter $V_{ee}$ both for the total interaction potential, a function on $\R^{dN}$, and the associated quadratic form, a functional on the wavefunction space $\WN$.} Typically, 
\begin{equation} \label{eqn:typicalw}
    w(\bfr) = \vert \bfr\vert^{-1}
\end{equation}
is the Coulomb electron repulsion and $v$ is the Coulomb potential generated by $M$ nuclei which are at positions $\bfR_\nu$ with charges $Z_\nu$,
\begin{equation}\label{eqn:typicalv}
v(\bfr) = - \sum_{\nu=1}^M \frac{Z_\nu}{\vert \bfr - \bfR_\nu\vert}.
\end{equation}
If additional fields are present, the external potential $v$  contains extra terms. 
 
The central quantity of interest is the {\it ground state energy} of the system. By the {\it Rayleigh-Ritz variational principle}, it is given by 
\begin{equation}\label{GSE}
E_0[v] = \inf \lbrace E[\Psi,v]
: \Psi \in \WN \rbrace 
\end{equation}
where the infimum is taken over the class $\WN$ of wavefunctions which are antisymmetric and have finite kinetic energy,
\begin{equation} \label{WN}
\WN = \left\lbrace \Psi \in \bigwedge_{i=1}^N H^1(\R^d\times\Z_2;\C) \, : \, \sum_{s_1,...,s_N\in\Z_2}\int_{\R^{dN}} \vert \nabla \Psi\vert^2 d\bfr_1\dots d\bfr_N  < +\infty, \; ||\Psi|| = 1  \right\rbrace.
\end{equation}
Here $H^1$ is the usual Sobolev space of square-integrable functions with square-integrable gradient, and $||\Psi||$ denotes the $L^2$ norm of $\Psi$.  
The ground state energy \eqref{GSE} is well defined whenever the potentials $v$ and $w$ are sufficiently regular so that the functional $E$ is well defined on $H^1((\R^d\times\Z_2)^N)$. A simple sufficient condition in dimension $d=3$ which encompasses \eqref{eqn:typicalv}, \eqref{eqn:typicalw} is 
$v$, $w\in L^{3/2}(\R^3) + L^\infty(\R^3)$. 

Whether or not the infimum in \eqref{GSE} is actually a minimum, that is, a minimizing $\Psi$ exists, is much more subtle. For neutral or positively charged molecules in dimension $d=3$ (\eqref{eqn:typicalw}, \eqref{eqn:typicalv} with $Z =\sum^M_{i=1}Z_i > N-1$) the answer is yes, as was proved by  Zhislin \cite{Zhi-TMMO-60} via a careful spectral analysis of the underlying Hamiltonian operator. For an alternative proof based on variational methods see Friesecke \cite {Fri-ARMA-03}. 

\subsection{Universal density functional}\label{sec:DFT}

\quad In \textcolor{black}{quantum mechanics}, the absolute value squared $\vert \Psi(\bfr_1,s_1,\dots,\bfr_N,s_N)\vert^2$ of a wave function $\Psi \in \WN$ corresponds to an $N$-point probability distribution: it gives the probability density of finding the electrons at positions $\bfr_i\in\R^d$ with spins $s_i~\in~\Z_2, ~i\in \lbrace 1,\dots,N\rbrace$.

\quad By integrating the $N$-point probability distribution over the spins, we obtain the $N$-point position density,  
\begin{equation}\label{eq:Nppd}
\pi_N^{\Psi}(\bfr_1,\dots,\bfr_N) := \sum_{s_1,\dots,s_N\in\Z_2} \vert \Psi(\bfr_1,s_1,\dots,\bfr_N,s_N)\vert^2, \quad \Psi \in \WN.
\end{equation}

\quad The single particle density $\rho_{\Psi}(\bfr_j)$ is then obtained by integrating out all but one electron position $\bfr_j\in\R^d$, 
\begin{equation}\label{eq:rho1}
\rho_{\Psi}(\bfr_j) := \textcolor{black}{N}\int_{\R^{d(N-1)}}  \pi^{\Psi}_N(\bfr_1,\bfr_2,\dots, \bfr_j,\dots,\bfr_N) \prod_{i\neq j} d\bfr_i, \quad ~\forall~ j \in \{1,...,N\}.
\end{equation}
We denote by $\Psi \mapsto \rho$ the relation between $\Psi$ and $\rho$ given by equations  \eqref{eq:Nppd}, \eqref{eq:rho1}. This means that the wave function $\Psi$ has single-electron density $\rho$.

\quad Following the work of Hohenberg and Kohn \cite{HohKoh-PR-64}, Levy \cite{Lev-PRA-82} and Lieb \cite{Lie-IJQC-83} showed that the electronic ground state problem \eqref{GSE} can be recast as a minimization over single-electron densities $\rho$ instead of many-electron wavefunctions $\Psi$:  
\begin{equation}\label{eq.groundstatedensity}
E_0[\vne] = \inf_{\rho \in \mathcal{D}^N} \bigg\lbrace F_{\rm LL}[\rho] +
N\int_{\R^d}\vne(\bfr)\rho(\bfr)d\bfr \bigg\rbrace,
\end{equation}
with
\begin{equation}\label{eq.FHK}
F_{\rm LL}[\rho] = \min\bigg\lbrace \Te[\Psi] + \Vee[\Psi] : \Psi \in \WN, \Psi \mapsto \rho \bigg\rbrace, \medskip
\end{equation}
where $F_{\rm LL}[\rho]$ is the Levy-Lieb functional. The above direct definition of $F_{\rm LL}$ by a constrained search replaced an earlier, indirect existence proof of a universal functional satisfying \eqref{eq.groundstatedensity} \cite{HohKoh-PR-64}. The space $\mathcal{D}^N$ is defined as the set of densities $\rho$ coming from a wave function $\Psi \in \WN$ (i.e, $\Psi\mapsto\rho$), i.e., the $N$-representable one-particle densities. It can be fully characterized \cite{Lie-IJQC-83} and is given by
\begin{equation} \label{eq.DN}
\mathcal{D}^N = \lbrace \rho \in L^1(\R^d) : \rho\geq 0, \sqrt{\rho} \in H^1(\R^d), \int_{\R^d} \rho  = N \rbrace.
\end{equation}
Also, is known that the minimum in \eqref{eq.FHK} is attained. For more details about these matters see the chapter by Lewin, Lieb and Seiringer.

%\input{sections/sil}
%
%
%
% !TEX root = ../main.tex

\section{Strictly correlated electrons (SCE) functional}

\subsection{Constrained-search definition}

From the early days of DFT it has been clear that a useful approximation to the kinetic energy contribution in \eqref{eq.FHK} is given by the functional
\begin{equation} \label{eq.TsLL} 
      T_{\rm s,LL}[\rho] = \min_{\Psi\in{\cal W}^N, \, \Psi\mapsto\rho} \langle \Psi | T | \Psi\rangle
\end{equation}
and by its further approximation $T_{\rm S}[\rho]$ obtained by Kohn and Sham \cite{KS65} via restricting the above search to Slater determinants built from orthonormal spin orbitals, 
\begin{equation} \label{eq.Ts}
\begin{split}
     T_{\rm S}[\rho] = \min \left\{ \sum_{i=1}^N \sum_{s\in\Z_2}\int_{\R^3} |\nabla\phi_i(\bfr,s)|^2 \; : \; \phi_i\in H^1(\R^3\times\Z_2;\C) \, \forall i, \, \right. \\
    \left. \langle \phi_i|\phi_j\rangle = \delta_{ij} \, \forall i,j, \;
     \sum_{i=1}^N \sum_{s\in\Z_2} |\phi_i(\bfr,s)|^2=\rho(\bfr) \, \forall \bfr \right\}.
\end{split}
\end{equation}

The natural analogue of $T_{s,LL}$ for the interaction energy contribution in \eqref{eq.FHK} is the SCE functional
\begin{equation} \label{eq.SIL-WF}
    V_{ee}^{\rm SCE}[\rho] = \inf_{\Psi\in{\cal W}^N, \, \Psi\mapsto\rho} \langle \Psi | V_{ee} | \Psi\rangle
\end{equation}
which was introduced by Seidl \cite{Sei-PRA-99}. 
The acronym SCE stands for {\it strictly correlated electrons}, and will be explained shortly. As detailed in the next section, the functional \eqref{eq.SIL-WF} is a rigorous leading-order asymptotic limit of $F_{\rm LL}[\rho]$ in the low-density regime, where interaction dominates, just as the kinetic functional \eqref{eq.TsLL} is a leading-order asymptotic limit at high density, where the kinetic energy dominates. 

What is more, there also exists a natural analogue to $T_{\rm S}$ for interaction, which approximates the high-dimensional minimization over wavefunctions on $3N$ dimensional space in \eqref{eq.SIL-WF}  by a minimization over just $N$ maps on $\R^3$; see subsection \ref{sec:SCEansatz}.

\subsection{Derivation as low-density or strong-interaction limit of the Levy-Lieb functional}

For any given $N$-particle density $\rho$ on $\R^d$, consider its dilation obtained by uniform coordinate scaling
$$
   \rho^\gamma(\bfr) = \gamma^d \rho(\gamma \bfr)
$$
where $\gamma>0$ is a scaling factor. Note that this scaling preserves the total density, 
$$
    \int_{\R^d} \rho^\gamma(\bfr) = \int_{\R^d} \rho(\bfr) = N.
$$
We are interested in the small-$\gamma$ regime, which corresponds to a low-density limit. 

If $\Psi$ is a wavefunction with density $\rho$, then the scaled wavefunction
$$
  \Psi^\gamma(\bfr_1,s_1,....,\bfr_N,s_N) = \gamma^{\frac{dN}{2}} \Psi(\gamma\bfr_1,s_1,...,\gamma\bfr_N,s_N)
$$
has density $\rho^\gamma$. But as first noticed by Levy and Perdew \cite{Lev-Per-85}, scaling does not commute with constrained search. Instead, by an elementary change of variables, 
$$
  T[\Psi^\gamma] = \gamma^2 T[\Psi], \;\;\; V_{ee}[\Psi^\gamma] = \gamma V_{ee}[\Psi],
$$
and therefore
\begin{eqnarray}
   F_{\rm LL}[\rho^\gamma] & = & \min_{\Psi^{^\gamma}\in{\cal W}_N, \, \Psi^{^\gamma}\mapsto\rho^{^\gamma}} \Bigl\langle \Psi^\gamma|T+V_{ee}|\Psi^\gamma\Bigr\rangle \nonumber \\
   & = & \gamma \min_{\Psi\in{\cal W}_N, \, \Psi\mapsto\rho} \Bigl\langle \Psi |\gamma T + V_{ee}|\Psi\Bigr\rangle  \nonumber \\[1.8mm]
   & = & \gamma^2 F^{{1}/{\gamma}}[\rho], \label{eq.scaling}
\end{eqnarray}
where 
\begin{equation} \label{eq.LLalpha}
   F^\lambda[\rho] = \min_{\Psi\in{\cal W}_N, \, \Psi\mapsto\rho}
   \Bigl( T[\Psi] + \lambda V_{ee}[\Psi] \Bigr)
\end{equation}
is a Levy-Lieb functional with coupling constant $\lambda$. This suggests, assuming that the minimization in the second line of \eqref{eq.scaling} commutes with taking the limit $\gamma\to 0$, \begin{equation} \label{eq.asy1}
   F_{\rm LL}[\rho^\gamma] \underset{\gamma \to 0}{\sim}   \gamma\, V_{ee}^{\rm SCE}[\rho] 
\end{equation}
or equivalently, by starting from the Levy-Lieb functional with coupling constant, eq.~\eqref{eq.LLalpha}, as done in  \cite{Sei-PRA-99, SeiGorSav-PRA-07}
\begin{equation} \label{eq.asy2}
   \lim_{\lambda\to\infty} \tfrac{1}{\lambda}\, F^\lambda[\rho] = V_{ee}^{\rm SCE}[\rho].
\end{equation} 
Mathematically, as pointed out in \cite{SeiGorSav-PRA-07} it is not obvious whether the minimization in the second line of \eqref{eq.scaling} commutes with passing to the limit $\gamma\to 0$ since the optimal wavefunction depends on $\gamma$. Nevertheless the above leading-order asymptotics can be rigorously justified; see Theorem \ref{T:asy} in the next section. 

Repeating the calculation in \eqref{eq.scaling} without the kinetic energy and replacing ``min'' by ``inf'' shows that 
\begin{equation}
    V_{ee}^{\rm SCE}[\rho^\gamma] = \gamma V_{ee}^{\rm SCE}[\rho],
\end{equation}
whence the asymptotic result \eqref{eq.asy1} can also be re-written as
\begin{equation} \label{eq.asy3}
     F_{\rm LL}[\rho^\gamma] \underset{\gamma \to 0}{\sim}   V_{ee}^{\rm SCE}[\rho^\gamma].
\end{equation}
Off the low-density limit, we remark that $V_{ee}^{\rm SCE}$ still provides a rigorous lower bound for the Levy-Lieb functional, 
\begin{equation}\label{eq:VSILlowerboundFLL}
    F_{{\rm LL}}[\rho] \ge V_{ee}^{\rm SCE}[\rho]  \;\;  ~\forall\,\rho\in\mathcal{D}^N.
\end{equation} 
This is a trivial consequence of the constrained-search definitions \eqref{eq.FHK} and \eqref{eq.SIL-WF} and the nonnegativity of the kinetic energy functional $T$.
For typical atomic densities on $\R^3$, this lower bound is a significant improvement over the Lieb-Oxford bound with best known constant. 
%\begin{equation}
%    F_{{\rm LL}}[\rho] \ge c \int_{\R^3} \rho(\bfr)^{4/3} d\bfr.
%\end{equation}
%\gero{Maybe we can put an example with a 'model' atomic density here which documents the improvement, such as the noninteracting Helium ground state $\rho(\bfr) = \frac{N}{\sqrt{\pi Z^{3/2}}}e^{-Z|\bfr|}$, $N=Z=2$?}

%
%
%
%

\subsection{Enlarging the constrained search to probability measures}\label{sec:Enl}

\begin{figure}[htbp]
	\begin{center} 
		\includegraphics[width=0.25\textwidth]{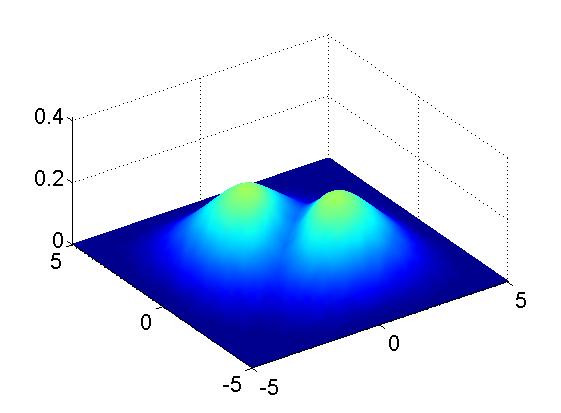} \hspace*{-3mm}
		\includegraphics[width=0.25\textwidth]{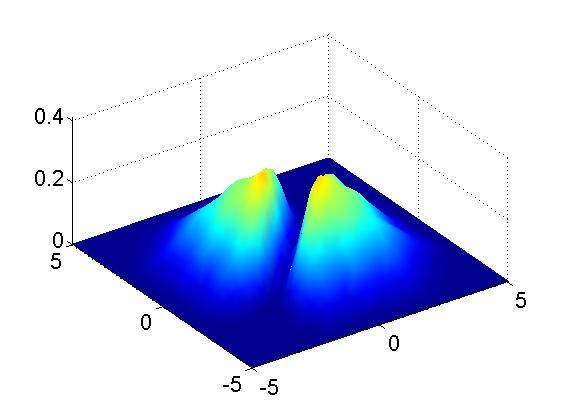} \hspace*{-3mm}
		\includegraphics[width=0.25\textwidth]{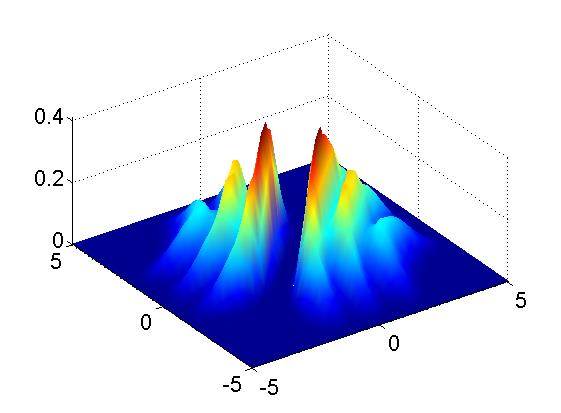} \hspace*{-3mm}
		\includegraphics[width=0.25\textwidth]{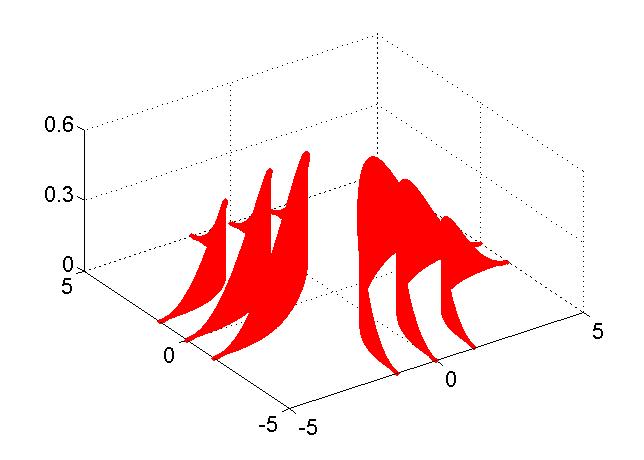}
		\vspace{-6mm}
	\end{center}
	\caption{Numerically computed ground state wave functions for $F^\lambda[\rho]$ in \eqref{eq.LLalpha} for $N=4$ and one-body density $\rho(r) = \frac{1}{2L}(1+\cos(\frac{\pi}{L} r)), r\in[-L,L]$ ($L=5$) for different values of $\lambda$: $\lambda=0.1$, $1$, $10$, $\infty$. Shown: pair density $\sum_{s_1,s_2,s_3,s_4}\int\!  dr_3\, dr_4 |\Psi^{\lambda}(r_1,s_1,r_2,s_2,r_3,s_3,r_4,s_4)|^2$. Picture from \cite{CheFri-MMS-15}, see also \cite{MorCoh-JPCL-17} for a numerical approximation of \eqref{eq.LLalpha} with $N=2$. The pair density on the left is governed by exchange effects, whereas the one on the right is governed purely by Coulombic correlations.}
		\label{fig:FcexSGS}
\end{figure}

The variational principle underlying the definition of $V_{ee}^{\rm SCE}[\rho]$ in \eqref{eq.SIL-WF}, 
\begin{equation} \label{VP.preSIL}
  \mbox{Minimize }\langle \Psi | V_{ee}|\Psi\rangle = \int_{\R^{dN}} V_{ee}(\bfr_1,...,\bfr_N) \, \pi_N^{\Psi}(\bfr_1,...,\bfr_N)\, d\bfr_1 ... d\bfr_N \mbox{ over }\{\Psi\in{\cal W}^N\, : \, \Psi\mapsto\rho\}, 
\end{equation}
with $N$-point density $\pi_N^{\Psi}$ as in \eqref{eq:Nppd}, typically has no minimizer. That is, no minimizing wavefunction $\Psi\in{\cal W}^N$ exists and the infimum in \eqref{eq.SIL-WF} is not attained.\footnote{This is not cured by dropping the  requirement in \eqref{WN} that $\Psi$ must have square-integrable gradient and requiring mere square-integrability, i.e. replacing $\bigwedge_{i=1}^N H^1(\R^d\times\Z_2;\C)$ by $\bigwedge_{i=1}^N L^2(\R^d\times\Z_2;\C)$.} 
Physically, this reflects the phenomenon that if $\Psi_\lambda[\rho]$ is a sequence of square-integrable functions depending on a parameter $\lambda>0$ such that $\langle \Psi_\lambda | V_{ee} | \Psi_\lambda\rangle$ approaches the infimum in \eqref{eq.SIL-WF} as $\lambda$ tends to infinity -- prototypical is the $\Psi_\lambda$ that minimizes $\langle \Psi|T+ \lambda V_{ee}|\Psi\rangle$ subject to $\Psi\mapsto\rho$ -- then $|\Psi_\lambda|^2$ integrates to $1$ but is typically concentrating on a lower dimensional subset, as depicted in Figure Fig.~\ref{fig:FcexSGS}.

This basic shortcoming of \eqref{VP.preSIL} -- that wavefunctions which are closer and closer to being optimal in the constrained search \eqref{eq.SIL-WF} do not converge to any proper wavefunction -- can be overcome as follows \cite{ButDepGor-PRA-12, CotFriKlu-CPAM-13}. First, interpret the variational principle \eqref{VP.preSIL} as variational principle for the $N$-point density as suggested by the second expression in \eqref{VP.preSIL}; second, enlarge 
the space of admissible $N$-point densities is enlarged to the space ${\cal P}(\R^{dN})$ of {\it probability measures} on $\R^{dN}$ with density $\rho$. Then the constrained search becomes well-posed, that is, optimizers exist.  
 See Theorem \ref{T:basic} below. This enlargement allows $N$-point densities to concentrate on lower dimensional subsets as in Figure 	\ref{fig:FcexSGS}. The condition that a probability measure $\Pi\in {\cal P}(\R^{dN})$ has density $\rho$ now means that $\Pi$ has marginals equal to the density divided by the particle number, $\frac{\rho}{N}$: 
\begin{equation} \label{eq.margs}
   \int_{(\R^d)^{j-1} \times A_j \times (\R^d)^{N-j}} d\Pi = \int_{A_j} \frac{\rho}{N} \;\mbox{ for all }j=1,...,N \mbox{ and all open sets }A_j \mbox{ in }\R^d.
\end{equation}
We denote the relation given by eq.~\eqref{eq.margs} by $\Pi\mapsto\rho$. This yields the variational principle
\begin{equation} \label{VP.SIL}
    \mbox{Minimize } \int_{\R^{dN}} \Vee(\bfr_1,...,\bfr_N) \, d\Pi(\bfr_1,...,\bfr_N)\,  \mbox{ over }\{ \Pi\in{\cal P}(\R^{dN})\, : \, \Pi\mapsto\rho\} 
\end{equation}
and the following enlarged-constrained-search definition of the SCE functional 
\begin{equation} \label{eq.SIL2}
   V_{ee}^{\rm SCE}[\rho] = \min_{\Pi\in{\cal P}(\R^{\textcolor{blue}{d}N}), \, \Pi\mapsto\rho} \int_{\R^{dN}} \Vee(\bfr_1,...,\bfr_N) \, d\Pi(\bfr_1,...,\bfr_N).
\end{equation}
This alternative definition of $V_{ee}^{\rm SCE}$ and the underlying enlarged variational problem \eqref{VP.SIL} were introduced by Buttazzo, DePascale, Gori-Giorgi, Cotar, Friesecke, and Kl\"uppelberg \cite{ButDepGor-PRA-12, CotFriKlu-CPAM-13}, along with the insight that minimizers now exist (see Theorem \ref{T:basic} (1) below) and \eqref{eq.SIL2} is mathematically an  {\it optimal transport problem} and can be usefully analyzed with methods from optimal transport theory (see section \ref{sec:OT}).  We call \eqref{VP.SIL} the {SIL variational principle}, the acronym SIL standing for strong-interaction limit. %\gero{Paola, Augusto, do you think that this is a good terminology, in the spirit of our discussion that the variational principle is different, but the ensuing functional is the same? How about simply ``enlarged constrained search''?}\augusto{I like as it is rather than "enlarged constraint search", since the word relaxation is already used above.}

The notation in \eqref{eq.SIL2} (``min'' instead of ``inf''; using the same notation for the ensuing density functional even though a priori the right hand side of \eqref{eq.SIL2} could be lower than that in \eqref{eq.SIL-WF} since the minimization is over a larger set) is justified because of: 

\begin{theorem} \label{T:basic} Let $\rho$ be any $N$-particle density in the class ${\cal D}^N$ (see \eqref{eq.DN}), and let $w(\bfr)=|\bfr|^{-1}$ be the Coulomb interaction. \\[1mm]
{\rm (1)} The minimum in \eqref{eq.SIL2} is attained; that is, there exists a minimizing probability measure $\Pi$. \\[1mm]
{\rm (2)} {\rm \cite{CotFriKlu-CPAM-13, BinDep-JEP-17, CotFriKlu-ARMA-18}} The minimum value in \eqref{eq.SIL2} is equal to the infimum in \eqref{eq.SIL-WF}.
\end{theorem}

Statement (1) is a special case of general existence theorems in optimal transport theory. For a textbook account see \cite{Fri-Book-22}. Proofs of such results rely on Prokhorov's theorem from probability theory as well as on approximation and lower semi-continuity results for functionals of the form $\Pi\mapsto \int V_{ee} d\Pi$.
  
Statement (2), although plausible, is mathematically much more subtle. It rests on the nontrivial result that arbitrary symmetric probability measures $\Pi\in{\cal P}(\R^{dN})$ with marginal $\rho$ can be approximated by $N$-point densities of quantum wavefunctions $\Psi\in{\cal W}^N$ with the {\it same} marginal. Note that such wavefunctions must be antisymmetric and must have a square-integrable gradient; but applying standard smoothing techniques from mathematics -- such as mollification -- to a given probability measure with marginal $\rho$ does not preserve the marginal, nor does it yield the $N$-point density of an antisymmetric function. This result, and the ensuing statement (2), was first proved for $N=2$ \cite{CotFriKlu-CPAM-13}, and later extended to $N=3$ \cite{BinDep-JEP-17} and general $N$ \cite{CotFriKlu-ARMA-18} (see also \cite{Lew-CRM-18} for a similar extension to general $N$ allowing mixed states). 

\begin{remark} \label{R:symmetrization} {\rm (Symmetrization)}
The minimum value in \eqref{eq.SIL2} is unchanged, and still attained, when the minimization over arbitrary probability measures with marginal $\rho/N$, $\{\Pi\in{\cal P}(\R^{dN})\, : \, \Pi\mapsto\rho\}$, is restricted to {\it symmetric} probability measures with marginal $\rho/N$, where a probability measure $\Pi\in{\cal P}(\R^{dN})$ is said to be symmetric if
\[
  \int_{A_1\times ... \times A_N} d\Pi = \int_{A_{\sigma(1)}\times ... \times A_{\sigma(N)}} d\Pi \mbox{ for all open sets }A_1,...,A_N \mbox{ in }\R^d \mbox{ and all permutations }\sigma.
\]
This is because whenever $\Pi$ is a probability measure in ${\cal P}(\R^{dN})$ with marginals $\rho$, eq.~\eqref{eq.margs}, then so is its symmetrization $S_N{\Pi}$ defined by  
\begin{equation} \label{eq.symmetrize}
   (S_N{\Pi})(A_1\times ... \times A_N) = \frac{1}{N!} \sum_{\sigma} \Pi(A_{\sigma(1)}\times ... \times A_{\sigma(N)}), 
\end{equation}
the sum being over all permutations of $\{1,...,N\}$; and the integral on the r.h.s. of \eqref{eq.SIL2} for $\Pi$ agrees with that for $S_N{\Pi}$, thanks to the permutation symmetry of $V_{ee}$. 
\end{remark}

Next we rigorously justify the asymptotic relations \eqref{eq.asy1}, \eqref{eq.asy2}, \eqref{eq.asy3} and complement them with an  asymptotic result on the associated constrained-search wavefunctions.

\begin{theorem} \label{T:asy} {\rm \cite{CotFriKlu-ARMA-18}} For any $N$-electron density $\rho$ in the class  ${\cal D}^N$ (see \eqref{eq.DN}), and with $w(\bfr)=|\bfr|^{-1}$ being the Coulomb interaction, the asymptotic results \eqref{eq.asy1}, \eqref{eq.asy2}, \eqref{eq.asy3} hold. 
Moreover if $\Psi_\lambda[\rho]$ is any minimizer in the constrained-search definition of $F^\lambda[\rho]$ (see \eqref{eq.LLalpha}), then every limit point\footnote{By a limit point $\Pi$ of a sequence $\Pi_\lambda$ of probability measures we mean a limit point in the sense of narrow convergence, that is, convergence of the integrals $\int f \, d\Pi_\lambda$ to $\int f \, d\Pi$ for any bounded continuous function $f$.} $\Pi$ of the sequence of $N$-point densities $\pi^{\Psi_\lambda[\rho]}$ is a minimizer in the enlarged-search definition \eqref{eq.SIL2} of $V_{ee}^{\rm SCE}[\rho]$. 
\end{theorem}

{\bf Proof of \eqref{eq.asy1}, \eqref{eq.asy2}, \eqref{eq.asy3}} The proof, taken from \cite{CotFriKlu-ARMA-18}, is easy, so we include it. We show  \eqref{eq.asy2}, the other statements being equivalent. Fix $\rho$. First, pick any minimizer $\Psi^\lambda[\rho]$ in the constrained-search definition of $F^\lambda[\rho]$, then
\begin{equation} \label{eq.lowbd}
  \frac{1}{\lambda}F^\lambda[\rho] = \frac{1}{\lambda}\Bigl( T[\Psi^\lambda[\rho]] + \lambda V_{ee}[\Psi^\lambda[\rho]]\Bigr) \ge V_{ee}[\Psi^\lambda[\rho]] \ge V_{ee}^{\rm SCE}[\rho],  
 \end{equation}
that is, the SCE functional is a lower bound of the left hand side. To show that it is also an asymptotic upper bound for large $\lambda$, we fix any positive number $\epsilon$ and 
pick a wavefunction $\tilde{\Psi}[\rho]$ in ${\cal W}^N$ such that $V_{ee}[\tilde{\Psi}[\rho]]\le V_{ee}^{SCE}[\rho] + \epsilon$. It follows that 
$$
  \frac{1}{\lambda}F^\lambda[\rho] \le \frac{1}{\lambda}\Bigl( T[\tilde{\Psi}[\rho]] + \lambda V_{ee}[\tilde{\Psi}[\rho]]\Bigr).
$$
Since $\tilde{\Psi}$ belongs to ${\cal W}^N$, its kinetic energy $T[\tilde{\Psi}]$ is finite, and so
$$
 \limsup_{\lambda\to\infty}  \frac{1}{\lambda}F^\lambda[\rho] \le V_{ee}[\tilde{\Psi}[\rho]] \le V_{ee}^{\rm SCE}[\rho] + \epsilon.
$$
Since $\epsilon>0$ was arbitrary, 
\begin{equation} \label{upbd}
    \limsup_{\lambda\to\infty}  \frac{1}{\lambda}F^\lambda[\rho] \le V_{ee}^{\rm SCE}[\rho].
\end{equation}
Combining \eqref{eq.lowbd} and \eqref{upbd} yields \eqref{eq.asy2}. 

The above simple argument only shows that the asymptotic error  in \eqref{eq.asy2} is $o(1/\lambda)$, but does not give its order, which turns out to be $O(1/\lambda^{1/2})$, see section \ref{sec:next}.

\subsection{The SCE ansatz} \label{sec:SCEansatz}
The SIL variational principle \eqref{VP.SIL} still requires minimization over a high-dimensional space of $N$-point probability measures. 

Seidl \cite{Sei-PRA-99} (see also \cite{SeiGorSav-PRA-07})  proposed the following low-dimensional ansatz: we restrict minimization over $N$-point probability measures to minimization over singular probability measures of the special form 
\begin{equation}\label{eq.psiSCE} 
d\Pi(\bfr_1,...,\bfr_N) = \frac{\rho(\bfr_1)}{N} 
 \prod_{n=2}^N \delta\bigl(\bfr_n-f_{n-1}(\bfr_1)\bigr) d\bfr_1 ... d\bfr_N
\end{equation} 
where, for any $\bfr_1\in\R^d$, $\delta\big(\bfr_n - f_{n-1}(\bfr_1)\big)$ denotes the delta function of $\bfr_n$ (alias Dirac measure) centered at $f_{n-1}(\bfr_1)$, and $f_1,...,f_{N-1}$ are maps from $\R^d$ to $\R^d$. The singular densities \eqref{eq.psiSCE} are concentrated on the d-dimensional set
\begin{equation} \label{eq:Omega0}
     \Omega_0 = \{ (\bfr_1,...,\bfr_N)\in\R^{dN} \, : \, \bfr_2 = f_1(\bfr_1),...,\bfr_N=f_N(\bfr_1)\}.
\end{equation}
From a physical point of view, such a density describes a state in which the position of one of the electrons, say $\bfr_1$, can be freely chosen according to the density $\rho$, but this then uniquely fixes the position of all the other electrons through the functions $f_2,...,f_N$, that is, $\bfr_2=f_1(\bfr_1)$ etc. Thus states of form \eqref{eq.psiSCE} are  called {\it strictly correlated states}, or SCE states for short. The $f_i$ are called {\it co-motion functions} or {\it transport maps}.
%\gero{I think it is best if we use delta function notation here, not OT notation. In delta function notation the ansatz is simpler and more intuitive. Moreover, to derive the SCE variational principle in the form \eqref{SCE'}
%no change of variables is needed as one straightforwardly integrates over the delta functions. In OT notation, deriving the ensuing interaction energy is extremely complicated as one needs a change-of-variables formula for non-diffeomorphisms, which is unknown not just in physics but also in applied and computational maths. For those who do like OT notation, we can mention it as an alternative in section \ref{sec:OT}.}\augusto{For me is more than ok}

The marginal constraint that $\Pi$ must have marginals $\rho$, eq.~\eqref{eq.margs}, turns into the following constraint on the maps $f_n$: the $f_n$ must transport the density $\rho$ to itself, 
\begin{equation} \label{eq.pfw}
   f_i{}_\sharp \rho = \rho \;\;\;\forall i\in\{2,...,N\}
\end{equation}
where, for any measurable map $f\, : \, \R^p\to\R^q$ and any measure $\mu$ on $\R^p$, the push-forward $f_\sharp\mu$ is the measure on $\R^q$ defined by 
\begin{equation} \label{eq.push}
    (f_\sharp\mu)(B) = \mu(f^{-1}(B))
    \mbox{ for all open sets $B$ in $\R^q$}. 
\end{equation}
More explicitly, if $p=q$, $\mu$ is absolutely continuous with density $\rho$, $f$ is a diffeomorphism, and the density of the push-forward $f_\sharp\mu$ is denoted by $f_\sharp\rho$, we have
$$
  (f_\sharp \rho)(\bfr') = |\det Df^{-1}(\bfr')| \, \rho\bigl(f^{-1}(\bfr')\bigr) .
$$

By substituting this formula for the push-forward into \eqref{eq.pfw} and changing variables $f^{-1}(\bfr')=\bfr$, the constraint \eqref{eq.pfw} turns -- provided the $f_n$ are diffeomorphisms -- into the following nonlinear first-order partial differential equation:
$$
  \rho(f_i(\bfr)) = \frac{\rho(\bfr)}{|\det Df_i(\bfr)|} \;\;\; \forall i\in\{2,...,N\}.
$$

Plugging the ansatz \eqref{eq.psiSCE} into the SIL variational principle \eqref{VP.SIL} and integrating out the variables $\bfr_2,...,\bfr_N$ yields the SCE variational principle
\begin{equation} \label{SCE'}
       \mbox{Minimize }\int_{\R^d} V_{ee}\bigl(\bfr_1,f_1(\bfr_1),...,f_{N-1}(\bfr_1)\bigr) \, \frac{\rho(\bfr_1)}{N} \, d\bfr_1 \; \mbox{ over maps }f_1,...,f_{N-1} \in {\cal T}_\rho,
\end{equation}
with the minimization being over maps in the admissible class 
\begin{equation} \label{eq.Trho}
     {\cal T}_\rho = \{ f \, : \, \R^d\to\R^d \, : \, f \mbox{ measurable}, \; f_\sharp \rho = \rho \}.
\end{equation}
Thanks to Theorem \ref{T:SCE} (1) below, this yields a third  construction of the SCE functional, 
\begin{equation} \label{SCEfunctional3}
       V_{ee}^{\rm SCE}[\rho] = \inf_{f_1,...,f_{N-1}\in {\cal T}_\rho} \int_{\R^d} V_{ee}\bigl(\bfr,f_1(\bfr),...,f_{N-1}(\bfr)\bigr) \, \frac{\rho(\bfr)}{N} \, d\bfr. 
\end{equation}

In the Coulomb case, \eqref{eqn:typicalw}, and denoting $f_0(\bfr)=\bfr$, we thus have
\begin{equation} \label{SCE3'}
   V_{ee}^{\rm SCE}[\rho] = \inf_{f_1,...,f_{N-1}\in {\cal T}_\rho} \sum_{0\le i<j\le N-1} \int_{\R^d} \frac{1}{|f_i(\bfr)-f_j(\bfr)|} \, \frac{\rho(\bfr)}{N} \, d\bfr. 
\end{equation}
Physically, this means that one needs to minimize the mutual Coulomb repulsion of the co-motion functions. 
\color{black}
This construction of the SCE functional was introduced by Seidl \cite{Sei-PRA-99}. A priori it is not clear, but was conjectured by Seidl, that it is equivalent to the original construction \eqref{eq.SIL-WF}. This is now rigorously known (see Corollary \ref{C:basic} below). 

The construction \eqref{SCEfunctional3} should be considered the analogue for interaction of the classical Kohn-Sham kinetic energy functional $T_S$. Just as $T_S$ is determined by $N$ low-dimensional functions (the Kohn-Sham spin orbitals $\phi_1,...\phi_N\, : \, \R^3\times\Z_2\to\C$), $V_{ee}^{\rm SCE}$ is determined by $N-1$ low-dimensional maps (the co-motion functions or transport maps $f_1,...,f_{N-1}\, : \, \R^3\to\R^3$) which can be easily stored on a computer. Moreover -- like the Kohn-Sham orbitals -- the co-motion functions are obtained by just minimizing a $3$-dimensional integral.

The reader is warned, however, that the behaviour of the SCE variational principle and its relationship to the SIL variational principle is subtle, and open questions remain. In particular, it is not known -- except in special cases -- whether minimizers in \eqref{SCE3'} exist. The following results have been rigorously proved.

\begin{theorem} \label{T:SCE} Let $\rho \, : \, \R^d\to\R$ be any $N$-particle density in the class ${\cal D}^N$ (see \eqref{eq.DN}), and let $w(\bfr)=|\bfr|^{-1}$ be the Coulomb interaction. \\[1mm]
{\rm (1)} The infimum in \eqref{SCEfunctional3} is equal to the minimum in \eqref{eq.SIL2}. 
\\[1mm]
{\rm (2)} For two electrons ($N=2$), and in arbritary space dimension $d$, the infimum in \eqref{SCEfunctional3} is attained; that is, there exists a minimizing map $f_1$. Moreover $f_1$ is unique, and the induced probability measure \eqref{eq.psiSCE} is the unique minimizer of the SIL variational principle \eqref{VP.SIL}. 
\\[1mm]
{\rm (3)} In one space dimension ($d=1$), and for arbitrary $N$, the infimum in \eqref{SCEfunctional3} is attained; that is, there exist minimizing maps $f_1,...,f_{N-1}$. Moreover %the maps are unique up to relabelling \gero{Augusto, can you please check this?}, and 
the symmetrization (see Remark \ref{R:symmetrization}) of the associated probability measure \eqref{eq.psiSCE} is the unique symmetric minimizer of the SIL variational principle \eqref{VP.SIL}.  
\end{theorem}

%\augusto{I need to understand what you mean by relabelling, but I think you are correct because they are all "equivalent" to one. Look at Figure 4 in page 29.}

Statement (1) is a consequence, pointed out in \cite{ColDiM-INC-13}, of a general theorem by Ambrosio \cite{Ambrosio-03} and Pratelli \cite{Pra-AIHP-07} in optimal transport theory. 
%which holds for very general densities and interaction potentials. 
For $N=2$ or $d=1$, use of the Ambrosio-Pratelli theorem can be avoided since the assertion follows from (2) respectively (3).

The existence of optimal maps in (2) and (3) is subtle and depends on special Coulombic features. For non-Coulombic counterexamples see
Remark \ref{R:nonattainment} below.
In the Coulomb case, it is an open question whether the infimum in \eqref{SCEfunctional3} is attained for general (physically reasonable)  densities $\rho$ when $d>1$ and $N\ge 3$. 

Statement (2) completely justifies Seidl's SCE ansatz for $N=2$: the SCE problem
$$
   \mbox{Minimize }\int_{\R^d} \frac{1}{|\bfr-f_1(\bfr)|} \mbox{ over maps }f_1\in {\cal T}_\rho
$$
has a unique minimizer and the associated SCE state
\begin{equation} \label{eq.psiSCE2_2body}
  d\Pi(\bfr_1,\bfr_2) = \frac{\rho(\bfr_1)}{2}\delta\bigl(\bfr_2-f_1(\bfr_1)\bigr) \, d\bfr_1 d\bfr_2 
\end{equation}
is the unique minimizer of the SIL problem
$$
  \mbox{Minimize }\int_{\R^d\times\R^d} \frac{1}{|\bfr_1-\bfr_2|} d\Pi(\bfr_1,\bfr_2) \mbox{ over }\Pi\mapsto\rho.
$$
This was proved in \cite{CotFriKlu-CPAM-13}, by modifying the analysis by Gangbo and McCann \cite{Gangbo-McCann-96} of optimal transport with costs $w(\bfr,\bfr')$ which are convex or concave in the displacement ${\bf z}=\bfr-\bfr'$. Note that the Coulomb cost is neither:  near any ${\bf z_0}\neq 0$, it is convex in radial direction and concave in all perpendicular directions. A simpler  proof using Kantorovich duality (see section \ref{sec:Kantdual}) was suggested in \cite{ButDepGor-PRA-12}, and made rigorous in \cite{DMaGerNen-TOOAS-17}. The SCE map is given by
\begin{equation}\label{eq:SCEMap2electrons}
f_1(\bfr) = \bfr + \dfrac{\nabla u(\bfr)}{\vert \nabla u(\bfr)\vert^{3/2}},
\end{equation}
for some function $u:\R^d\to\R$  (Kantorovich potential). The notion of Kantorovich potential will be  explained in section \ref{sec:Kantdual}.  Eq.~\eqref{eq:SCEMap2electrons} follows by solving   eq.~\eqref{eq.gradpot2} for $f_1$. 

Statement (3), together with an explicit construction of the optimal maps given in section \ref{sec:1D}, was suggested in the original paper by Seidl \cite{Sei-PRA-99} on grounds of physical arguments, and was rigorously proved in \cite{ColDepDim-CJM-15} with the help of cyclical monotonicity methods from optimal transport theory. See section \ref{sec:1D} for more information.
%An alternative proof using Kantorovich duality will appear in \cite{Fri-Book-22}. 

The uniqueness statements in (2) and (3) are somewhat surprising: the optimal $N$-point densities arising from Levy-Lieb constrained search in the strongly interacting limit  are always {\it unique} when either $N=2$ or $d=1$! No analogue holds off the strongly interacting limit. 

\begin{figure}\label{F:jump} 
\begin{center}
   {\includegraphics[height=60mm]{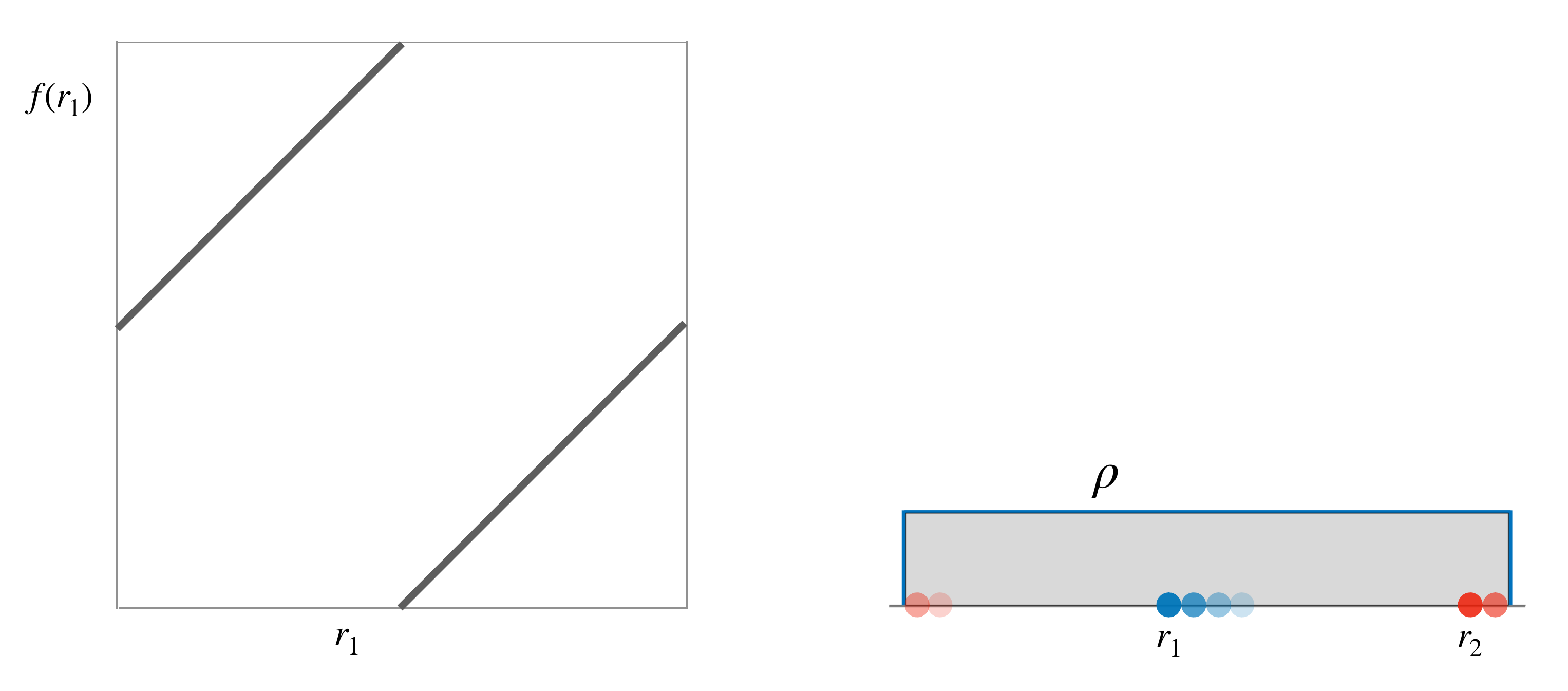}}
\end{center}
\vspace*{-3mm}

\caption{SCE state of a two-electron system with homogeneous  density in a one-dimensional domain. By Theorems \ref{T:asy} and \ref{T:SCE}, this state is an asymptotically exact approximation to the true quantum ground state at low density. Left: optimal co-motion function or transport map $f$. Right: position of the two electrons in the one-dimensional domain. The position of the second electron, $r_2$, is determined by that of the first electron, $r_1$,  through the equation $r_2=f(r_1)$, with the optimal $f$ keeping the electrons at a constant distance, of half the domain size. The position of the first electron varies over the whole domain according to the density $\rho$ (see eq.~\eqref{eq.psiSCE2_2body}). As the first electron (depicted in blue) passes through the mid-point, the position of the second electron (depicted in red) jumps from the right end to the left end, causing a discontinuity of $f$. 
%\gero{I've re-drawn the figure on the right, also showing the density governing the position of the first electron (in line with the new figure a la Seidl which I've drawn for our 1D section), and revised the caption accordingly}
}

\end{figure}

\begin{example} \label{ex.1Dhomogeneous} Consider a two-electron system with uniform density in a one-dimensional interval $[0,L]$. The unique minimizer $f_1=f$ of the SCE variational principle \eqref{SCE'} can be shown (see section \ref{sec:1D})  to be 
\begin{equation} \label{eq:1Dhom}
    f(r_1) = \begin{cases} r_1 + \tfrac{L}{2} & \mbox{if }r_1 \le \tfrac{L}{2} \\ r_1 - \tfrac{L}{2} & \mbox{if }r_1 > \tfrac{L}{2}.
    \end{cases}
\end{equation}
See Figure \ref{F:jump}. 
\end{example}

By combining Theorems \ref{T:SCE} and \ref{T:basic} we obtain:

\begin{corollary} \label{C:basic} Let $\rho \, : \, \R^d\to\R$ be any $N$-particle density in the class ${\cal D}^N$ (see \eqref{eq.DN}), and let $w(\bfr)=|\bfr|^{-1}$ be the Coulomb interaction. Then the infimum in \eqref{SCEfunctional3} is equal to that in \eqref{eq.SIL-WF}. 
\end{corollary}

{\bf Proof} This follows from the fact that both quantities are equal to the minimum value of the SIL variational principle \eqref{VP.SIL}, by Theorem \ref{T:basic} (2) respectively Theorem \ref{T:SCE} (1). 

We remark that no proof is known which bypasses the SIL variational principle, even though the corollary was conjectured before the latter was introduced.

We close this introductory section on the SCE ansatz with some remarks. 

\begin{remark} \label{R:nonattainment} {\rm (Nonattainment)}
For simple non-Coulombic counterexamples to attainment of the infimum in \eqref{SCEfunctional3} for $N=3$ even in one space dimension see {\rm \cite{Fri-SIAMJMA-19,GerKauRaj-SIAMJMA-19}}. For instance, one can take the uniform density in the interval $[0,3]$ and the interaction potential  $w(r)=r^4/4-r^3/3$ {\rm \cite{Fri-SIAMJMA-19}}. Earlier more intricate counterexamples can be found in {\rm \cite{MoaPas-ESAIMCOCV-17}}. Such a nonattainment has the undesirable consequence that numerically computed optimal maps will necessarily exhibit wilder and wilder oscillations as the mesh is refined or the basis set approaches completeness, and fail to converge in any pointwise sense to actual optimal maps. 
\end{remark}

\begin{remark} {\rm (Existence of non-SCE minimizers)}
For a Coulombic example for $N=3$ in three space dimensions showing that the SIL variational principle can possess minimizers which are not of SCE form see {\rm \cite{Pas-NL-13}}. This example exhibits nonuniqueness and it is not known whether it also admits minimizers which {\rm are} of SCE form.
\end{remark} 

\begin{remark} {\rm (Alternative formulations of the SCE ansatz)} By denoting  $f_0(\bfr)=\bfr$, one can write the SCE ansatz \eqref{eq.psiSCE} in the following form in which all coordinates $\bfr_1,...,\bfr_N$ appear on an equal footing:
\begin{equation} \label{eq.psiSCE2} 
  d\Pi(\bfr_1,...,\bfr_N) = \int \frac{\rho(\bfr)}{N} \prod_{n=1}^N \delta\bigl( \bfr_n - f_{n-1}(\bfr)\bigr) \, d\bfr.
\end{equation}
Also, one can work with the symmetrized form of this ansatz,
\begin{equation} \label{eq.psiSCE3} 
  d\Pi(\bfr_1,...,\bfr_N) = \frac{1}{N!} \sum_\sigma \int \frac{\rho(\bfr)}{N} \prod_{n=1}^N \delta\bigl( \bfr_n - f_{\sigma(n-1)}(\bfr)\bigr) \, d\bfr
\end{equation}
(where $\sigma$ runs over the permutations of the SCE map indices $0,...,N-1$); the symmetrization doesn't change the energy $\int V_{ee}\,  d\Pi$, and the symmetrized form \eqref{eq.psiSCE3} minimizes the SIL problem  \eqref{eq.SIL2} if and only if the unsymmetrized form \eqref{eq.psiSCE2} does, as was explained in Remark \ref{R:symmetrization}.
\end{remark}

\begin{remark} {\rm (Nonsmoothness of optimal maps)} The reader might wonder why, in SCE theory, no differentiability and not even continuity is imposed on the competing maps (the maps in the admissible class \eqref{eq.Trho} are merely required to be measurable). This is because optimal maps, when they exist, are typically discontinuous. This important effect can be understood  intuitively from simple examples as in Figure \ref{F:jump}.  As the first electron passes through the midpoint of the domain, the position of the second electron jumps from the right end of the domain to the left end, yielding the discontinuous map depicted in the Figure. For a radial density in three dimensions ($d=3$), an analogous discontinuity occurs in that spheres near zero are mapped to spheres near infinity \cite{CotFriKlu-CPAM-13}. For general densities and general $N$, the presence of discontinuities across unknown surfaces makes eq.~\eqref{SCEfunctional3} very challenging for numerical computations.
\end{remark}

\subsection{Next leading term} \label{sec:next}

We have treated so far the limit of the Levy-Lieb functional at infinite coupling strength $\lambda$ (or, equivalently, at extreme low density). One could ask how is this limit approached, or, in other words, what is the next leading term in equations~\eqref{eq.asy1}-\eqref{eq.asy2}. 

% I have commented this paragraph and introduce in the end.
%In his seminal work, Seidl \cite{Sei-PRA-99} had the intuition that this next term should be given by zero-point oscillations around the manifold parametrized by the co-motion functions. He also carried out explicit calculations in 3D for the spherically-symmetric case with $N=2$ electrons, using the co-motion function of Sec.~\ref{sec:RadSymm}. This idea was extended to the general many-electron case in \cite{GorVigSei-JCTC-09}, where it was also found that the original calculation of Seidl had a wrong factor 2. 

The strategy employed in \cite{GorVigSei-JCTC-09} to compute this next leading term relies on the assumption that the minimizer in \eqref{VP.SIL} is of the SCE or Monge type, see the detailed discussion in the previous section \ref{sec:SCEansatz}. Under this assumption, as shown in section \ref{sec:optimality} the classical potential energy
\begin{equation}\label{eq:epotmin}
    E_{\rm pot}(\bfr_1,\dots,\bfr_N)=V_{ee}(\bfr_1,\dots,\bfr_N)-\sum_{i=1}^N v_{\rm SCE}(\bfr_i),
\end{equation}
with $v_{\rm SCE}(\bfr)$ defined by Eqs.~\eqref{eq.gradpot2} and \eqref{vsce}, is minimum on the manifold $\Omega_0$ parametrised by the co-motion functions, 
\begin{equation}
\Omega_0=\{(\bfr_1,\dots,\bfr_N)\in\R^{dN} \, : \, \bfr_1=\bfr,\bfr_2={\bf f}_2(\bfr),\dots ,\bfr_N={\bf f}_N(\bfr)\}.
\end{equation}
When $\lambda$ in Eq.~\eqref{eq.LLalpha} is very large but finite, we can expect that the support of the minimizer in  Eq.~\eqref{eq.LLalpha} be strongly localised around $\Omega_0$, as illustrates Figure~\ref{fig:FcexSGS} in Sec~\ref{sec:Enl}. We can then expand $E_{\rm pot}$ around its minimum through second order. The corresponding hessian matrix $\mathbb{H}(\bfr)$ evaluated on $\Omega_0$ for any fixed $\bfr$, will have $d$ zero eigenvalues (along the manifold $\Omega_0$) and $dN-d$ positive eigenvalues. By using curvilinear coordinates along the manifold $\Omega_0$ and orthogonal to it, the sought next leading term is determined by adding the kinetic energy to the second-order expansion of $E_{\rm pot}$, which corresponds to the hamiltonian of zero-point oscillations in the space orthogonal to $\Omega_0$ \cite{GorVigSei-JCTC-09}. The final result is that Eqs.~\eqref{eq.asy1}-\eqref{eq.asy2} are extended to \cite{GorVigSei-JCTC-09,GorSei-PCCP-10} 
\begin{align} \label{eq:nexttermtot}
F_{\rm LL}[\rho^\gamma] & \; \underset{\gamma \to 0}{\sim} \; \gamma\,V_{ee}^{\rm SCE}[\rho]+\gamma^{3/2}F^{\rm ZPE}[\rho] \\
    F_\lambda & \;\underset{\lambda \to \infty}{\sim}  \; \lambda V_{ee}^{\rm SCE}[\rho]+\sqrt{\lambda}\,F^{\rm ZPE}[\rho],
\end{align}
where
\begin{equation}
    F^{\rm ZPE}[\rho]=\frac{1}{2}\int_{\mathbb{R}^d} \frac{\rho(\bfr)}{N}{\rm Tr}\left(\sqrt{\mathbb{H}(\bfr)}\right). \label{eq:ZPOfunc}
\end{equation}
In \cite{GroKooGieSeiCohMorGor-JCTC-17} this term has been computed explicitly for $N=2$ electrons in 1d and it has been compared with accurate numerical calculations for the Levy functional at very large $\lambda$, finding excellent agreement. %Very recently, a rigorous proof for Eqs.~\eqref{eq:nexttermtot}-\eqref{eq:ZPOfunc} for the many-electron 1d case has been provided by Colombo, Di Marino and Stra \cite{ColDiMStra-arxiv-21}.

The intuition that the next term of the Levy-Lieb functional at infinite coupling strength $\lambda$ should be given by zero-point oscillations around the manifold parametrized by the co-motion functions appeared for the first time in Seidl's seminal work \cite{Sei-PRA-99}. He also carried out explicit calculations in 3D for the spherically-symmetric case with $N=2$ electrons, using the co-motion function introduced in Sec.~\ref{sec:RadSymm}. This idea was extended to the general many-electron case in \cite{GorVigSei-JCTC-09}, where it was also found that the original calculation of Seidl had a wrong factor 2. Very recently, a rigorous proof for Eqs.~\eqref{eq:nexttermtot}-\eqref{eq:ZPOfunc} for the many-electron $1d$ case has been provided by Colombo, Di Marino and Stra \cite{ColDMaStra-arxiv-21}.

\subsubsection{The fermionic statistics}
Equations~\eqref{eq:nexttermtot}-\eqref{eq:ZPOfunc} are the first-order correction due to kinetic energy in the large-$\lambda$ (or $\hbar\to 0$) limit of the Levy-Lieb functional. This correction is still independent of the particle statistics. A natural question to ask is then at which order will the fermionic antisymmetry enter. 

In Refs.~\cite{GorVigSei-JCTC-09,GorSeiVig-PRL-09} it has been conjectured that the particle statistics enters in the $\lambda\to\infty$ limit at orders $\sim e^{-\sqrt{\lambda}}$. The physical intuition behind this idea is simply that the effect on the energy of antysmmetrization vanishes as the overlap between gaussians centerd at each set of strictly-correlated positions (each $\bfr$ value in $\Omega_0$). The scaling $\sqrt{\lambda}$ of such gaussians comes from the zero-point hamiltonian. This conjecture has been confirmed numerically \cite{GroKooGieSeiCohMorGor-JCTC-17} for the case of $N=2$ electrons in 1D, again by comparison with accurate numerical calculations of the exact Levy functional at large $\lambda$.

\subsection{The strongly interacting limit of DFT from the point of view of optimal transport} \label{sec:OT}

We now introduce a fruitful interpretation of the strongly interacting limit of DFT as ``optimal transport with Coulomb cost''. 

Optimal transport theory (see \cite{RacRus-BOOK-98, Vil-BOOK-03, San-Book-15, Fri-Book-22} for textbook accounts) is concerned with the following two problems, introduced in special cases in fundamental work by Kantorovich \cite{Kan-DAN-42} respectively Monge \cite{Mon-BOOK-1781}: \vspace*{2mm}

a) {\it Kantorovich optimal transport problem:} For given probability measures $\mu_1,...,\mu_N$ defined on closed subsets $X_1,...,X_N$ of $\R^d$, find a joint probability measure $\Pi$ on the product space $X=X_1\times ... \times X_N\subseteq\R^{Nd}$ which minimizes a cost functional
$$
   {\cal C}[\Pi] = \int_{X} c(\bfr_1,...,\bfr_N) \, d\Pi(\bfr_1,...,\bfr_N)
$$
subject to the marginal constraints 
$$
   \int_{X_1\times ... \times X_{i-1} \times A_i \times X_{i+1}\times ... \times X_N} d\Pi = \int_{A_i} 
   d\mu_i\mbox{ for all measurable sets }A_i\subseteq X_i \mbox{ and all }i\in \{1,...,N\}.
$$
Here $c \, : \, X_1\times ... \times X_N\to\R\cup\{+\infty\}$ is some given cost function, and validity of the above constraint is denoted $\Pi\mapsto \mu_1,...,\mu_N$. 
\vspace*{2mm}

b) {\it Monge optimal transport problem:} For given probability measures $\mu_1,...,\mu_N$ defined on measurable subsets $X_1,...,X_N$ of $\R^d$ of positive volume which possess integrable densities $p_1,...,p_N$ (i.e. $p_i\in L^1(X_i)$), and a cost function $c$ as above, find measurable maps $f_1,...,f_{N-1}$ with $f_i \, : \, X_1\to X_{i+1}$ which minimize
$$
   I[f_1,...,f_{N-1}] = \int_{X_1} c\bigl(\bfr_1,f_1(\bfr_1),...,f_{N-1}(\bfr_N)\bigr) \, d\mu_1
$$
subject to the marginal constraints
$$
   f_i{}_\sharp p_1 = p_{i+1} \;\;\; \mbox{ for }i\in\{1,...,N-1\}.
$$
This corresponds to making the ansatz
\begin{equation} \label{eq.Mongeans}
   d\Pi(\bfr_1,...,\bfr_N) = d\mu_1(\bfr_1) \delta\bigl(\bfr_2-f_1(\bfr_1)\bigr) \cdots \delta\bigl(\bfr_N-f_{N-1}(\bfr_1)\bigr) d\bfr_2 ... d\bfr_N
\end{equation}
or equivalently -- using the notion of push-forward introduced in \eqref{eq.push} --  
\begin{equation} \label{eq.Mongeans2}
  \Pi = (id,f_1,...,f_N)_\sharp \mu_1
\end{equation}
in the Kantorovich problem, where $id$ denotes the identity map $id(\bfr_1)=\bfr_1$.
\vspace*{1mm}

\begin{example}  \label{ex:coul} {\rm ($N$ equal marginals, Coulomb cost)} If we take  
$$
   X_1=...=X_N=\R^d, \;\;\; \mu_1=...=\mu_N=\frac{\rho}{N}, \;\;  c(\bfr_1,...,\bfr_N)=\sum_{1\le i<j\le N}\frac{1}{|\bfr_i-\bfr_j|} 
$$
the Kantorovich optimal transport problem is precisely the SIL variational problem, \eqref{VP.SIL}, and the Monge optimal transport problem is precisely the SCE variational problem, \eqref{SCE'}.
\end{example}
Thus the strongly interacting limit of DFT can be viewed as {\it  optimal transport with Coulomb cost}. This viewpoint, introduced by Buttazzo, DePascale, Gori-Giorgi, Cotar, Friesecke, and Kl\"uppelberg \cite{ButDepGor-PRA-12, CotFriKlu-CPAM-13},  opened the door to much of the current understanding of the strong-interaction limit of DFT. 

%The subtle and incompletely understood relationship between the SIL and the SCE problem is a special case of the subtle and incompletely understood relationship between the Kantorovich and the Monge formulation of optimal transport. 
%
%

\begin{example} \label{ex:proto} {\rm (Two unequal marginals, positive power cost)} The prototype problem of classical optimal transport theory going back to {\rm \cite{Kan-DAN-42, Mon-BOOK-1781}} is to instead take
$$
  N=2, \;\; X_1=X_2=\R^d, \;\; c(\bfr_1,\bfr_2)=|\bfr_1-\bfr_2|^p, \;\; p\ge 1.
$$
That is, one considers \\
-- only two marginals \\
-- unequal instead of equal marginals \\
-- a positive instead of a negative power of the euclidean distance as cost. \\
Denoting $\mu_1=\mu$, $\mu_2=\nu$, $f_1=T$, $\bfr_1=x$, $\bfr_2=y$, the Kantorovich problem then becomes 
\begin{equation} \label{eq.Kant}
   \mbox{Minimize }{\cal C}[\Pi] = \int_{\R^d\times\R^d} |x-y|^p d\Pi(x,y) \mbox{ over }\Pi\in {\cal P}(\R^{2d}) \mbox{ subject to }\Pi\mapsto\mu,\,\nu
\end{equation}
and the Monge problem becomes
\begin{equation} \label{eq.Monge}
  \mbox{Minimize } I[T] = \int_{\R^d} |x-T(x)|^p d\mu(x) \mbox{ over measurable maps }T\, : \, \R^d\to\R^d \mbox{ subject to }T_\sharp \mu = \nu.
\end{equation}
The analogon of the SCE functional is the optimal cost as a functional of the two prescribed marginals,
$$
   C_{\rm opt}[\mu,\nu] = \min\{{\cal C}[\Pi] \, : \, \Pi\mapsto\mu,\,\nu\} = \inf \{ I[T]\, : \, T_\sharp\mu = \nu\}.
$$
Its $p$-th root, $W_p(\mu,\nu)=(C_{\rm opt}[\mu,\nu])^{1/p}$, is the celebrated $p$-Wasserstein distance, which is a metric on the space of probability measures.  

\end{example}
Thus the SCE functional can be thought of as the {\it Coulomb analogue of the Wasserstein distance}.

We remark that the motivation of Monge and Kantorovich for considering Example \ref{ex:proto} came from civil engineering respectively economics, and explains the name {\it optimal transport}: Monge thought of moving a given pile of sand on a construction site into a given hole in a way that minimizes  the overall distance of transport, with $T(x)$ describing the target position of sand originally located at $x$ and with pile and hole modelled, respectively, by $\mu$ and $\nu$. Kantorovich thought of transporting some economic good, say steel, from producers (steel mines) to consumers (factories), at minimal transportation cost; $\Pi(x,y)$ then describes the density of goods transported from location $x$ to location $y$, and is called a {\it transport plan}. In the latter context it is natural {\it not} to make the Monge ansatz
$$
        d\Pi(x,y)=d\mu(x)\delta\bigl( y-T(x) \bigr)dx
$$
but instead allow one producer located at $x$ to supply several consumers located at different positions $y$, i.e. consider the general problem \eqref{eq.Kant}. 

The general question for which costs and marginals the Monge and Kantorovich problems are equivalent, i.e. the Kantorovich problem admits minimizers of Monge form, is not well understood. A sufficient condition \cite{San-Book-15} for $N=2$ (and, say, compact convex sets $X_1$ and $X_2$ and continuously differentiable costs $c$) is that the marginal measure $\mu_1$ is absolutely continuous and $c$ satisfies the so-called twist condition that the map $\bfr_2 \mapsto \nabla_{\bfr_1}c(\bfr_1,\bfr_2)$ be injective for every $r_1$. 
%This almost holds for the Coulomb cost except for the singularity at $\bfr_1=\bfr_2$.
For $N>2$, generalized twist conditions have been studied by Pass \cite{Pas-Thesis-11,Pas-DCDS-14,Pas-ESAIM-15}; unfortunately these are not satisfied for the Coulomb cost.

%Optimal transport problems also arise naturally in many other fields. For instance, in probability theory and statistics one is interested -- given $N$ $\R^d$-valued random variables $Z_i$ with known laws $\mu_i$ -- in the underlying joint law (or ''coupling'') $\Pi$ which is maximally correlated in the sense of minimizing the expectation %$$
%      {\mathbb E}\Bigl(\sum_{1\le i<j\le N}|Z_i-Z_j|^2\Bigr).
%$$
%This corresponds to the general Kantorovich problem with harmonic oscillator cost $c(\bfr_1,...,\bfr_N)=\sum_{1\le i<j\le N} |\bfr_i-\bfr_j|^2$. 

%From the point of view of the strongly correlated limit of DFT, the following general insights from optimal transport theory are particularly useful: \\
%-- Kantorovich duality \\
%-- optimality conditions for optimizers. \\
%-- the twist condition. \\
%Let us first state these insights in general form, and then see what they yield for the SCE functional.  

%We state here only fundamental versions which are available in textbook form, and put the reader on notice that a great many modifications in numerous directions, often requiring highly technical proofs, were worked out in the mathematics literature. 
%-- ranging from replacing the $X_i$ by abstract spaces and the $\mu_i$ to finitely additive set functions, modifying the admissible class of potentials, weakening the assumptions on the cost function at the expense of strengthening those on the $X_i$ and $\mu_i$, and often worked out only for two marginals even though the analysis could be extended to $N$ marginals. 

\subsection{Dual construction of the SCE functional} \label{sec:Kantdual}
We now introduce a fourth -- dual -- construction of the SCE functional.

A cornerstone principle of optimal transport theory, {\it Kantorovich duality}, says that the minimum of a given Kantorovich optimal transport problem (see section \ref{sec:OT}) equals the supremum of an associated explicit dual problem. The general form of the dual is recalled in Appendix \ref{App:Kantdual}. For the SIL problem \eqref{VP.SIL}, the dual problem is the following (see Appendix \ref{App:Kantdual} for a quick derivation from general OT theory): maximize the functional 
\begin{equation} \label{eq:dualfctnal}
      J[u] = \sum_{i=1}^N \int_{\R^d} u(\bfr) \, \rho(\bfr) \,  d\bfr
\end{equation}
over potentials $u \, : \, \R^d\to\R$ which must satisfy the pointwise constraint
\begin{equation} \label{eq.dualconstr2}
      \sum_{i=1}^N u(\bfr_i) \le V_{ee}(\bfr_1,...,\bfr_N) \;\; \forall (\bfr_1,...,\bfr_N)\in\R^{dN}.
\end{equation}
Maximization is over the admissible class 
\begin{equation} \label{eq.A}
  {\mathcal A} = \{ u \, : \, \R^d\to\R \, \Big| \, u \mbox{ bounded and measurable}, \, u \mbox{ satisfies \eqref{eq.dualconstr2}} \}.
\end{equation}
This yields the following alternative definition of the SCE functional: 
\begin{equation} \label{eq.SCEdual}
    V_{ee}^{\rm SCE}[\rho] = \sup_{u\in{\mathcal A}} \int_{\R^d}\, u(\bfr) \, \rho(\bfr) \, d\bfr.
\end{equation}
This construction is due to Buttazzo, DePascale, and Gori-Giorgi \cite{ButDepGor-PRA-12}. Note that the optimization here is not over $N$-point densities, but over (suitable) external potentials $u$. Optimizers are called {\it Kantorovich potentials}. Heuristically, they can be thought of as Lagrange multipliers associated with the marginal constraints in the original problem \eqref{VP.SIL}. This is explained in our discussion of optimality conditions in section \ref{sec:optimality}.

It can be rigorously shown that the new construction yields, again, the SCE functional, and that optimal potentials exist: 

\begin{theorem} \label{T:SCE-dual} Let $\rho \, : \, \R^d\to\R$ be any $N$-particle density in the class ${\cal D}^N$ (see \eqref{eq.DN}), and let $w(\bfr)=|\bfr|^{-1}$ be the Coulomb interaction. Then: \\[1mm]
{\rm (1) \cite{ButDepGor-PRA-12}}~The supremum in \eqref{eq.SCEdual} is equal to the minimum in \eqref{eq.SIL2}. 
\\[1mm]
{\rm (2) \cite{ButDepGor-PRA-12, Dep-ESAIMMMNA-15}}~The supremum in \eqref{eq.SCEdual} is attained; that is, there exists a maximizing potential $u$ in the class \eqref{eq.A}. 
\\[1mm]
{\rm (3)\cite{DMaGerNen-TOOAS-17, ButChaDeP-AMO-18}}~If, in addition, $\rho>0$ everywhere, there exists a maximizing potential which is in addition Lipschitz continuous.
\end{theorem}
Statement (1) follows directly from the general Kantorovich duality theorem of OT theory; see Appendix \ref{App:Kantdual}. The question of existence and regularity of optimal potentials is more delicate. Note that the Coulomb potential $V_{ee}$ which upper-bounds $u(\bfr_1)+...+u(\bfr_N)$ tends to plus infinity as the distance $\bfr_i-\bfr_j$ between any two position coordinates goes to zero; so one might a priori think that $u$'s are favourable which also tend to plus infinity at certain places. But statement (2) in the  above theorem says that this does not happen; the existence proof of bounded optimal potentials is due to \cite{ButDepGor-PRA-12} for $N=2$ and to \cite{Dep-ESAIMMMNA-15} for general $N$.

{\bf Quantum analogue.} We remark that the dual construction of the SCE functional in eq.~\eqref{eq.SCEdual} admits a quantum analogue. In \cite{Lie-IJQC-83}, Lieb proposed an extension of the Levy-Lieb functional \eqref{eq.FHK} to mixed states, i.e. $\mathrm{F}_{\rm{L}}:\mathcal{D}_N\to\R$, 
\begin{equation}\label{eq:FL}
\mathrm{F}_{\rm{L}}[\rho] = \min\left\lbrace \rm{Tr}\left(-\frac{1}{2}\sum^N_{j=1}\Delta_{\bfr_j} + V_{ee}(\bfr_1,...,\bfr_N)\right)\Gamma \, : \, \Gamma=\Gamma^{*}\geq 0, \rm{Tr}(\Gamma)=1, \Gamma\mapsto\rho \right\rbrace, 
\end{equation}
where $\Gamma$ is an operator acting on the fermionic Hilbert space and, similarly to \eqref{eq:rho1}, $\Gamma\mapsto\rho$ denotes the relation $\rho = N \int_{\R^{dN}} \Gamma(\bfr,\bfr_2,\dots,\bfr_N;\bfr,\bfr_2,\dots,\bfr_N)\,  d\bfr_2\dots d\bfr_N$. 
In \cite{Lev-PRA-82}, M. Levy introduced a similar functional requiring in addition that $\Gamma = |\psi\rangle\langle \psi |$ be a rank-one operator. An advantage of the Lieb functional $F_{\rm{L}}$ is that it is convex. Ignoring issues of rigor, \eqref{eq:FL} admits a dual formulation
\begin{equation}\label{eq:FLdual}
F_{\rm{L}}[\rho] = \sup\left\lbrace \int_{\R^{3}}u(\bfr)\rho(\bfr)\, d\bfr \, : \, \sum^N_{i=1}u(\bfr_i) \le -\frac{1}{2}\sum^N_{j=1}\Delta_{\bfr_j} + V_{ee}(\bfr_1,...,\bfr_N) \right\rbrace,
\end{equation}
with the above inequality understood in the sense of self-adjoint operators. For a rigorous discussion of eq.~\eqref{eq:FLdual} see the Chapter by Lewin, Lieb, and Seiringer. This equation is the quantum analogue (for mixed states) of the dual construction of the SCE functional. Note that because the right hand side of the constraint on $u$ now contains an additional positive term, the value of the supremum will be higher than in \eqref{eq.SCEdual}, as it should.

\subsection{Optimality conditions} \label{sec:optimality} With the help of Kantorovich duality one obtains very interesting necessary conditions for solutions to the SIL variational principle \eqref{VP.SIL}. 
%, as pointed out in \eqref{eq:epotmin}. 
In particular, for optimizers of SCE (alias Monge) form one can express the gradient of the Kantorovich potential $u$ in terms of the co-motion functions (alias transport maps). 

We follow the rigorous presentation for general OT problems in \cite{Fri-Book-22}, but specialize throughout to the SIL problem. For the benefit of less mathematically minded readers, we also include a heuristic derivation at the end of this section.

\begin{theorem} \label{T:Optimality} {\rm (Optimality conditions \cite{Fri-Book-22})} Let $\rho \, : \, \R^d\to\R$ be any $N$-particle density in the class ${\cal D}^N$ (see \eqref{eq.DN}). Let $V_{ee} \, : \, \R^{dN}\to \R\cup\{+\infty\}$ be any interaction potential which is symmetric, bounded from below, lower semi-continuous, and has the property that the minimum in \eqref{eq.SIL2} is finite. Suppose $\Pi$ is a solution to the SIL problem \eqref{eq.SIL2}, and $u$ is a solution to the dual problem, i.e. a maximizer of $J$ in the class ${\cal A}$. 
\\[1mm]
{\rm (1)} $\Pi$ is zero outside the set 
$$
   {\mathcal M}=\{ (\bfr_1,...,\bfr_N)\in \R^{dN} \, : \, V_{ee}(\bfr_1,...,\bfr_N) - \sum_{i=1}^N u(\bfr_i) = \mbox{\rm min}\}.
$$
{\rm (2)} $\Pi$ is an unconstrained minimizer (i.e., a minimizer on ${\cal P}(\R^{dN})$) of the modified functional 
$$
  {\cal L}[\Pi] = \int_{\R^{dN}}\Bigl( V_{ee}(\bfr_1,...,\bfr_N)-\sum_{i=1}^N u(\bfr_i)\Bigr) \, d\Pi(\bfr_1,...,\bfr_N).
$$
{\rm (3)} At any point $(\bfr_1,...,\bfr_N)$ in ${\cal M}$ where the function in {\rm (1)} is differentiable with respect to $\bfr_1$, 
\begin{equation} \label{eq.gradpot1}
    \nabla u(\bfr_1) = \nabla_{\bfr_1}V_{ee}(\bfr_1,...,\bfr_N).
\end{equation}
In particular, if $\Pi$ is of SCE form, \eqref{eq.psiSCE}, and $V_{ee}(\bfr_1,...,\bfr_N)$ is the Coulomb interaction $\sum_{1\le i<j\le N}\tfrac{1}{|\bfr_i-\bfr_j|}$, 
\begin{equation} \label{eq.gradpot2}
    \nabla u(\bfr) = - \sum_{i=1}^{N-1} 
    \frac{\bfr - f_i(\bfr)}{|\bfr - f_i(\bfr)|^3} \;\; \mbox{ at any point }\bfr\mbox{ where $u$ is differentiable and  }\rho(\bfr)>0.
\end{equation}
\end{theorem}

The physical and mathematical meaning of these results is as follows.

(1) says that the classical potential energy 
$$
   E_{\rm pot}(\bfr_1,...,\bfr_N)=V_{ee}(\bfr_1,...,\bfr_N)-\sum_i u(\bfr_i)
$$
is minimal on the manifold of configurations which occur with nonzero probability under the optimal plan $\Pi$. In particular, when $\Pi$ is of the SCE or Monge type, the classical potential energy is minimal on the manifold \eqref{eq:Omega0} parametrized by the co-motion functions. Besides being interesting in its own right, this underlies the derivation of the next leading term of the Levy-Lieb functional outlined in section \ref{sec:next}.  

(3) says that the Kantorovich potential $u$ is an {\it effective one-body potential emulating the many-body system}, in the following sense: its gradient at the point $r$ is precisely the classical repulsive force exerted on an electron at $\bfr$ by the other electrons at positions $f_i(\bfr)$. Eq.~\eqref{eq.gradpot2} is called the {\it force equation}.

(2) can be viewed as an infinite-dimensional Lagrange multiplier rule, with any Kantorovich potential (i.e. any optimizer of the dual variational principle \eqref{eq.SCEdual}) playing the role of a Lagrange multiplier associated with the constraint $\Pi\mapsto\rho$.

We remark that results of the above form have a long history in OT theory; for the two-marginal problem with interaction potential  $|\bfr_1-\bfr_2|$ respectively $|\bfr_1-\bfr_2|^2$, (1) goes back to Kantorovich himself \cite{Kan-DAN-42}, while the differential version (3) and its usefulness were first realized by Knott and Smith \cite{Knott-Smith-84}. 
%As in the case of Kantorovich duality, optimality conditions were subsequently revisited and generalized by many workers over a long time, with the complete picture given above emerging more recently. 

{\bf Proof} The following proof, taken from \cite{Fri-Book-22}, is simple and illuminating, so we include it. By Kantorovich duality (in the form of Theorem \ref{T:SCE-dual}~(1)), 
$$
     0 = \int_{\R^{dN}} V_{ee} \, d\Pi - \sum_{i=1}^N  \int_{\R^d} u(\bfr_i) \frac{\rho(\bfr_i)}{N} \, d\bfr_i.
$$
Since $\Pi$ has equal marginals $\frac{\rho}{N}$, $\int_{\R^d} u(\bfr_i)\, \frac{\rho(\bfr_i)}{N} \, d\bfr_i =\int_{\R^{dN}} u(\bfr_i) \, d\Pi(\bfr_1,...,\bfr_N)$, and so 
$$
    0 = \int_{\R^{dN}} \Bigl( V_{ee}(\bfr_1,...,\bfr_N) - \sum_i u(\bfr_i)\Bigr) d\Pi(\bfr_1,...,\bfr_N).
$$
But since $u$ satisfies the constraint \eqref{eq.dualconstr} at every point in $\R^{dN}$, the integrand is nonnegative. So the minimum value of the integrand must be zero and attained, and $\Pi$ must vanish wherever the integrand is positive. This establishes (1) and (2). The elementary calculus fact that the gradient of a differentiable function vanishes at minimum points now yields \eqref{eq.gradpot1}. Finally, \eqref{eq.gradpot2} follows since the point $\bigl(\bfr_1,f_1(\bfr_1),...,f_{N-1}(\bfr_1)\bigr)$ belongs to ${\cal M}$ whenever the density $\rho$ is positive at $\bfr_1$. 

We complete this section with a more heuristic derivation of the optimality conditions. 

{\bf Heuristic derivation of Theorem \ref{T:Optimality}} Let us re-write the SIL variational principle \eqref{VP.SIL} in the form
\begin{equation}\label{eq:LM}
   \mbox{Minimize }{\cal C}[\Pi] = \int_{\R^{dN}} V_{ee} \,  d\Pi \mbox{ subject to the constraints }G^{(\bfr_1)}[\Pi]=\rho(\bfr_1) \, \forall \bfr_1\in\R^d
\end{equation}
where $G^{(\bfr_1)}[\Pi]$ is the functional which assigns to an $N$-point probability measure $\Pi$ the value of its single-particle density at the point $\bfr_1$, and where the minimization is over symmetric probability measures (see Remark \ref{R:symmetrization}). Let us now postulate the existence of a family of Lagrange multipliers $(\lambda(\bfr_1))_{\bfr_1\in\R^d}$, one for each $G^{(\bfr_1)}$, such that minimizers of ${\cal C}$ subject to the constraints $G^{(\bfr_1)}[\Pi]=\rho(\bfr_1)$ are unconstrained minimizers of the Lagrangian
$$
  {\cal L}[\Pi] = {\cal C}[\Pi] - \int_{\R^d} \lambda(\bfr_1) \, G^{(\bfr_1)}[\Pi] d\bfr_1.
$$
But since $G^{(\bfr_1)}[\Pi]$ is the one-body density of $\Pi$,  and $\Pi$ is symmetric, 
\begin{eqnarray}
  {\cal L}[\Pi] &=&  \int_{\R^{dN}} \Big[ V_{ee}[\bfr_1,...,\bfr_N] - \lambda(\bfr_1) \, N \Big] d\Pi(\bfr_1,....,\bfr_N) \nonumber \\
  &=&
  \int_{R^{dN}} \Big[ V_{ee}(\bfr_1,...,\bfr_N) - \sum_{i=1}^N\lambda(\bfr_i)\Big] d\Pi(\bfr_1,...,\bfr_N), \label{eq:LM2}
\end{eqnarray}
so the Lagrangian coincides with the functional in (2) with $\lambda=u$. It is clear that minimizers of the Lagrangian must be concentrated on the set of pointwise minimizers of the integrand, yielding (1). Statement (3) now follows as in the rigorous proof.

The above argumentation obtains the Kantorovich potential $u$ quickly but non-rigorously as a Lagrange multiplier. In fact, with such a more heuristic construction of $u$, statements (1) and (3) were already derived  in \cite{SeiGorSav-PRA-07} before the discovery of the {\it SCE theory/optimal transport} connection. 

But readers are put on notice that there is no such thing as a general and rigorous Lagrange multiplier rule which would guarantee existence of Lagrange multipliers for infinite-dimensional non-smooth problems like Levy-Lieb constrained search or its strongly interacting limit \eqref{VP.SIL}. In DFT (in its original form with both kinetic energy and electron repulsion present), the existence problem for Lagrange multipliers -- i.e., the existence of one-body potentials which, when added to the Hamiltonian $T+V_{ee}$, reduce a constrained search to an unconstrained search -- is known as the {\it $v$-representability problem}. This is a longstanding open problem, see e.g.  \cite{Hel-book-16,Koh-PRL-83,Lev-PNAS-79,Lie-IJQC-83, Lee-AQC-03}.  
For variants of the problem at positive temperature respectively quantum lattices see \cite{ChaCha-JSP-84, ChaChaLie-CMP-84}; $v$-representability for a regularization of the exact Levy-Lieb functional is discussed in the chapter by Kvaal.  

%It is then quite remarkable that in the strongly interacting limit, the $v$-representability problem, alias the problem of existence of Lagrange multipliers, can be completely solved, by combining Theorem \ref{T:Optimality} (2) with an existence theorem for the dual variational principle \eqref{eq.SCEdual}. This is explained in the next section.

\subsection{Solution of the purely-interacting $v$-representability problem} \label{sec:vrep}

We now show that in the strongly interacting limit the $v$-representablity problem, alias the problem of existence of Lagrange multipliers for density functionals defined by constrained search, can be completely solved. As we will see, this fact follows by combining known results. 
We assume in this section that $w(\bfr)=|\bfr|^{-1}$ is the Coulomb interaction. 

Recall that a density $\rho \, : \, \R^d\to\R$ is called 
\begin{itemize}
    \item {\it $N$-representable} if it comes from a wave function $\Psi\in {\cal W^N}$ (i.e. $\Psi\mapsto\rho$)
    \item {\it $v$-representable} 
    if it comes from a minimizer of $\langle \Psi | T + V_{ee} + \sum_i v(\bfr_i) | \Psi \rangle$ on ${\cal W^N}$ for some potential $v\, : \, \R^d\to\R$
    \item {\it non-interacting $v$-representable} if it comes from a minimizer of $\langle \Psi | T + \sum_i v(\bfr_i) | \Psi \rangle$ on ${\cal W^N}$ for some potential $v\, : \, \R^d\to\R$
    \item {\it purely-interacting $v$-representable} if it comes from a minimizer of $\int_{\R^{dN}} \bigl(V_{ee} + \sum_i v(\bfr_i)\bigr)d\Pi$ on ${\cal P}(\R^{dN})$ for some potential $v\, : \, \R^d\to\R$.  
\end{itemize}    

\begin{theorem} \label{T:vrep} ($N$-representability implies purely-interacting $v$-representability) Any $N$-representable $\rho$, i.e. any $\rho$ belonging to the class ${\cal D}^N$ (see \eqref{eq.DN}), is purely-interacting $v$-representable by some bounded measurable potential $v\, : \, \R^d\to\R$. Explicitly, the following choice will do: 
\begin{equation} \label{eq.vrep}
  v = -u
\end{equation}
where $u$ is any bounded Kantorovich potential for $\rho$ (see Theorem \ref{T:SCE-dual} for existence of the latter). 
%If $\rho$ is in addition everywhere positive, $u$ and hence the above $v$ can be taken to be continuous.
\end{theorem}

%\gero{Augusto, is this also true for Lipschitz potentials? See my query below the next theorem.}\augusto{Let is discuss}

This result is quite remarkable, given that -- to our knowledge --
% nothing interesting
not much
is known on the rigorous level off the strongly interacting limit.

% It is an open question whether, under reasonable conditions on $\rho$, the potential $v$ is unique up to an additive constant.

%In general, we do not expect maximizers to be unique. \gero{Why not? Does anyone have a counterexample with everywhere-positive $\rho$?} 

{\bf Proof of Theorem \ref{T:vrep}} By Theorem \ref{T:SCE-dual}~(2), there exists a bounded maximizer $u$ of the dual functional, i.e. an associated Kantorovich potential.
% which can moreover be taken Lipschitz if $\rho$ is everywhere positive
Let $v=-u$. 
By Theorem \ref{T:basic} (1), there exists a minimizer $\Pi[\rho]$ of the SIL variational problem \eqref{VP.SIL}. By Theorem \ref{T:Optimality} (2), this $\Pi[\rho]$ is a minimizer of $\int_{R^{dN}}\bigl(V_{ee}+\sum_i v(\bfr_i)\bigr)d\Pi$ on ${\mathcal P}(\R^{dN})$. 
Since by construction $\Pi$ has density $\rho$, it follows that $v$ represents $\rho$. 

If in addition $\rho>0$ everywhere, the above proof together with Theorem \ref{T:SCE-dual} (3) shows that $\rho$ is even purely-interacting $v$-representable by some Lipschitz continuous potential.

\subsection{Functional derivative and SCE potential} \label{sec:vSCE}
It is not difficult to deduce from Theorem \ref{T:vrep} that when the density $\rho$ is  sufficiently nice (say, continuous and everywhere positive) and the Kantorovich potential $u[\rho]$ (i.e. the maximizer of the dual problem \eqref{eq.SCEdual}) is unique, the SCE functional is functionally differentiable at $\rho$ with functional derivative
\begin{equation} \label{SCEpot}
     \frac{\delta V_{ee}^{\rm SCE}[\rho]}{\delta \rho}[\rho] = u[\rho] + const
\end{equation}
where $const$ is an arbitrary additive constant. Here for any functional $F$ on ${\cal D}^N$ the functional derivative $\frac{\delta F}{\delta \rho}[\rho]$ at some density $\rho$ (if it exists) is defined by the requirement
$$
    \frac{d}{dt}F[\rho + t\,\eta]\Big|_{t=0} = \int_{\R^d} \frac{\delta F}{\delta\rho}[\rho](\bfr)\, \eta(\bfr) \, d\bfr 
$$
for all smooth mass-preserving localized perturbations $\eta \, : \, \R^d\to\R$ (mathematically: $\eta\in C_0^\infty(\R^d)$, $\int \eta = 0$), and is unique up to an additive constant. For an informal derivation of eq.~\eqref{SCEpot} see e.g. \cite{CheFriMen-JCTC-14}, and for a rigorous proof under suitable assumptions see \cite{DMaGer-arXiv-2020}. 

As for any Hartree-exchange-correlation functional, the  Hartree-exchange-correlation {\it potential} associated to the SCE functional is the functional derivative with additive constant chosen so that the potential vanishes at infinity, in our case
\begin{equation} \label{vsce}
    v_{\rm SCE}[\rho](\bfr) = u[\rho](\bfr) + C[\rho], \;\;\;\;
    C[\rho]\mbox{ a constant that ensures }\lim_{|\bfr|\to\infty}v_{\rm SCE}[\rho](\bfr) = 0.
\end{equation}
This functional derivative is called the SCE potential. 

To summarize: {\it the SCE potential for the strongly correlated limit of DFT agrees up to a shift with the Kantorovich potential from optimal transport theory}. 

Assume now that the density is everywhere positive, that the ground state of \eqref{VP.SIL} is an SCE state, and that 
\begin{equation} \label{eq.staybdd}
    \mbox{the values  $f_1(\bfr)$,...,$f_{N-1}(\bfr)$ stay in a bounded region as $|\bfr|\to\infty$}. 
\end{equation} 
It then follows from \eqref{eq.gradpot2} that the SCE potential has the correct asymptotic behaviour
\begin{equation} \label{eq.vSCEasy}
   v_{\rm SCE}[\rho](\bfr) \sim_{|\bfr|\to\infty} \frac{N-1}{|\bfr|}.
\end{equation}
By contrast, Hartree-exchange-correlation potentials for all semilocal functionals (LDA, GGAs) are well known to have the wrong asymptotics on physical (i.e. exponentially decaying) densities,
\begin{equation} \label{eq.vLDAasy}
   v_{\rm Hxc}^{\rm semiloc}[\rho](\bfr) \sim_{|\bfr|\to\infty} \frac{N}{|\bfr|}.
\end{equation}
{\bf Open problem.} Rigorously justify \eqref{eq.staybdd}, and hence \eqref{eq.vSCEasy}, for general densities $\rho$. Note that for $N\! =\! 2$ and radial densities, or any $N$ and arbitrary densities in one dimension, assumption \eqref{eq.staybdd} follows from the explicit formulae for the $f_i$ in \cite{CotFriKlu-CPAM-13} respectively \cite{ColDepDim-CJM-15}. 

\begin{figure}[http] 
\begin{center}
   {\includegraphics[height=50mm]{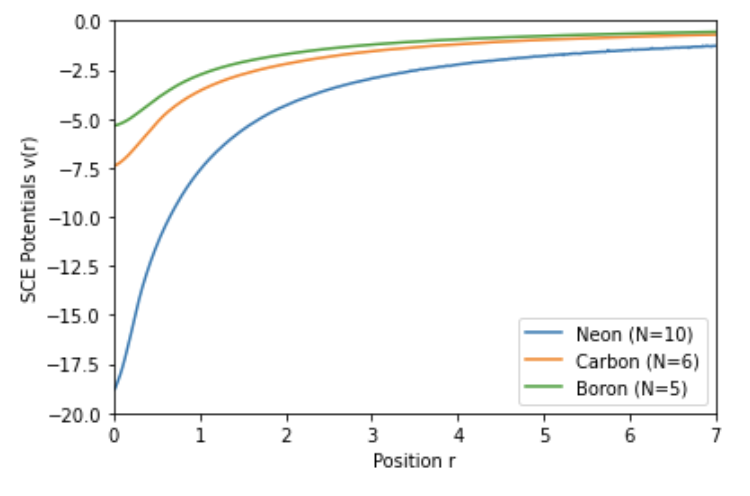}}
\end{center}
\vspace*{-3mm}

\caption{SCE potentials $v_{SCE}(r)$ corresponding to the (radially symmetric) densities of Neon ($N=10$), Carbon ($N=6$) and Boron $(N=5)$, Fig. 9 in \cite{SeiGorSav-PRA-07}. Data obtained by the following steps: (i)~compute the density $\rho(r)$ by an accurate full CI or quantum Monte Carlo computation; (ii)~compute the SGS maps corresponding to $\rho$ as described in section \ref{sec:RadSymm} below; (iii) obtain the corresponding SCE potentials $v_{SCE}$ via eq.~\eqref{eq.gradpot2} and \eqref{vsce}.}
\label{F:SCEpotentials}
\end{figure}

\begin{example}\label{eq:scepot} %\gero{I re-wrote this example, correcting the normalization of $\rho$ which was off by a factor 2, adjusting the notation ($v_{SCE}$ instead of $u$), and moving the computation of the co-motion function to Example 2.17, where it makes more sense as it needs the Seidl construction.} 

Let $N=2$, and let $\rho(r) = 2/\pi(1+r^2)$ be the one-dimensional Lorenzian density, normalized so that $\int\rho = 2.$ The co-motion function $f_1=f$ can be computed explicitly and is given by $f(r) = -1/r$, see Example \ref{ex:SCELor} in section \ref{sec:1D}. The SCE potential must satisfy the differential equation \eqref{eq.gradpot2} which in our case reads
\[
v_{{\rm SCE}}'(r)\;=\;\frac{{\rm sgn}(r)}{\big[r-f(r)\big]^2}\;=\;{\rm sgn}(r)\,\frac{r^2}{(r^2+1)^2}.
\]
The boundary condition $v_{SCE}(r)\to0$ for $r\to\infty$ (eq.~\eqref{vsce}) yields the solution 
\[
v_{{\rm SCE}}(r)\;=\;\frac{{\rm sgn}(r)}2\left[\arctan r\,-\,\frac{r}{r^2+1}\right]\,-\,\frac{\pi}{4}.
\]
\end{example}

\subsection{Strictly correlated electrons in one dimension} \label{sec:1D}

In one dimension the strong interaction limit (eq.~\eqref{VP.SIL}) can be solved exactly. The minimizing probability measure is given by an SCE state \eqref{eq.psiSCE} with explicit co-motion functions alias transport maps. The minimizer was found by Seidl himself in the original paper \cite{Sei-PRA-99}, on grounds of physical intuition. A proof of its optimality was found much later by Colombo, De Pascale and Di Marino \cite{ColDepDim-CJM-15}. 

\begin{figure}[http!]
\begin{center}
    \includegraphics[width=\textwidth]{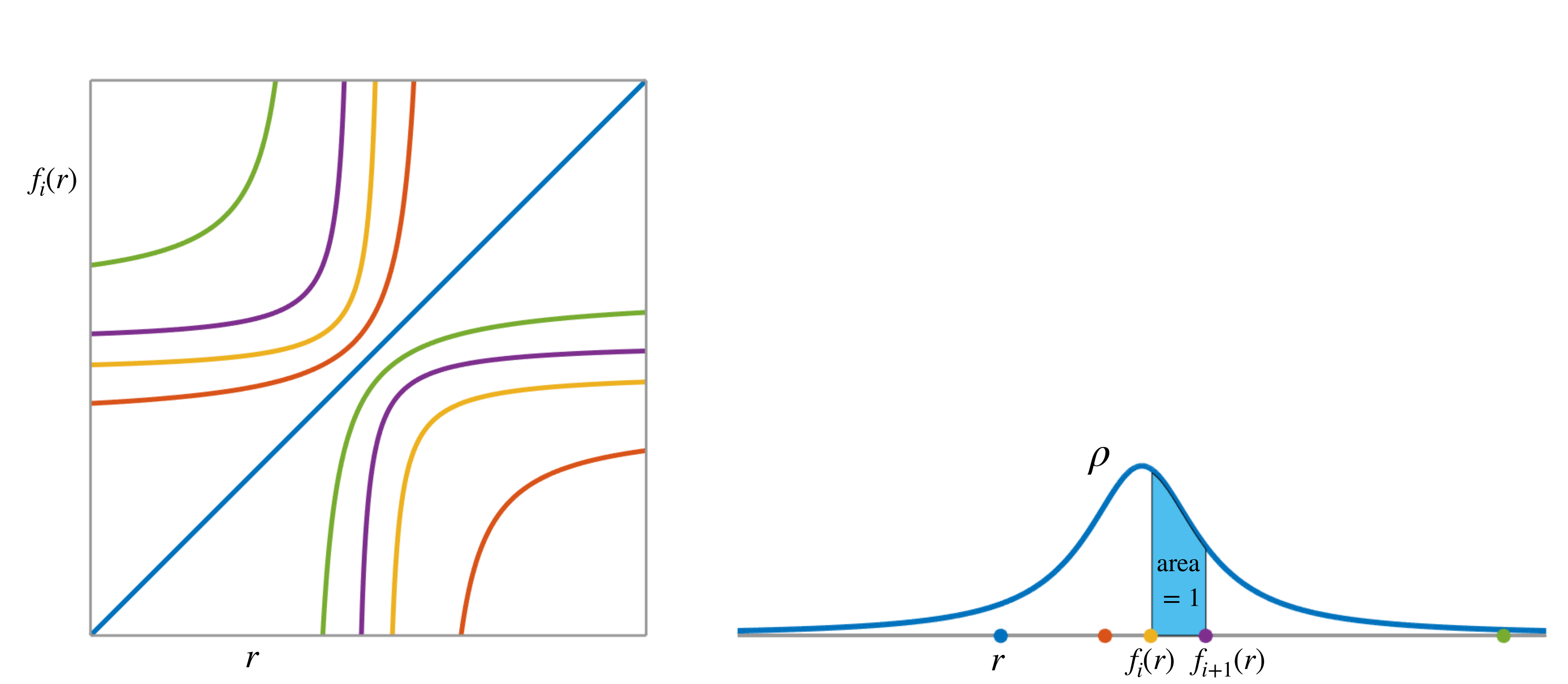}
\end{center}
\caption{SCE state of $N$ electrons in one dimension. {\it Right:} Electron positions. Given that the first electron (depicted in blue) is at $r$, the positions of the other electrons are completely determined by the requirement that neighbouring electrons are separated by an amount of density of $1$ (blue area). The co-motion functions or transport maps $f_i$ ($i=1,...,N-1$) of SCE theory are defined as the positions of the other electrons as a function of the first position $r$. The latter is distributed according to the given single-particle density $\rho$. {\it Left:} Graphs of the maps $f_i$, with $f_0(r)=r$ also shown. The figure corresponds to the Lorenzian density (Example \ref{ex:SCELor}) and $N=5$.}
\label{F:Seidl}
\end{figure}

{\bf Seidl's construction.} For a given integrable density $\rho \, : \, \R\to\R$ with $\rho\ge 0$ and $\int \rho = N$, begin by choosing $f_1 \, : \, \R\to\R$ so that the amount of density between $r$ and $f_1(r)$ is $1$. Now choose $f_2$ so that the amount of density between $f_1(r)$ and $f_2(r)$ is again $1$, and so on, i.e., denoting $f_0(r)=r$, 
\begin{equation} \label{sei1}
    \int_{f_i(r)}^{f_{i+1}(r)} \rho(r') \, dr'= 1 
\end{equation}
for all $i=0,...,N-1$. For equation \eqref{sei1} to possess a solution $f_{ i+1}(r)$ in $\R\cup\{+\infty\}$ we must have $\int_{f_i(r)}^\infty \rho \ge 1$; otherwise one needs to integrate first up to $+\infty$ and then onwards from $-\infty$ so as to obtain a total value of $1$, 
\begin{equation} \label{sei2}
     \int_{f_i(r)}^\infty \rho(r') \, dr' + \int_{-\infty} ^{f_{i+1}(r)} \rho(r')\, dr'  = 1.
\end{equation}
Physically this means that, given that the first electron is at some position $x_1=r$, all the other electrons at  $x_2=f_1(r),...,x_N=f_{N-1}(r)$ are separated by an equal amount of density between nearest neighbours. See Figure \ref{F:Seidl}, right panel. As always for SCE states, the first electron position is distributed according to the given density $\rho$. 

The above construction can be expressed concisely in terms of the cumulative distribution function 
\begin{equation} \label{CDF}
   \Frho(r) = \int_{-\infty}^r \frac{\rho(r')}{N} \, dr'
\end{equation}
and its generalized inverse 
\begin{equation} \label{cdfi}
   \Frho^{-1}(y) = \inf \{ r\in\R \, : \, \Frho(r) > y\}.
\end{equation}
(When $\rho$ is continuous and everywhere positive, $\Frho^{-1}$ is just the usual inverse function; the above definition has the virtue that it works for any nonnegative integrable $\rho$ with integral $N$.) Equations \eqref{sei1}, \eqref{sei2} now take the form $G_\rho(f_{i+1}(r))-G_\rho(f_i(r)) = 1/N$ respectively $1 - G_\rho(f_i(r)) + G_\rho(f_{i+1}(r)) = 1/N$, so by solving for $f_{i+1}$ in terms of $f_i$ and using $f_0(r)=r$
\begin{equation} \label{sei3}
   f_i(r) =\begin{cases}  \Frho^{-1} \bigl(\Frho(r) + \tfrac{i}{N}\bigr) \qquad & \text{ if }\Frho(r) \leq \tfrac{N-i}{N} \\ 
   \Frho^{-1} \bigl( \Frho(r) + \tfrac{i}{N} - 1 \bigr) & \text{ otherwise,} \end{cases} 
\end{equation}
for $i=1,...,N-1$. 

{\bf Optimality.} This construction is indeed optimal: 

\begin{theorem} \label{teo:1DN} Let $w(r)=|r|^{-1}$. For any nonnegative integrable density $\rho \, : \, \R\to\R$ with $\int \rho = N$, the SCE state \eqref{eq.psiSCE} with $f_1,...,f_{N-1}$ given by the Seidl construction \eqref{sei3} is a minimizer of the SIL problem
$$
  \text{Minimize } \int_{\R^N} V_{ee}(r_1,...,r_N)\, d\Pi(r_1,...,r_N) \text{ over }\{ \Pi\in{\mathcal P}(\R^N) \, : \, \Pi\mapsto\rho \}. 
$$
Moreover when $\rho$ is everywhere positive, this miminizer is unique for $N=2$, and its symmetrization (see Remark \ref{R:symmetrization}) is the unique symmetric minimizer for arbitrary $N$.
\end{theorem}
This theorem is due to \cite{CotFriKlu-CPAM-13} for $N=2$ and to \cite{ColDepDim-CJM-15} for arbitrary $N$. Despite the intuitive nature of the optimizer, the proof is not elementary. It is based on a careful analysis of the structure of $\Vee$-cyclically monotone sets in $\R^N$, and strongly relies on both optimal transport theory and the ordering properties of the real line.
Note that uniqueness cannot hold for $N\ge 3$ unless symmetry is required, as re-labelling the $f_i$ then yields another solution. This is purely a mathematical, not a physical effect since  solutions to the SIL problem arising as low-density limits of $N$-point densities of quantum wavefunctions (as described by Theorem \ref{T:asy}) are always symmetric, corresponding to the symmetrization of the state \eqref{eq.psiSCE}, \eqref{sei3}. 

{\bf Group law.} Formula \eqref{sei3} implies an interesting group law for the co-motion functions, already noticed in \cite{Sei-PRA-99}: the $i^{th}$ function is the $i$-fold composition of the first function with itself,
$$
   f_i = \underbrace{f_1\circ ... \circ f_1}_{i \text{ times}},
$$
and the $N$-fold composition of the first function gives the identity $f_0(r)=r$. 
  
%Example \ref{ex:SCE1d} for $N\geq 2$. %Theorem 1.1

{\bf Explicit examples.} The following examples further illustrate the nonlinear governing equations \eqref{sei1}--\eqref{sei2}, 
and may serve as useful benchmarks for numerical simulations in the strongly interacting limit (or close to it). 

{\bf Example \ref{ex.1Dhomogeneous}, ctd.} {\it Consider a two-electron system with $\rho$ being the uniform density in a one-dimensional interval $[0,L]$. In this case we have $\Frho(r)=r/L$, and formula \eqref{sei3} readily yields the co-motion function \eqref{eq:1Dhom}. For its graph see Figure \ref{F:jump}. Mathematically this map switches the right and left half of the interval; note that its composition with itself indeed gives the identity, as it must by the group law.} 

\begin{example}[\cite{GorVigSei-JCTC-09,GroKooGieSeiCohMorGor-JCTC-17}]\label{ex:SCELor} Let $\rho(r) = {N}/\pi(1+r^2)$ be the Lorenzian density, normalized so that $\int\rho = N$. In this case $\Frho(r)=\tfrac{1}{\pi}\arctan r + \tfrac{1}{2}$ and so eq.~\eqref{sei1} for $f_1$ in the region $G_\rho(r)\le \tfrac{N-1}{N}$ is, recalling the notation $f_0(r)=r$, 
\begin{equation} \label{Sei3a}
  \arctan f_1(r) = \arctan r + \tfrac{\pi}{N}.
\end{equation}
When $N=2$ it follows that 
\begin{equation} \label{Sei3b}
   f_1(r) = - \frac{1}{r}
\end{equation}
(note that then the derivatives of both sides of \eqref{Sei3a} agree, as do their values at $r=0$). From now on let us assume $N\ge 3$. In this case we can use the addition formula for the tangent, $\tan(x+y) = (\tan x + \tan y)/(1 - \tan x \tan y)$ for $x,\, y,\, x\! + \! y\neq \pi/2 + \Z$, and obtain
\begin{equation} \label{Sei4}
  f_1(r) = \frac{r+t_1}{1-t_1 r}, \;\;\;\; t_1 = \tan \tfrac{\pi}{N}.
\end{equation}
In the region $G_\rho(r)>\tfrac{N-1}{N}$, or equivalently $\arctan r > \tfrac{\pi}{2}-\tfrac{\pi}{N}$, or equivalently (because $\tan (\tfrac{\pi}{2}-x)=1/\tan x$) $r>1/t_1$, eq.~\eqref{sei2} for $f_1$ is
$$
   \arctan f_1(r) - \bigl( - \tfrac{\pi}{2}\bigr) = \arctan r - \frac{\pi}{2} + \frac{\pi}{N},
$$
that is to say $\arctan f_1(r) = \arctan r + \tfrac{\pi}{N} - \pi$. Using the addition formula for the tangent and $\tan x = \tan(x-\pi)$ we again find that $f_1$ is given by \eqref{Sei4}, so this formula describes $f_1$ on the whole real line. It remains to compute its $i$-fold composition $f_i$. Here we give a different derivation as compared to \cite{GorVigSei-JCTC-09, GroKooGieSeiCohMorGor-JCTC-17}. Note that mathematically $f_1$ is a Moebius map, i.e. a map of the form $M_a(r)=(r+a)/(1-ar)$. Using the (elementary to check) composition formula $M_a\circ M_b = M_{\frac{a+b}{1-ab}}$ and the addition formula for the tangent we find 
\begin{equation} \label{Sei5}
    f_i(r) = \frac{r+t_i}{1-t_i r}, \;\;\; t_i = \tan \tfrac{i\pi}{N} \;\;\; (i=1,...,N-1). 
\end{equation}
Moreover setting $i=N$ in the above formula we recover the abstract fact that the $N$-fold composition of $f_1$ must be the identity. Hence the co-motion functions for the Lorenzian density form a discrete subgroup of the Moebius group. 
For the graph of these functions when $N=5$ see Figure \ref{F:Seidl}.  
\end{example}

\subsection{Radially symmetric densities}\label{sec:RadSymm}

When the one-body density $\rho$ is radially symmetric, Seidl, Gori-Giorgi and Savin \cite{SeiGorSav-PRA-07} conjectured an explicit minimizing probability measure in \eqref{VP.SIL} of a radial-symmetry-preserving SCE form which is related to the explicit SCE state of one-dimensional systems.\footnote{The original conjecture concerned the physical case $d=3$. Subsequently, two-dimensional models have also being considered in the literature \cite{SeiDiMGerNenGieGor-arxiv-17,SeiGorSav-PRA-07}.}
Let us describe their conjecture in detail. 

Starting point is the following reduction to a 1d problem with effective interaction.

\begin{lemma}[Reduction to a 1d problem, \cite{BenCarNen-SMCISE-16,Pas-IOP-13}]\label{lemma:radialcase} Let $\rho \, : \, \R^d\to\R$ be an integrable density with $\rho\ge 0$ and $\int\rho = N$ which is radially symmetric, that is, $\rho(\bfr)=\rho_0(|\bfr|)$ for some function $\rho_0$, and let 
$$
 \rho_{rad}(r) = \omega_d r^{d-1} \rho_0(r)
$$
where $\omega_d$ is the area of the unit sphere in $\R^d$ (for $d=3$, $\omega_d=4\pi$). Then the SCE functional defined by \eqref{eq.SIL2} reduces to
\begin{equation}\label{eq:red_rad}
	V_{ee}^{\rm SCE} [\rho] = \min_{ \eta \in {\mathcal P}([0,\infty)^N), \; \eta \mapsto \rho_{rad} } \int_{[0,\infty)^N} \Veered(r_1, \ldots, r_N ) \, d \eta(r_1,...,r_N),
	\end{equation}
where $\Veered$ is the reduced Coulomb cost 
\begin{equation}\label{eq:c1reduced}
\Veered(r_1, \ldots, r_N ) = \min \left\{ \sum_{1\leq i<j \leq N}  \frac 1{| \bfr_j - \bfr_i |} \; : \; |\bfr_i|=r_i \; \forall i=1, \ldots, N \right\}.
\end{equation}
Moreover $\Pi\in{\mathcal P}(\R^{dN})$ is a minimizer for the full SIL variational principle \eqref{VP.SIL} in $d$ dimensions if and only if its radial projection $\Pi_{\rm rad}$, defined by 
$$
     \int_{A_1\times ... \times A_N} d\Pi^{\rm rad}(r_1,...,r_N)
      = \int_{ \{|\bfr_1|\in A_1\} \times ... \times \{ |\bfr_N|\in A_N\} } d\Pi(\bfr_1,...,\bfr_N) \mbox{ for all intervals }A_1,...,A_N,
$$
is a minimizer for the right hand side of \eqref{eq:red_rad} and $V_{ee}(\bfr_1,...,\bfr_N)=V_{ee}^{\rm rad}(|\bfr_1|,...,|\bfr_N|)$ $\Pi$-a.e.
\end{lemma}

In \cite{Sei-PRA-99,SeiGorSav-PRA-07}, the following interesting explicit state was conjectured to be optimal for the reduced problem in \eqref{eq:red_rad}:
\begin{equation} \label{SGSans1}
    d\eta(r_1,...,r_N) = \rho_{\rm rad}(r_1)\prod_{n=2}^N\delta(r_n - S^{(n)}(r_1))
\end{equation}
where $S^{(n)}$ denotes the $n$-fold composition $S\circ ... \circ S$ and $S\, : \, [0,\infty)\to [0,\infty)$ is defined as follows. Let  $0=a_0 <a_1< \ldots < a_{N-1}< a_N= \infty$ be such that the intervals $A_n=[a_{n-1},a_n)$ between successive $a_n$'s carry equal mass, that is, $\int_{A_n} \rho^{\rm rad} = 1$ for all $n$, and let $S|_{A_n}$ be the unique function such that
\begin{equation} \label{SGSans2}
   S|_{A_n} \mbox{ decreasing}, \;\;\; 
   S \mbox{ transports } \rho^{\rm rad}|_{A_n} \mbox{ to }\rho^{\rm rad}|_{A_{n+1}}
\end{equation}
(with the convention $A_{N+1}=A_1$). In terms of the original SIL problem \eqref{VP.SIL}, this ansatz corresponds to 
the SCE ansatz \eqref{eq.psiSCE} with maps satisfying the additional property
$$
  |f_n(\bfr)| = S^{(n)}(|\bfr|)
$$
for the above explicit $S$ and with suitably chosen angles so that $\Pi\mapsto\rho$ and $V_{ee}(\bfr_1,...,\bfr_N)=V_{ee}^{\rm rad}(|\bfr_1|,...,|\bfr_N|)$ $\Pi$-a.e. 
%
%Let $\rho \in \mathcal{P}(\R^3)$, $\rho = \rho\mathcal{L}^3$ be an absolutely continuous  measure with respect to the Lebesgue measure, with radial symmetry and let $\radi{\rho}$ as in \eqref{eqn:def_rhorad}.  Let $A_i=[a_i,a_{i+1}[$ have all the same radial measure $\radi{\rho}(A_i) = 1/N$. Then let $\Frho(r)= \radi{\rho} (]0,r])$ be the cumulative radial function and let $S: [0, \infty[ \to [0,\infty[$ be defined on the interval $A_i$ as the unique anti monotone function such that $S_{\sharp}\mu_r|_{A_i} = \mu_r|_{A_{i+1}}$:
%\[
% S(r) =\Frho^{-1} (  2i/N-\Frho(r)) \qquad \text{ if }a_{i-1} \leq r < a_i \text{ and }i <N
%\]
%\[
%\Frho(S(r)) = \begin{cases} \Frho^{-1}(\Frho(r) + 1/N -1) \quad & \text{ if $N$ is even}\\ \Frho^{-1}(1 - \Frho(r)) & \text{ if $N$ is odd, } \end{cases} \qquad \text{ if }a_{N-1} \leq r < a_N.
%\]
We call $S$ the {\it SGS map}, and the probability measure  $\eta$ given by \eqref{SGSans1}, \eqref{SGSans2} the {\it SGS state}.  

\begin{figure}[h]\label{eq:SGSpic}
    \centering
    \includegraphics[scale=0.5]{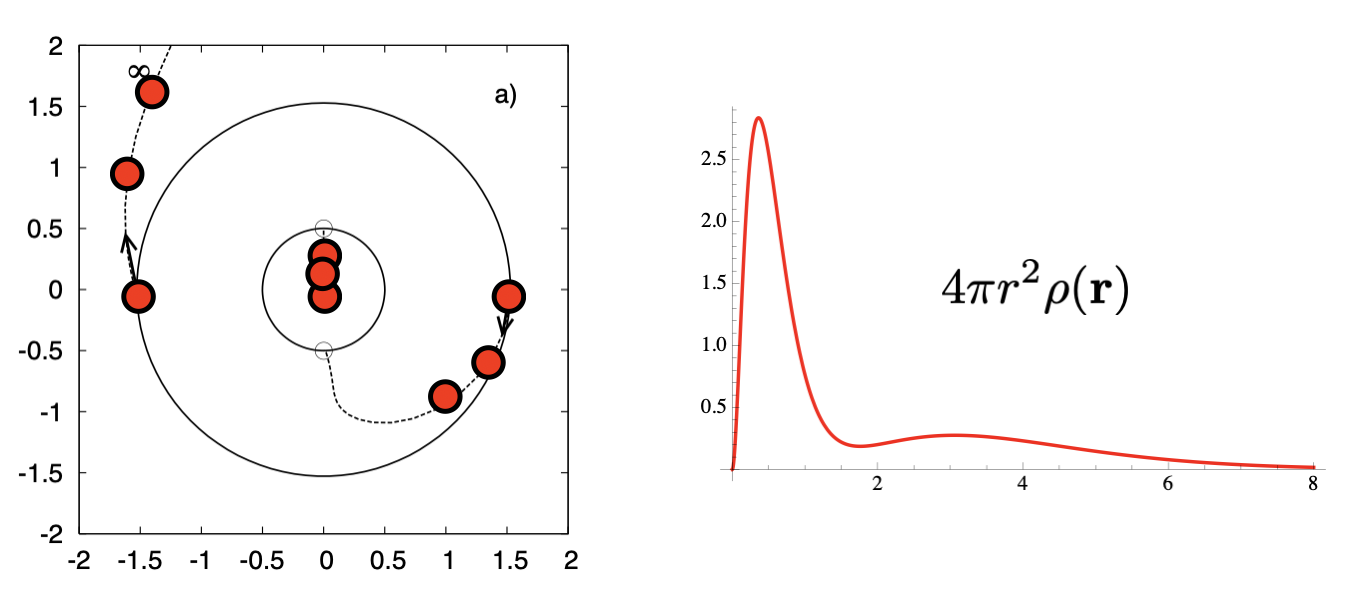}
    \caption{SGS state (left panel)  for the Lithium atom density (right panel). The exteme angular and radial correlation exhibited by this state is  illustrated here while sending one of the electrons (the leftmost) to infinity.}
\end{figure}

\begin{figure}[http]
\begin{center}
\begin{tikzpicture}[scale=3]
\draw [smooth, samples=100, domain=0:4.5,thick] plot(\x,{2*(\x)/(((\x)*(\x)+1)*((\x)*(\x)+1))});
\draw[->] (-0.3,0) -- (4.7,0);
\draw[->] (0,-0.1) -- (0,0.7);

  \def\offset{0.05};
  \def\offsets{1/7-\offset};

\foreach \x in {1,...,6} {\draw[ thick] ( {sqrt(\x/(7-\x)) )  },0) -- ( {sqrt(\x/(7-\x)) )  },{2*sqrt(\x*(7-\x))*(7-\x)/49});}

\foreach \x in {1,3,5,7} 
	{
	 \draw[\mycolor(\x)] ( { 1/sqrt(1/(\offset+(\x-1)/7) -1)  },0) node[below] {$x_{\x}$}-- ({ 1/sqrt(1/(\offset+(\x-1)/7)-1)  },{2*(1-(\offset+(\x-1)/7))*(1-(\offset+(\x-1)/7))/sqrt(1/(\offset+(\x-1)/7)-1)});
	 
	 \draw[fill=\mycolor(\x)!50, opacity=0.3] plot[smooth, samples=100, domain= {sqrt((\x-1)/(8-\x)) }:{1/sqrt(1/(\offset+(\x-1)/7) -1)}] ( \x,{2*(\x)/(((\x)*(\x)+1)*((\x)*(\x)+1))}) |- ({sqrt((\x-1)/(8-\x))},0) -- cycle;
 }
\foreach \x in {2,4,6} 
	{
	  \draw[\mycolor(\x)] ( { 1/sqrt(1/(\offsets+(\x-1)/7) -1)  },0) node[below] {$x_{\x}$}-- ({ 1/sqrt(1/(\offsets+(\x-1)/7)-1)  },{2*(1-(\offsets+(\x-1)/7))*(1-(\offsets+(\x-1)/7))/sqrt(1/(\offsets+(\x-1)/7)-1)});
	  
	  \draw[fill=\mycolor(\x)!50, opacity=0.3] plot[smooth, samples=100, domain= {1/sqrt(1/(\offsets+(\x-1)/7) -1)}:{sqrt((\x)/(7-\x)) }] ( \x,{2*(\x)/(((\x)*(\x)+1)*((\x)*(\x)+1))}) |- ({1/sqrt(1/(\offsets+(\x-1)/7) -1)},0) -- cycle;
	   }
\end{tikzpicture}
\end{center}

%\begin{figure}[http]
\begin{minipage}{.49\textwidth}
\begin{tikzpicture}[scale=1.2]

%\onslide<17->{ 
\draw[thick] (0,0) -- (4.5,4.5) node[right] {$x_1$};
 \draw[->] (-0.3,0) -- (4.7,0) node[below] {$x_1$};
  \draw[->] (0,-0.3) -- (0,4.7);
  
  \def\xmini{0};
  \def\xmaxi{4.5};
 \def\tmaxi{0.95294};

 \foreach \y in {2,4,6}
	{
	  \pgfmathtruncatemacro{\z}{\y+1};
	  \draw[text=\mycolor(\y)]  ({\xmini},{\xmini  +sqrt((\y)/(7-\y))}) node[left] {$x_{\y}=x_{\z}$};
	}
  \foreach \y in {3,5,7}
	{
	  \draw[text=\mycolor(\y)]  ({\xmini},{\xmini  +sqrt((\y-1)/(8-\y))}) node[left] {$x_{\y}$};
	}

  \foreach \y in {3,5,7}
	{
	   \draw[text=\mycolor(\y)]  ({\xmaxi},{  sqrt(abs(7-7*\tmaxi+7-\y+1)/abs(7-15+7*\tmaxi +\y))  }) node[right] {$x_{\y}$};
	}

  \clip (0,0) rectangle (4.5,4.5);
\foreach \y in {3,5,7}
	{
	  \draw [smooth, samples=100, domain=\xmini:{(8-\y)/7}, \mycolor(\y)] plot({sqrt(\x/(1-\x))},{sqrt((7*\x+\y-1)/(8-7*\x -\y))});
	  \draw [smooth, samples=100, domain={(8-\y)/7+0.01}:1, \mycolor(\y)] plot({sqrt(\x/(1-\x))},{sqrt(abs(7-7*\x+7-\y+1)/abs(7-15+7*\x +\y))});
	}
\foreach \y in {2,4,6}
	{
	  \draw [smooth, samples=100, domain=\xmini:{(\y)/7}, \mycolor(\y)] plot({sqrt(\x/(1-\x))},{sqrt((-7*\x+(\y))/(7+7*\x -(\y)))});
	  \draw [smooth, samples=100, domain={(\y)/7}:1, \mycolor(\y)] plot({sqrt(\x/(1-\x))},{sqrt(abs(7*\x-(\y))/(7-7*\x +(\y)))});
	}
%}

%\draw [smooth, samples=100, domain=-6.5:6.5] plot(\x,{(\x  + tan(180*(1)/7))/(1-(\x)*tan(180*(1)/7))});
%\draw [smooth, samples=100, domain=-6.5:6.5, green] plot(\x,{(\x  + tan(180*(2)/7))/(1-(\x)*tan(180*(2)/7))});
%\draw [smooth, samples=100, domain=-6.5:6.5, blue] plot(\x,{(\x  + tan(180*(3)/7))/(1-(\x)*tan(180*(3)/7))});
%\draw [smooth, samples=100, domain=-6.5:6.5, red] plot(\x,{(\x  + tan(180*(4)/7))/(1-(\x)*tan(180*(4)/7))});
%\draw [smooth, samples=100, domain=-6.5:6.5, orange] plot(\x,{(\x  + tan(180*(5)/7))/(1-(\x)*tan(180*(5)/7))});
%\draw [smooth, samples=100, domain=-6.5:6.5, violet] plot(\x,{(\x  + tan(180*(6)/7))/(1-(\x)*tan(180*(6)/7))});

\end{tikzpicture}
\end{minipage}
\begin{minipage}{.15\textwidth}

\end{minipage}
\begin{minipage}{.35\textwidth}
\begin{tikzpicture}[scale=5.0]

%\onslide<18->{
 \draw[thick] (0,0) -- (1,1) node[right] {$y_1$};
 \draw[->] (-0.1,0) -- (1.1,0) node[below] {$y_1$};
  \draw[->] (0,-0.1)  -- (0,1.1);
  \draw (1,-.01) node[below]{1} -- (1,0.01);
\foreach \y in {1,...,6} { \draw ({\y/7},-.01) node[below]{$\frac {\y}7$} -- ({\y/7},0.01); }

% \foreach \y in {2,...,7}
%	{
%	  \draw[text=\mycolor(\y)]  (0,{(\y-1)/7}) node[left] {$y_{\y}$};
%	  \draw[text=\mycolor(\y)]  (1,{(\y-1)/7}) node[right] {$y_{\y}$};
%	}

\foreach \y in {3,5,7}
	{
	  \draw [smooth, samples=100, domain=0:{(8-\y)/7)}, \mycolor(\y)] plot(\x,{\x+ (\y-1)/7});
	  \draw [smooth, samples=100, domain={(8-\y)/7}:1, \mycolor(\y)] plot(\x,{2-(\x +  (\y-1)/7)});
	}	
\foreach \y in {2,4,6}
	{
	  \draw [smooth, samples=100, domain=0:1, \mycolor(\y)] plot(\x,{abs(-\x+ (\y)/7)});
	}
%}
\end{tikzpicture}
\end{minipage}
\caption{SGS state when $N=7$. Top: a radial measure $\rho_{\rm rad}(r)$. Bottom left: the maps $S,S^{(2)},\dots,S^{(6)}:[0,\infty)\to[0,\infty)$, plotted with colors green, blue, red, violet, yellow, and brown. Bottom right: the same graphs under a change of variables $y_1 = \int_0^{x_1} \tfrac{\rho_{\rm rad}}{N}$ which transports $\rho_{\rm rad}$ to the uniform density on the interval $[0,1]$. Picture from \cite{SeiDiMGerNenGieGor-XX-XX}.}
\end{figure}
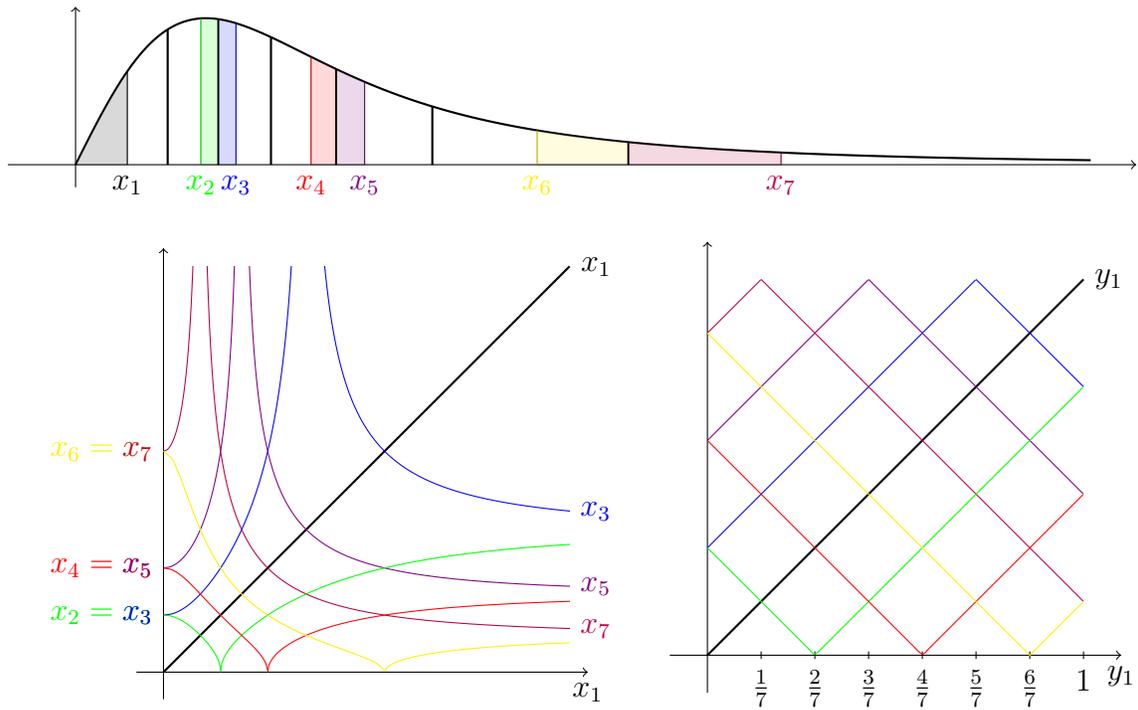

%\fi

The SGS state has been rigorously proved to be optimal in some specific cases.

\begin{example}{(SCE for radially symmetric densities, \cite[Theorem 4.10]{CotFriKlu-CPAM-13})} \label{ex:SCEN2Seidl}
Let $N=2$, and let $\rho$ be a radially symmetric density on $\R^d$ such that $\rho(\bfr) > 0$ for all $\bfr$. Then the optimal co-motion function $f$ is given by
\[
f(\bfr) = s(\vert \bfr\vert)\dfrac{\bfz}{\vert \bfz\vert} 
\]
for some function $s:[0,+\infty)\to\R$ such that $s\leq 0$, $s$ is increasing,  $\lim_{r\to+\infty}s(z) = 0$, and $\lim_{r\to 0+}s(r) = -\infty$. 
\end{example}
The function $s$ in the above example corresponds to minus the SGS map, i.e. $s=-S$ in the $N=2$ case.

Optimality of the SGS state has also been proved for some special class of densities $\rho$ when $N=d=3$ \cite{ColStr-M3AS-15,GerThesis,SeiDiMGerNenGieGor-arxiv-17} and $N=3$ and $d=2$ \cite{BinDepKau-arxiv-20}.

Recently, counterexamples of radially symmetric probability densities were found for which the SGS state is not optimal. 
The simplest one is a uniform density on a thin annulus: 
%
%
%
%$\tilde{\mu}$ in $\R^3$ such that 
%\[
%\VSIL[\tilde{\mu}] < \int_{\R^{3}}\Veered(r_1,r_2,r_3)d\pi_S(r_1,r_2,r_3), 
%\]
%where $\gamma_S = (Id,S,\dots,S^{(N-1)})_{\sharp}\tilde{\mu}$ and $S$ is the SGS map defined in the SGS conjecture \eqref{conj.Seidl}, have been studied in the literature \cite{BinDepKau-arxiv-20, ColStr-M3AS-15,SeiDiMGerNenGieGor-arxiv-17,SeiDiMGerNenGieGor-XX-XX}. 
%
%\quad The main idea of the constructions of these counterexamples is to consider a measure $\tilde{\mu}_{\varepsilon} = \rho_{\varepsilon}(\bfz)\mathcal{L}^3$, concentrated on a very thin annulus and depending on a positive parameter $\varepsilon$, which is chosen \textit{a posteriori}, such that the above inequality holds. 

\begin{example}[\cite{ColStr-M3AS-15}, see also \cite{BinDepKau-arxiv-20,GerThesis,SeiDiMGerNenGieGor-XX-XX,SeiDiMGerNenGieGor-arxiv-17} for related examples]
Let $N=3$. For sufficiently small $\ep>0$, and the density 
\begin{equation}\label{eq:muep}
\rho^{\rm rad}_{\ep} = c_\ep 1_{[1,1+\ep]} 
\end{equation}
(with the constant $c_\ep$ chosen such that $\int \rho^{\rm rad}=3$), the SGS state $\eta_{\ep}$ defined by \eqref{SGSans1}--\eqref{SGSans2} is not optimal for the variational problem \eqref{eq:red_rad}. 
\end{example}

This example illustrates that guessing the optimal  SCE states can be a tricky business even for 1d problems, and makes it all the more remarkable that optimality of Seidl's guess for the 1d Coulomb problem is a rigorous theorem (Theorem \ref{teo:1DN}). The proof of nonoptimality relies on a Taylor expansion of the reduced interaction $\Veered$ (defined in equation \eqref{eq:c1reduced}) at the point $(1, 1, 1)$ and on cyclical monotonicity methods from optimal transport theory.

%proving that the support of the SGS coupling $\pi_{S}$ is not $\Vee-$monotone for $\mu_{\ep}$ defined in \eqref{eq:muep}.
%
%\quad A second approach developed by M. Seidl, S. Di Marino, A. Gerolin \textit{et al} \cite{SeiDiMGerNenGieGor-arxiv-17}, relies on the same Taylor expansion but reduce the disprove of SGS conjecture to the Strong Interaction Limit with repulsive harmonic interaction. % $V_{ra}(\bfz_1,\brz_2,\bfz_3) = -\sum_{1\leq i<j \leq 3}\vert \bfz_i-\bfz_j\vert^2$ \cite{SeiDiMGerNenGieGor-arxiv-17}, where the structure of the couplings are concentrated on hyperplane $\lbrace (\bfz_1,\bfz_2,\bfz_3) \in \R^9 \, : \, \bfz_1+\bfz_2+\bfz_3 = k\rbrace$, $k\in\R$. 
%
%\quad A more explicit connection between the density $\tilde{\mu}_{\ep}$ and the optimality
%or non-optimality of the SGS map has been addressed in \cite{BinDepKau-arxiv-20} in the plane $\R^2$ with $N=3$ electrons. In particular, a new class of examples and counterexamples when the support of $\pi_S$ is $\Vee-$monotone are presented.

While this counterexample disproves optimality of the SGS state in general, the density \eqref{eq:muep} is quite different from typical atomic densities and the following remains an interesting mathematical problem.
\\[2mm]
{\bf Open problem.} Find sufficient conditions on radial densities $\rho$ such that the SGS state is optimal for \eqref{eq:red_rad}. 

\subsection{An example with  irregular co-motion functions for repulsive harmonic interactions}

One of the more challenging properties of co-motion functions is that they are typically discontinuous. Here we give an extreme example 
with modified electron-electron interaction which is discontinuous {\it everywhere}, due to Di Marino, Gerolin and Nenna \cite{DMaGerNen-TOOAS-17}. 

\begin{example} \label{ex:fractal} Let $d$ be arbitrary, $V_{ee}(\bfr_1,...,\bfr_N)= - \sum_{1\le i<j\le N}|\bfr_i-\bfr_j|^2$ (repulsive harmonic interaction), and $N=3$. Let $\rho=3 \, \cdot \,   1_{[0,1]^d}$ (uniform density on a cube in $\R^d$). Then there exists a nowhere continuous map $T \, : \, [0,1]^d \to [0,1]^d$ which transports $\rho$ to itself such that 
\begin{equation} \label{eq:fractmeas}
   d\Pi(\bfr_1,\bfr_2,\bfr_3) = \frac{\rho(\bfr_1)}{N} \delta(\bfr_2 - T(\bfr_1)) \delta(\bfr_3 - T(T(\bfr_1))
\end{equation}
is an optimal probability measure for the SIL problem \eqref{VP.SIL}. 
\end{example}

\begin{figure}[http!]
\includegraphics[width=1\textwidth]{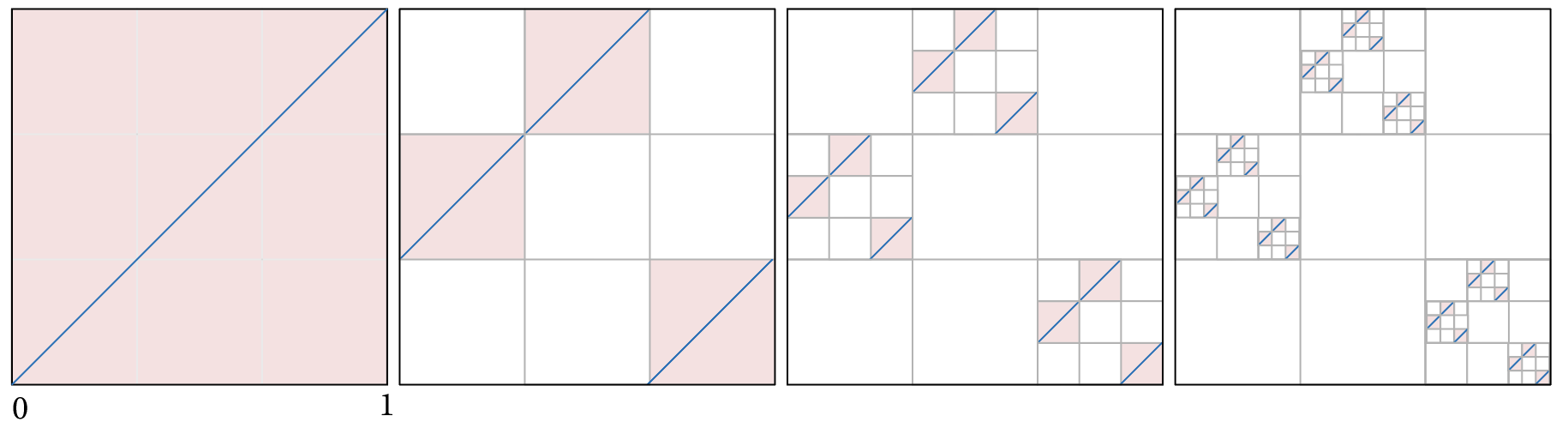}
\caption{Construction of the optimal map $T$ in Example \ref{ex:fractal}. The picture shows the graph (in blue) of the first few iterations of eq.~\eqref{eq:iteration}; each graph consists of three scaled copies of the previous one. The exact map is reached in the limit of infinitely many iterations.}
\label{F:fractal}
\end{figure}

The map $T$ is an explicit fractal map. For $d=1$ it is depicted in Figure \ref{F:fractal} and constructed as the unique fixed point of the iteration
\begin{equation} \label{eq:iteration}
f \mapsto 
g(x) = \begin{cases} \frac 13 f(3x) + \frac 13 \qquad &\text{ for } 0 \leq x <  \frac 13 \\
       \frac 13 f(3x - 1) + \frac 23 &\text{ for } \frac 13 \le x < \frac 23 \\
\frac 13 f(3x - 2)  & \text{ for } \frac {2}3 \leq x < 1,
\end{cases}
\end{equation}
starting with $f(x) = x$. 
To see what the iteration is doing, divide $[0,1]^2$ into a $3\times 3$ grid of squares and put scaled copies of the graph of the original function into the two squares directly above the diagonal and the bottom right square. Optimality of the resulting fractal SCE state \eqref{eq:fractmeas} is easy to see from the following special property of the repulsive harmonic cost which was first observed by Pass \cite{Pas-CVPDE-12}: thanks to the identity $V_{ee}=|\bfr_1 + ... + \bfr_N|^2 - N \sum_{i=1}^N|\bfr_i|^2$ and the fact that the integral of the second term against a  probability measure $\Pi$ only depends on its marginal, the minimizers of the SIL problem are precisely the probability measures supported on the surface $\bfr_1+...+\bfr_N=0$.

The above example and construction works for arbitrary $N$, see \cite{DMaGerNen-TOOAS-17}.

{\bf Open problem.} Do such extreme examples also occur for the Coulomb interaction? Note that the repulsive harmonic interaction arises by locally Tayor-expanding the Coulomb interaction in angular direction.

\subsection{Minimizers of the discretized SIL variational principle are quasi-Monge states} \label{sec:quasi}

We now come back to the important issue that the SIL variational principle \eqref{VP.SIL} still requires minimization over a high-dimensional space of $N$-point probability measures, whereas the low-dimensional SCE ansatz \eqref{eq.psiSCE} can fail to yield an optimizer of \eqref{VP.SIL}. One can ask whether some modified low-dimensional ansatz is enough to solve \eqref{VP.SIL} exactly. In other words, can one achieve Seidl's original goal of solving the strongly interacting limit of DFT with a low-dimensional ansatz that can be easily stored on a computer? 

For the discretization of \eqref{VP.SIL} on a grid, Friesecke and V\"ogler \cite{FriVog-SIAMJMA-18} found a modified ansatz which achieves this, for arbitrary space dimensions, densities and interaction potentials:   
\begin{align} \label{eq:QM}
    d\Pi(\bfr_1,...,\bfr_N) & = S_N \int_{\R^d} \alpha(\bfr) \prod_{n=1}^N \delta\bigl(\bfr_n-f_{n-1}(\bfr)\bigr) \, d\bfr  \\
    &= \frac{1}{N!}\sum_{\sigma} \int_{\R^d} \alpha(\bfr) \prod_{n=1}^N \delta\bigl(\bfr_n - f_{\sigma(n-1)}(\bfr)\bigr) \, d\bfr \nonumber
\end{align}
where $\sigma$ runs over all permutations of the indices $0,...,n-1$, $\alpha$ is some (free to choose) probability density on the single-particle space $\R^d$, and the $f_n$ are maps from $\R^d$ to $\R^d$. States of this form are called {\it quasi-Monge states} or {\it quasi-SCE states}. With the specific choice 
$\alpha=\frac{\rho}{N}$, the quasi-Monge ansatz \eqref{eq:QM} reduces precisely to the SCE (alias Monge) ansatz in its symmetric form \eqref{eq.psiSCE3}. The novelty is the additional freedom of choosing the auxiliary density $\alpha$. For the quasi-Monge ansatz, the marginal constraint $\Pi\mapsto\rho$ takes, instead of the conditions $f_n{}_\sharp \rho = \rho$ ($n=0,...,N-1$) (eq.~\eqref{eq.pfw}), the form of a single condition, 
\begin{equation} \label{eq.pfwquasi}
    \frac{1}{N} \sum_{n=0}^{N-1} f_n{}_\sharp\alpha = \frac{\rho}{N}.
\end{equation}
That is, the {\it average} push-forward of the auxiliary density $\alpha$ under the quasi-SCE maps must be the (suitably normalized) physical density. 

Plugging the ansatz \eqref{eq:QM} into the SIL variational principle \eqref{VP.SIL} and integrating out the variables $\bfr_2,...,\bfr_N$ yields the {\it quasi-Monge} or {\it quasi-SCE}  variational principle
\begin{equation} \label{eq:QMVP}
 \mbox{Minimize }\int_{\R^d} V_{ee}(f_0(\bfr),...,f_{N-1}(\bfr)) \, \alpha(\bfr) \, d\bfr
 \mbox{ over probability densities }\alpha\mbox{ and maps }f_0,...,f_{N-1},
\end{equation}
with the minimization being subject to the constraint \eqref{eq.pfwquasi}. 
%where $\alpha$ belongs to ${\cal P}(\R^d)\cap L^1(\R^d)$ and the quasi-Monge maps are measurable maps from $\R^d$ to $\R^d$ satisfy the constraint \eqref{eq.pfwquasi}.

\begin{theorem} \label{T:FV} [Justification of the quasi-Monge ansatz, \cite{FriVog-SIAMJMA-18}] 
Let $\rho$ be any discrete $N$-particle density on $\R^d$, that is to say $\rho(\bfr)=\sum_{i=1}^\ell \rho_i \delta(\bfr-\bfa_i)$ for some distinct discretization points $\bfa_i\in\R^d$ and some $\rho_i\ge 0$ with $\sum_i\rho_i=N$, and let $V_{ee}\, : \, \R^{dN}\to\R\cup\{+\infty\}$ be any interaction potential which is symmetric in the electron coordinates (e.g., the Coulomb interaction $V_{ee}(\bfr_1,...,\bfr_N)=\sum_{i<j}1/|\bfr_i-\bfr_j|$). Then the SIL problem \eqref{VP.SIL} possesses a minimizer which is a quasi-Monge state \eqref{eq:QM}. Equivalently, it possesses a minimizer of the form \eqref{eq:sparse}, i.e., a superposition of at most $\ell$ symmetrized Dirac measures.
\end{theorem}
This result rigorously reduces the number of unknowns from exponential to linear with respect to the number of electrons; more precisely, from $\ell^N$ (the dimension of the space of $N$-point probability measures supported on $\{\bfa_1,...,\bfa_\ell\}^N$) 
%(or, accounting for symmetry, $\binom{N+\ell-1}{\ell+1}$)) 
to $\ell\cdot (N+1)$ ($\ell$ unknowns for each of the $N$ quasi-Monge maps, and another $\ell$ unknowns for the auxiliary density $\alpha$). 

The above result fails if the class of quasi-Monge states is narrowed to Monge (alias SCE) states, see \cite{Fri-SIAMJMA-19}. For continuous $\rho$'s, it is an open question whether the SIL problem always (or at least in the Coulomb case) admits minimizers of quasi-Monge form. 

{\bf Proof of Theorem \ref{T:FV}} (following \cite{FriVog-SIAMJMA-18}). Let us explain the intuition and reasoning behind the quasi-Monge ansatz and Theorem \ref{T:FV}, which comes from convex geometry. Before passing to a geometric viewpoint, we note that by the symmetry of  $V_{ee}$ the minimization in \eqref{VP.SIL} can be restricted to symmetric probability measures (see  Remark \ref{R:symmetrization}); moreover any symmetric probability measure $\Pi\in{\cal P}(\R^{dN})$ with $\Pi\mapsto\rho$ must be of the form
$$
  d\Pi(\bfr_1,...,\bfr_N) = \sum_{i_1,...,i_N=1}^\ell \gamma_{i_1...i_N} \delta(\bfr_1-\bfa_{i_1}) \cdots \delta(\bfr_N-\bfa_{i_N})
$$
for some symmetric tensor $(\gamma_{i_1...i_N})\in\R^{\ell\times ... \times \ell}$ with nonnegative entries which sum to $1$. Now geometrically, for fixed discretization points $a_1,...,a_\ell$ the set of these probability measures is a finite-dimensional convex polytope; let us denote it ${\cal P}_{sym}(\{\bfa_1,...\bfa_\ell\}^N)$. The subset satisfying the marginal constraint
$\Pi\mapsto\rho$, i.e.
\begin{equation} \label{eq:discmarg}
  \sum_{i_2,...,i_N=1}^\ell \gamma_{i_1i_2...i_N} = \rho_{i_1} \;\; \forall i_1\in\{1,...,\ell\},
\end{equation}
is also a convex polytope called {\it Kantorovich polytope}; let us denote it ${\cal P}_{\rho}(\{\bfa_1,...\bfa_\ell\}^N)$. While general probability measures in these sets possess a huge number of coefficients which increases combinatorially with the number $N$ of particles, the key point is that the {\it extreme points}\footnote{These are the points that cannot be written as convex combinations of any other points in the set.} of these sets are very sparse, with only a small number of nonzero coefficients. The extreme points of ${\cal P}_\rho(\{\bfa_1,...,\bfa_\ell\})$ are easily seen to be symmetrized products of delta functions, 
$S_N\delta(\bfr_1-\bfa_{i_1})...\delta(\bfr_N-\bfa_{i_N})$, 
where $S_N$ is the symmetrization operator. Now consider a subset of a convex polytope satisfying {\it one} linear constraint, geometrically: the intersection of the polytope with a hyperplane. It is geometrically expected (and not difficult to prove) that all extreme points of this new set are convex combinations of just two extreme points of the original polytope. Analogously, by a well known result in convex geometry the intersection of a convex polytope with $k$ hyperplanes has extreme points given by convex combinations of just $k+1$ of the original extreme points. Since the marginal condition \eqref{eq:discmarg} imposes $\ell-1$ constraints (note that one of the $\ell$ constraints is redundant due to the sum of the $\gamma_{i_1...i_N}$ being $1$), the extreme points of the Kantorovich polytope are convex combinations of just $\ell$ symmetrized delta functions, i.e., probability measures of the form 
\begin{equation} \label{eq:sparse}
   \sum_{\nu=1}^\ell \alpha_\nu S_N \delta(\bfr_1 - \bfa_{i_1^{(\nu)}}) ... \delta(\bfr_N - \bfa_{i_N^{(\nu)}})
\end{equation}
for some nonnegative coefficients $\alpha_{\nu}$. 
Defining the maps $f_n$ by $f_{n-1}(\bfa_\nu) = \bfa_{i_n^{(\nu)}}$ yields that all extreme points are quasi-Monge states \eqref{eq:QM}. Theorem \ref{T:FV} now follows from the general principle that the minimum of a linear functional (such as $\int V_{ee} \, d\Pi$) over a convex polytope is always attained at some extreme point. 

The quasi-Monge (or quasi-SCE) ansatz and Theorem \ref{T:FV} underlie the numerical method described in section \ref{sec:gencol}. 

\subsection{Entropic Regularization of the SCE functional}
\noindent
We have seen in Figure \ref{fig:FcexSGS} and section \ref{sec:optimality} that in the strongly interacting limit, the $N$-body density concentrates on the lower-dimensional manifold on which the classical effective potential energy $V_{ee}(\bfr_1,...,\bfr_N)-\sum_i v_{\rm SCE}(\bfr_i)$ is minimal. A regularization of the SCE functional which has nice mathematical properties and smears out the $N$-body density is the following:
\begin{equation}\label{eq:EntrFunctional2}
\Vtau[\rho]= \inf_{\pi\in{\mathcal P}(\R^{dN})\cap L^1(\R^{dN}), \, \pi\mapsto\rho} {\cal V}_{ee}[\pi] + \tau  S[\pi].
\end{equation}
Here $\tau>0$ is a small parameter, ${\cal V}_{ee}$ is the usual electron interaction energy, and $S$ is (minus) the Shannon-Von Neumann entropy, 
\begin{equation} \label{eq:defEntropy2}
  S[\pi] = \int_{\R^{dN}} \pi(\bfr_1,\dots,\bfr_N)\Bigl(\log\pi(\bfr_1,\dots,\bfr_N)-1\Bigr)d\bfr_1\dots d\bfr_N.
\end{equation}

As shown in Lemma \ref{lem:entrdef} below, the negative part of the entropy density has finite integral under very mild conditions on $\rho$ (e.g., finite first moment suffices), and so definitions \eqref{eq:EntrFunctional2}--\eqref{eq:defEntropy2} make rigorous sense.  
The existence of a minimizer in \eqref{eq:EntrFunctional2} can be obtained assuming that $\rho\log\rho \in L^1(\R^d)$ \cite{GerKauRaj-CVPDE-20}. Physically, the right hand side in \eqref{eq:EntrFunctional2} can be viewed as the free energy of $N$ classical particles with  interaction potential $V_{ee}$ and density $\rho$ at inverse temperature $\tau$. But our goal here is not to model a physical system at finite temperature, but instead to approximate the SCE functional. 

\begin{figure}[htbp]
	\begin{center} 
		\includegraphics[width=0.25\textwidth]{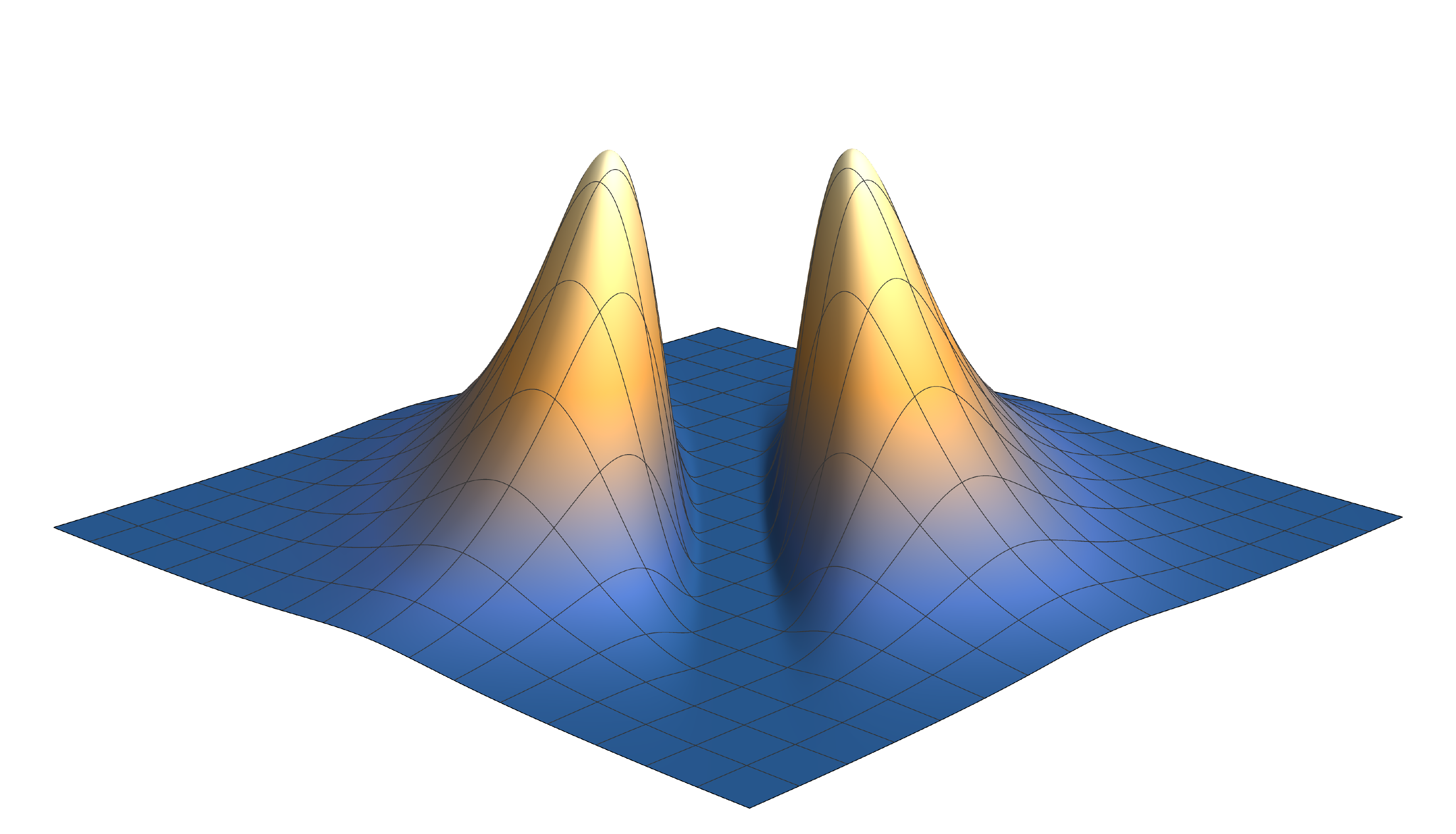} \hspace*{-3mm}
		\includegraphics[width=0.25\textwidth]{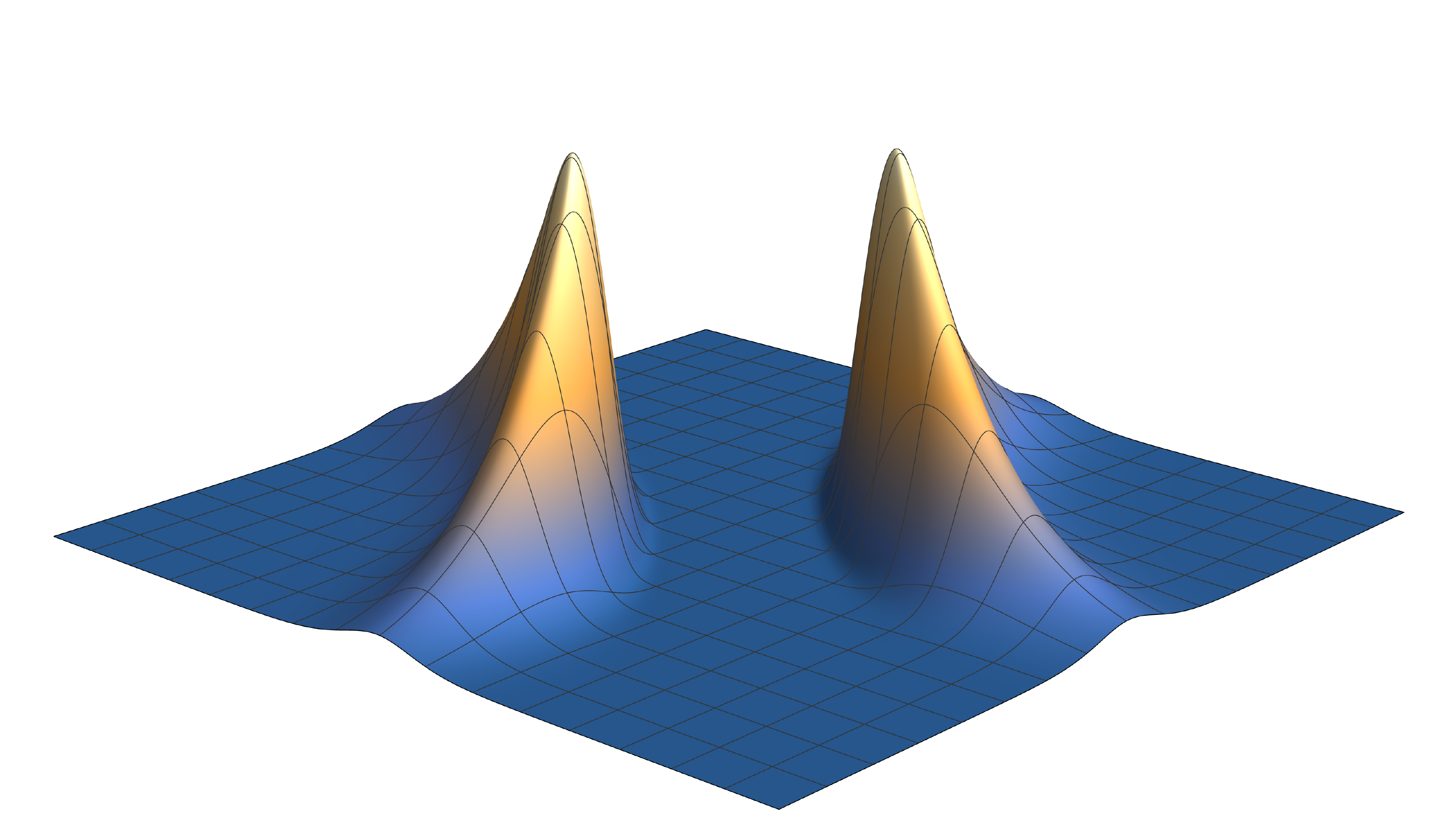} \hspace*{-3mm}
		\includegraphics[width=0.25\textwidth]{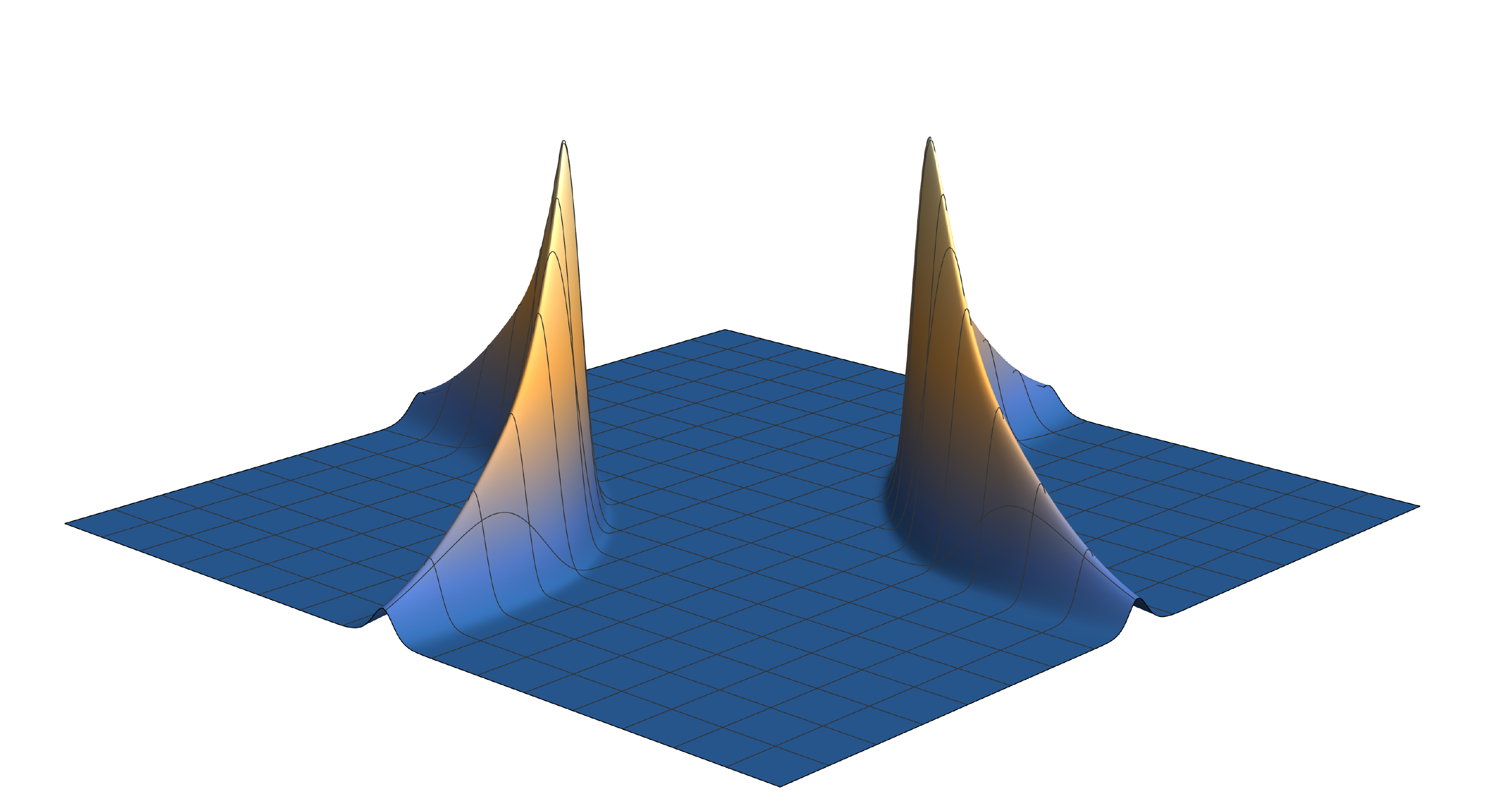} \hspace*{-3mm}
		\includegraphics[width=0.25\textwidth]{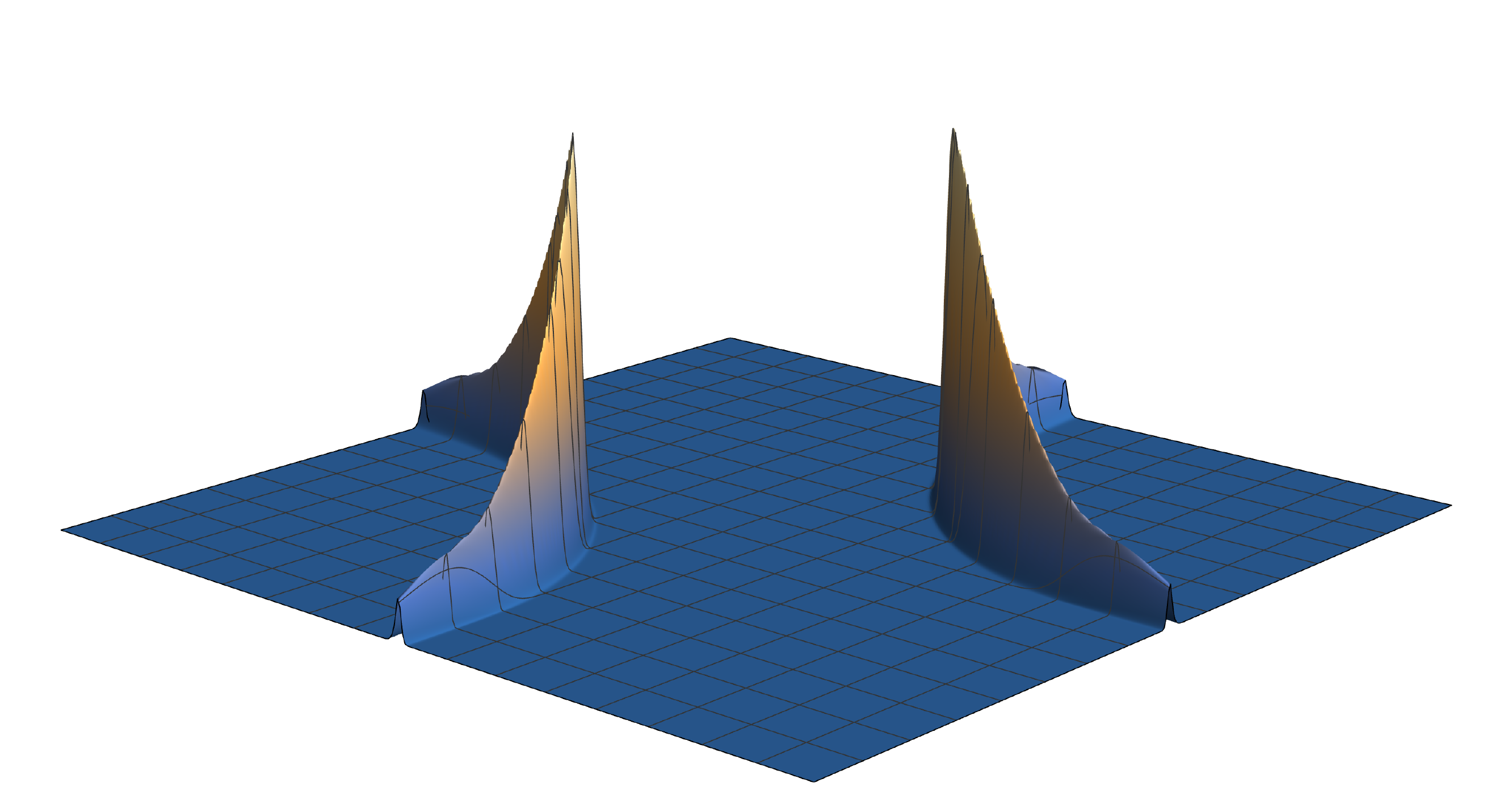}
		\vspace{-6mm}
		\caption{Numerically computed  optimal density  $\pitau(r_1,r_2)$ in \eqref{eq:EntrFunctional2} for two electrons in one dimension and the density $\rho(r)=c_0/\cosh r$,  $r\in[-10,10]$, for different values of the regularization parameter $\tau$. Here $c_0$ is a normalization constant, and we have used the effective Coulomb interaction~\cite{BakStoMilWagBurSte-PRB-15} $V_{ee}(r_1,r_2)=1.07\, e^{-| r_1-r_2|/2.39}$.}
\label{fig:gammacoul}
	\end{center}
	
\end{figure}

Figure \ref{fig:gammacoul} illustrates the effect of the entropy term in a two-electron example: the larger the regularization parameter $\tau$, the more the  minimizers $\pi$ are spread out around the support of the SCE state. (Recall that by Theorem \ref{T:SCE}, when $N=2$ the SIL variational principle is uniquely minimized by an SCE state.) 
%to which \eqref{eq:EntrFunctional2} reduces for $\tau=0$ 

The corresponding regularization for the Wasserstein distance squared instead of the SCE functional (Example \ref{ex:proto} instead of Example \ref{ex:coul}) in fact goes back to Erwin Schr\"odinger in 1931 \cite{Sch-VAWK-31}, and had a completely different motivation: Schr\"odinger was looking for models for the ``most likely'' evolution law between two probability distributions of particle positions which have been empirically observed at different times, perhaps hoping to re-discover his -- then still controversial -- quantum mechanics in a novel way.

An equivalent entropic problem has been consider by J. Chayes, L. Chayes and E. H. Lieb in \cite{ChaCha-JSP-84,ChaChaLie-CMP-84}. In their setting, the integration in the entropy functional $S$ is not against the Lebesgue measure but against the product measure $\prod_{i=1}^N \rho(\bfr_i)/N$, which constitutes a natural model in classical statistical mechanics. The role of the reference measure will be explained below and can be understood via equation \eqref{eq:roleofreference}.

In optimal transport, entropic regularization became a popular basis for computational methods following an influential paper by Cuturi \cite{Cut-ANIPS-13} in machine learning and Galichon and Salani\'e in economics \cite{Gal-CEPR-10}; see section \ref{sec:num} for a computational algorithm. Regularization by entropies other than the Shannon-Von Neumann one is considered in \cite{DMaGer-arXiv-2020,LorMah20cont}.

\paragraph{Basic properties.} Let us now informally discuss basic properties of \eqref{eq:EntrFunctional2}. %References to rigorous results for compactly supported $\rho$ (or rigorous results on related problems) are given at the end of this paragraph. 
%For a rigorous treatment which covers the case of compactly supported $\rho$'s and suitable classes of interaction potentials $V_{ee}$ we refer the reader to  \cite{references}. 

{\it Unique minimizer.} Assuming that $\rho\log\rho \in L^1(\R^d)$, a minimizer $\pi^\tau$ in \eqref{eq:EntrFunctional2} exists \cite{GerKauRaj-CVPDE-20}. This is expected from the convexity of the functional ${\cal V}_{ee}+\tau S$. Since the functional is strictly convex on the domain where it is finite, minimizers must be unique.   

{\it Euler-Lagrange equation; form of minimizer.} 
Assume a Lagrange multiplier rule as in \eqref{eq:LM2}. That is, assume the existence of Lagrange multipliers $(\lambda(\bfr_1))_{\bfr_1\in\R^d}$ such that $\pi^\tau$ is the unconstrained minimizer of the Lagrangian 
$$
 {\cal L}[\pi] = \int_{\R^{dN}} \Bigl( V_{ee}(\bfr_1,\ldots,\bfr_N) - \sum_{i=1}^N\lambda(\bfr_i)\Bigr) \pi(\bfr_1,\ldots,\bfr_N) d\bfr_1\ldots d\bfr_N + \tau S[\pi].
$$
Thus the function $\lambda(\bfr)$ has the usual physical interpretation of DFT as minus the potential that enforces the density constraint, and will in the following be denoted $u^\tau(\bfz)$. 
It follows that 
$$
   0= \frac{d}{dt}\Big|_{t=0}
 {\cal L} [\pi+t\eta] 
$$
for all variations $\eta$ with $\int\eta = 0$ and $\pi\pm\eta\ge 0$. That is to say, $0 = \int [\bigl( V_{ee} - \sum_i u^\tau(\bfr_i)) + \tau\log\pi ]\eta$ and therefore
$$
   V_{ee} - \sum\limits_i u^\tau(\bfr_i) + \tau \log\pi = const.
$$
By solving for $\pi$ and adjusting $u^\tau$ by an additive constant, it follows that 
\begin{equation}\label{eq:defai}
  \pitau(\bfr_1,\dots,\bfr_N) =  \prod_{i=1}^N a^\tau(\bfr_i) e^{-\frac{V_{ee}(\bfr_1,\dots,\bfr_N)}{\tau}} \mbox{ with } a^\tau(\bfr_i)=e^{\frac{1}{\tau}u^\tau(\bfr_i)}.
\end{equation}
The function $a^\tau$ can -- independently of its construction above with the help of Lagrange multipliers -- be interpreted as an entropic weight function which makes the probability density $\pi^\tau$ satisfy the constraint $\pi^\tau\mapsto \rho$. Note that by this constraint and eq.~\eqref{eq:defai}, $a^\tau$ must satisfy the following governing equation in which Lagrange multipliers no longer appear:
\begin{equation}\label{eq:entr3}
a^\tau(\bfz_j) \int_{\R^{d(N-1)}} \prod_{i\neq j} a^\tau(\bfz_i) e^{-\frac{V_{ee}(\bfz_1,\cdots,\bfz_N)}{\tau}}\mbox{$\prod\limits_{i\neq j} d\bfr_i$}  = \frac{\rho(\bfr_j)}{N} \quad \forall \, j =1,\dots,N.
\end{equation}

The above equations constitute the so-called (multi-marginal) Schr\"odinger system. When the density $\rho$ is Gaussian and the interaction potential $w$ is taken to be the repulsive or attractive harmonic interaction, the entropically regularized problem can be solved exactly, see  \cite{GerGroGor-JCTC-20}, for the one-dimensional case, and \cite{del2020statistical,Jan-NEURIPS-2020,Mal-InfoGeo-2021} for the general case. In \cite{Car-SIMA-19}, Carlier and Laborde showed the existence of a solution of the system \eqref{eq:entr3} via an inverse function theorem argument by assuming that the one-body density $\rho$ belongs to  $L^{\infty}(\R^d)$.

{\it Relative entropy formulation.} The functional ${\cal V}_{ee}+\tau S$ agrees up to an additive constant with the Kullback-Leibler divergence (or minus the relative entropy)\footnote{The KL divergence between two nonnegative densities with possibly unequal mass is formally defined as ${\rm KL}(f|g) = \int f\log\frac{f}{g}$.} between $\pi$ and a kernel function $\mathcal{K}$ of the electronic interaction $V_{ee}$~ \cite{Leo-DCDSA-14}: 
$$
   {\cal V}_{ee}[\pi] + \tau S[\pi] = \tau{\rm KL}(\pi|\mathcal{K}) - \tau \mbox{ with } \mathcal{K} = e^{-V_{ee}/\tau}.
$$
Thus the optimizer $\pi^\tau$ is the density with marginal $\rho$ which has minimal relative entropy with respect to the kernel $\mathcal{K}$. 

\textit{The role of the reference measure.} In the literature, the entropy functionals which are typically studied replace integration against the Lebesgue measure in \eqref{eq:defEntropy2} by integration against the product of the marginals $\mu^{\otimes N} = \otimes^N_{i=1}\mu$ (or any other finite reference measure), where $\rho/N=\mu$. As shown in Lemma 1.5 in \cite{DMaGer-JSC-20} (see also \cite{GerKauRaj-CVPDE-20} for the Coulomb case), both problems are equivalent since the following identity holds
\begin{equation}\label{eq:roleofreference}
\Vtau [\rho ]=\inf_{\pi\mapsto\rho}\left\lbrace {\cal V}_{ee}[\pi ] + \tau \int_{\R^{dN}} 
 \dfrac{d\pi}{d\mu}\Bigl(\log\dfrac{d\pi}{d\mu}-1\Bigr)d\mu\right\rbrace +\tau \int_{\R^{d}}\rho\log\frac{\rho}{N} d\bfr\, .
\end{equation}
Therefore, whenever at least one side of the equality above is finite, the original variational problem from the definition of $\Vtau[\rho]$ (eq.~\eqref{eq:EntrFunctional2}) and the variational problem defined on the right-hand side of \eqref{eq:roleofreference} have the same minimizers.

{\it Dual formulation.} As for the exact (unregularized) strongly interacting limit of DFT, there is a dual variational principle for the Lagrange multiplier and an associated dual construction of $V_{ee}^\tau[\rho]$. We have
\begin{equation}\label{eq:Vtaudual2}
 \Vtau[\rho] = \sup_{u} J[u],
\end{equation}
 where 
\begin{equation} \label{eq:Vtaudual3}
   J[u] = \int_{\R^d} u(\bfz)\rho(\bfz)d\bfr  - \tau\int_{\R^{dN}}e^{-\frac{1}{\tau}[V_{ee}(\bfz_1,\dots,\bfz_N) - \sum_i u(\bfr_i)]} d\bfz_1\dots d\bfz_N 
\end{equation}
and the supremum in \eqref{eq:Vtaudual2} is over a suitable class of potentials. The second term in \eqref{eq:Vtaudual3} can be viewed as a soft version of the inequality constraint $E_{\rm{pot}}(\bfr_1,\dots,\bfr_N) = V_{ee}(\bfr_1,...,\bfr_N)-u(\bfr_1)+...+u(\bfr_N)\ge 0$ in the unregularized theory (see \eqref{eq.dualconstr}), as it penalizes deviations from this inequality. Indeed, via the Laplace principle we have that, whenever the second term in \eqref{eq:Vtaudual2} is finite,
\[
\lim_{\tau\to 0^+}-\tau\log\left(\int_{\R^{dN}}e^{-\frac{1}{\tau}[V_{ee}(\bfz_1,\dots,\bfz_N) - \sum_i u(\bfr_i)]} d\bfz_1\dots d\bfz_N\right) = \inf_{\textcolor{black}{\bfr_1,\dots,\bfr_N} \in \R^d}\lbrace  E_{{\rm pot}}(\bfr_1,\dots,\bfr_N) \rbrace.
\]
In the discrete setting, this is precisely the LogSumExp formula. 
The existence of an optimizer $u^\tau$ for the dual problem and the representation formulae \eqref{eq:defai}, \eqref{eq:Vtauderi} with this $u^\tau$ were proved in \cite{DMaGer-JSC-20, DMaGer-arXiv-2020} under the assumption that $\rho\log\rho\in L^1(\R^d)$ and $V_{ee}$ is measurable and bounded. 

{\it Functional derivative.} As in exact SCE theory, the functional derivative of the energy functional is formally given by the optimal potential in the dual problem, that is to say
\begin{equation} \label{eq:Vtauderi}
  \frac{\delta \Vtau[\rho]}{\delta \rho} = u^\tau + const,
\end{equation}
where $u^\tau$ is the maximizer of \eqref{eq:Vtaudual3} (assuming such a maximizer exists and is unique). As in SCE theory, a natural choice of the additive constant is to require $\lim_{|\bfr|\to\infty} \bigl(u^\tau(\bfr) + const\bigr) = 0$. The ensuing potential $v^\tau = u^\tau + const$ is then an approximation to the SCE potential.

%\quad One can show that the maximum in \eqref{eq:Vtaudual} is reached if and only if  $\pitau \in\mathcal{P}(\R^{dN})$ and has the form \eqref{eq:gammawithu}. Therefore, $\pitau$ is an optimal coupling in \eqref{eq:EntrFunctional}. The existence of the maximum in  \eqref{eq:Vtaudual} was proved by S. Di Marino and A. Gerolin in \cite{DMaGer-JSC-20,DMaGer-arXiv-2020}, assuming that $V_{ee}$ is bounded or $w$-continuous.

%We now briefly discuss rigorous versions of the above findings. Existence of a unique minimizer in \eqref{eq:EntrFunctional2} \aug{in for the Coulomb interaction was obtained in \cite{GerKauRaj-CVPDE-20} assuming $\rho\log\rho\in L^1(d\bfr)$}. \gero{Augusto, didn't they do N=2 only?}\augusto{I have cited now my works, but they are really based on ideas from the two marginals case} \augusto{Probably we need to rephrase the following paragraph in light of the footnote 6 in page 36?}  %It is an open problem to rigorously cover the setting of Lemma \ref{}
 
\paragraph{Relation with the Levy-Lieb functional.} Just like the SCE functional itself, its entropic regularization is a rigorous lower bound of the exact functional, provided the regularization parameter $\tau$ is chosen suitably. More precisely:
 
\begin{theorem}[\cite{SeiDiMGerNenGieGor-PRA-17}]
 \label{thm:entlow-bound}
Let $\Psi$ be any $N$-electron wavefunction in the space ${\cal W}^N$ (see \eqref{WN}), or alternatively any bosonic wavefunction in $H^1(\R^{dN})$, and suppose $\Psi\mapsto\rho$. Let $V_{ee}$ be the Coulomb interaction. Then the scaled Levy-Lieb functional defined in eq.~\eqref{eq.LLalpha} satisfies
\begin{equation} \label{eq.entrlowbd}
  \frac{F^\lambda[\rho]}{\lambda} \ge V_{ee}^\tau[\rho] \;\; \mbox{ with }\tau = \frac{\pi}{2 \lambda}.
\end{equation}
In particular, the original Levy-Lieb functional \eqref{eq.FHK} satisfies 
\begin{equation} \label{eq.entrlowbd2} F_{\rm LL}[\rho]\ge V_{ee}^{\pi/2}[\rho].
\end{equation}
\end{theorem}

This result is a consequence of the logarithmic Sobolev inequality (LSI). We include a proof, following Seidl \textit{et al.}~\cite{SeiDiMGerNenGieGor-arxiv-17}. We begin by recalling a standard version of the LSI. 

 \begin{theorem}
 [LSI, Corollary 7.3 in \cite{GozChr-MPRF-10}] 
 \label{lsi-gaussian}
 Let $\nu \in \mathcal{P}(\mathbb{R}^n)$ such that $\nu(\bfr)=e^{-V(\bfr)}$ with ${\rm D}^2 V \geq \kappa {\rm Id}$. 
 Then, for every locally integrable function $f \geq 0$ on $\R^n$ such that $f\nu \in \mathcal{P}(\mathbb{R}^n)$ we have that $\int_{\R^n} f\log f d\nu \leq \frac 2{\kappa} \int | \nabla \sqrt{f}|^2 \, d \nu$.
 \end{theorem}
 
%As noted by Seidl \textit{et al.}~in \cite{SeiDiMGerNenGieGor-arxiv-17},
This implies the following LSI for the Lebesgue measure:

 \begin{corollary}[LSI for the Lebesgue measure,~\cite{SeiDiMGerNenGieGor-arxiv-17}] 
 \label{ent:cor-lsi}
 Let $f\geq 0$ be a function such that $\sqrt{f} \in H^1(\mathbb{R}^n)$ and $f\in\mathcal{P}(\R^n)$. Then $\int_{\R^n} f\log f d\bfr \leq \frac 1{\pi} \int_{\R^n} | \nabla \sqrt{f}|^2d{\bfr}$.
 \end{corollary}
 
{\bf Proof of Corollary \ref{ent:cor-lsi}.} 1. In the LSI in Theorem \ref{lsi-gaussian}, the requirement on $f$ that $\int f \, d\nu=1$ can be relaxed to $0 < \int f \, d\nu \le 1$. This follows by applying the LSI to $f/\alpha$, $\alpha=\int f \, d\nu$, and noting that the extra term $-(1/\alpha)\int f \log \alpha \, d\nu$ on the left hand side is $\ge 0$. 

2. Take $\nu_{\bfr_2}(\bfr_1)=e^{-\pi|\bfr_1-\bfr_2|}$, then $\nu$ satisfies the assumption of the LSI with $\kappa=2\pi$, and moreover $\int f d\nu_{\bfr_2}\le 1$. Hence by the LSI, $$
   \int f \log f \, d\nu_{\bfr_2} \le \frac{1}{\pi} \int |\nabla\sqrt{f}|^2 d\nu_{\bfr_2}. 
$$
3. Integrate over $\bfr_2$ and use that $\int e^{-\pi |\bfr_1-\bfr_2|}d\bfr_2=1$. This yields the assertion.

\begin{proof}[Proof of Theorem \ref{thm:entlow-bound}]
 Let $\psi\in {\cal W}^N$, $\Psi\mapsto\rho$, and let $\pi$ be its $N$-point position density $\eqref{eq:Nppd}$. By a version of the Hoffmann-Ostenhof inequality \cite{Hof-77}\footnote{Strictly speaking, this inequality and related ones are proved in \cite{Hof-77} under the tacit assumption that $\sqrt{\pi}$ (or related reduced quantities) belong to $H^1$ and can be differentiated by the chain rule. For further discussion of this point see the chapter by Kvaal in this volume.}, $\sqrt{\pi}\in H^1(\R^{dN})$ and 
 $$
    T[\Psi] \ge \frac12 \int_{\R^{dN}} |\nabla\sqrt{\pi}|^2 d\bfr_1 ... d\bfr_N.
 $$
 This together with the LSI for the Lebesgue measure (Corollary \ref{ent:cor-lsi}) applied to $\pi$
 gives 
\begin{align}
 \frac{1}{\lambda}T[\Psi] + \Vee[\Psi] & \geq \frac{\pi}{2\lambda}  \int_{\R^{dN}}\!\pi\log \pi \,  d\bfr_1\dots d\bfr_N+\int_{\R^{dN}}V_{ee} \,  \pi \, d\bfr_1 \dots d\bfr_N \\
 & = \frac{\pi}{2\lambda}\Bigl( S(\pi) + 1 \Bigr) + {V}_{ee}[\pi].
\end{align}
Taking the infimum over $\Psi\in{\cal W}^N$ yields $F^\lambda[\rho]/\lambda \ge V_{ee}^\tau[\rho] + \tau$, with $\tau$ as in the theorem.
 \end{proof}

Although Theorem \ref{thm:entlow-bound} provides a lower bound for the Levy-Lieb functional \eqref{eq.FHK}, in practice this bound can be rather loose \cite{GerGroGor-JCTC-20}. 
%
%\paragraph{Existence of the minimizer for the entropy-SCE functional under finite first moments and convergence to the SCE functional} The condition on $L^1$-integrability of $\rho\log\rho$ can be relaxed as well as that the entropy-SCE functional \eqref{eq:defEntropy2} converges to the SCE problem.
%
\paragraph{Well definedness of entropy and convergence to the SCE functional.} We now show that the entropy is well defined under very mild conditions on $\rho$ (e.g., finite first moment suffices), and that the entropically regularized functional $\Vtau$ convergence to the SCE functional when the regularization parameter tends to zero.
%
%In fact, the existence of a minimizer for the entropic regularization $\Vtau$ of the SCE functional can be obtained under very mild conditions on $\rho$ (e.g., finite first moment suffices). Notice that, this implies, in particular, the equivalence described in \eqref{eq:roleofreference} does not necessarily holds.

Note that a priori both the positive and the negative part of the integral \eqref{eq:defEntropy2} could be divergent;
%in which case expression \eqref{eq:defEntropy2} does not make sense; 
Lemma \ref{lem:entrdef} excludes this for the negative part, and so the integral always has a well defined value in $\R\cup\{+\infty\}$.

%Before reviewing the interesting mathematical properties of \eqref{eq:EntrFunctional2}, let us  clarify that the integral \eqref{eq:defEntropy2} is well defined, \aug{see also Proposition 2.6 in \cite{GerKauRaj-CVPDE-20} for a similar setting}.

\begin{lemma}[Well-definedness of entropy and of the regularized SCE functional] \label{lem:entrdef} Let 
$\rho\in L^1(\R^d)$, $\rho\ge 0$, $\int\rho = N$, and assume $\rho$ has finite first moment, that is to say $\int |\bfr|\rho(\bfr)d\bfr<\infty$. Let $\pi\in {\cal P}(\R^{dN})\cap L^1(\R^{dN})$ with $\pi\mapsto\rho$. Then the negative part $(\pi \log \pi)_-$ has finite integral; more precisely, for some constant $A_\rho>0$ which depends only on $\rho$ but not on $\pi$ 
$$
    \int (\pi\log \pi)_- \ge - A_\rho >  -\infty, 
$$
where $f_-(\bfr) = \min\{f(\bfr),0\}$ denotes the negative part of a function $f$. Hence $S$ as defined by \eqref{eq:defEntropy2} is well defined as a functional $S \, : \, \{\pi\in {\cal P}(\R^{dN}) \cap L^1(\R^{dN}) \, : \, \pi \mapsto\rho\}\to \R\cup\{+\infty\}$, and $V_{ee}^\tau[\rho]$ as defined by \eqref{eq:EntrFunctional2} is well defined as an element of $\R\cup\{+\infty\}$.
\end{lemma}

The assumption that $\rho$ has finite first moment cannot be omitted. For instance, for $N=2$ and $d=1$ the $N$-body density
$$
  \pi(r_1,r_2) = c_0 \prod_{i=1}^2 \frac{1}{r_i(\log r_i)^2} \; \mbox{ on }[2,\infty)^2,
$$
continued by zero to $\R^2$ and 
with $c_0$ chosen such that $\int \pi = 1$,  
belongs to $L^1(\R^2)$ but satisfies $\int (\pi \log \pi)_- = -\infty$, as the interested reader can check using that $\int_2^\infty \tfrac{1}{z|\log z|^\alpha} dz = \infty$ for $\alpha=1$ but $<\infty$ for $\alpha>1$. In particular, in such a case the equivalence described in \eqref{eq:roleofreference} does not necessarily hold.

The lemma implies that for any interaction potential $V_{ee}$ on $\R^{dN}$ which is symmetric and bounded from below, such as the Coulomb interaction, $V^{\tau}_{ee}[\rho]$ is well defined as an element of $\R\cup\{+\infty\}$. 

{\bf Proof of Lemma \ref{lem:entrdef}} $\pi\log\pi$ is $\le 0$ precisely in the region $\Omega=\{\bfr\in\R^{dN} \, : \, \pi(\bfr)\le 1\}$. Split $\Omega$ into $\Omega_<=\{\bfr\in\Omega\, : \, 0\le \pi(\bfr) < e^{-(|\bfr_1|+...+|\bfr_N|)}\}$ and $\Omega_>=\{\bfr\in\Omega\, : \, \pi(\bfr)\ge  e^{-(|\bfr_1|+...+|\bfr_N|)}\}$. Since $g(z)=z\log z$ satisfies $|g(z)|\le C\sqrt{z}$ in $[0,1]$ for some constant $C$, 
\begin{eqnarray}
  & & \int_{\R^{dN}} |(\pi\log\pi)_-| \; = \;  
  \int_{\Omega_<} |(\pi\log\pi)_-| + \int_{\Omega_>} |(\pi\log\pi)_-| \nonumber \\
  & & \le C \int_{\Omega_<} e^{-(|\bfr_1|/2 + ... + |\bfr_N|/2)} d\bfr_1\ldots d\bfr_N + \int_{\Omega_>} \pi(\bfr_1,...,\bfr_N) \Bigl(|\bfr_1|+\dots+|\bfr_N|\Bigr) d\bfr_1 \ldots d\bfr_N \nonumber \\
  & & \le  C \Bigl( \int_{\R^d} e^{-|\bfr_1|/2} d\bfr_1\Bigr)^N + \int_{\R^d} \rho(\bfr_1)|\bfr_1| d\bfr_1 =: A_\rho. \label{eq:Arho}
\end{eqnarray} 
By the assumption that $\rho$ has finite first moment, the right hand side is finite, completing the proof of the lemma. 
\vspace*{2mm}

Finally, we prove that -- as intuitively expected -- the entropically regularized functional $V_{ee}^\tau$ converges to the exact SCE functional when the regularization parameter tends to zero. The corresponding $\Gamma$-convergence result was obtained in \cite{GerKauRaj-CVPDE-20}.

\begin{theorem} \label{T:entconv} Let $\rho$ be any $N$-electron density which belongs to the class ${\cal D}^N$ (see \eqref{eq.DN}) and has finite first moment, and let $V_{ee}$ be the Coulomb interaction. Then
\begin{equation} \label{eq:entconv}
   \lim_{\tau\to 0} V_{ee}^\tau[\rho] = V_{ee}^{\rm SCE}[\rho].
\end{equation}
\end{theorem}

\begin{proof} We combine the upper bound on $V_{ee}^\tau[\rho]$ from Theorem \ref{thm:entlow-bound}, the asymptotic result on $F^\lambda[\rho]/\lambda$ in eq.~\eqref{eq.asy2} (see Theorem \ref{T:asy}), and the lower bound from Lemma \ref{lem:entrdef}. By inequality \eqref{eq:Arho} we have for any $\pi\mapsto\rho$
$$
  \tau S[\pi] \ge  \tau\int_{\R^{dN}} (\pi\log\pi)_- \, - \; \tau 
  \ge - \tau \Bigl(A_\rho + 1 \Bigr)
$$
and hence, by adding ${\cal V}_{ee}[\pi]$ to both sides and taking the infimum over $\pi$
$$
   V_{ee}^\tau[\rho] \ge V_{ee}^{\rm SCE}[\rho] - \tau \Bigl(A_\rho + 1\Bigr).
$$
Obviously this lower bound converges to $V_{ee}^{\rm SCE}[\rho]$ as $\tau\to 0$. On the other hand, by Theorem \ref{thm:entlow-bound} we have $V_{ee}^\tau[\rho] \le F^\lambda[\rho]/\lambda$ and by Theorem \ref{T:asy} this upper bound also converges to $V_{ee}^{\rm SCE}[\rho]$; hence so must  $V_{ee}^\tau[\rho]$.
\end{proof}

\section{Numerical methods and approximations} \label{sec:num}

\quad The SCE functional can not at the moment be accurately and efficiently computed for general three-dimensional densities and large $N$. But accurate numerical methods are available for small $N$ or special situations, novel methods aimed at large $N$ are under  development, and less accurate approximations can already be computed for large $N$. We review these methods and approximations in this section, and their use within Kohn-Sham DFT in section \ref{sec:KSSCE}.

%Although many methods have been proposed to compute numerical realizations of the SCE functional (\eqref{eq.SIL2} or \eqref{SCEfunctional3} or \eqref{eq.SCEdual}) when $N=2$, the general theory for computing these functionals for $N\geq 3$ electrons is still nascent.

%\quad Unfortunately, repulsive interactions are the worst case for Strong Interaction Limit functional or, more generally, multi-marginal optimal transport problems. In fact, as shown in \cite{AltBoi-arXiv-20}, several problems of interest are NP-hard to solve even approximately. For instance, the standard ${\rm LP}$ representation of \eqref{eq:VSIL} has the worst-case complexity bound of $O(k^{3N})$ for the standard deterministic interior-point algorithms, where $k$ is the number of points in the support $\rho$.

%\quad Very recently, some numerical methods has been developed providing efficient numerical methods for (crude) approximations for SIL functional.

\subsection{Numerical methods based on co-motion functions}
Numerical implementations using co-motion functions were confined to the following cases:
\begin{itemize}
    \item the exact maps are known: general $N$ in one dimension (see Sec.~\ref{sec:1D});
    \item an explicit ansatz, able to get very close to the true minimum, exists: spherically symmetric (radial) case (see Sec.~\ref{sec:RadSymm}).
\end{itemize}
In addition, co-motion functions can be extracted from optimal plans in the case
\begin{itemize}
    \item $N=2$, for which the existence of the map is proven and there are 1-1 correspondences between map, optimal plan, and Kantorovich potential (see equations \eqref{eq.psiSCE2} and \eqref{eq:SCEMap2electrons}).
\end{itemize}
We review here and in the following section the implementation for these three classes of problems. Their use in combination with Kohn-Sham DFT is then discussed in Sec.~\ref{sec:KSSCE}.
\subsubsection{One-dimensional $N$-electron systems} \label{sec:1DnumSCE}
The SCE functional has been implemented for one-dimensional (1D) many-electron systems using the exact co-motion functions (maps) of Seidl \cite{Sei-PRA-99}, which we reported and illustrated in Sec.~\ref{sec:1D}. These applications typically aim at modeling physical systems in which electrons are confined in elongated traps (quantum wires): the interaction used is thus 3D Coulomb renormalized for small interparticle distances. The idea is that at long range the electrons feel the $1/|x|$ interaction, but at short range they can avoid each other due to the finite thickness of the wire, which is mimicked by removing the divergence at $x=0$. For example, a widely used effective quasi-1D interaction is obtained by integrating the 3D Coulomb interaction over normalized gaussians in two of the three spatial directions \cite{GiuVig-BOOK-05}, modeling harmonic confinement within a wire of thickness $b$,
\begin{equation}\label{eq:veewire}
    v_{ee}^{\rm wire}(x)=\frac{1}{4\pi\, b^2}\int_{-\infty}^{\infty}d y\int_{-\infty}^{\infty}d z \frac{e^{-\frac{1}{4b^2}(x^2+b^2)}}{\sqrt{x^2+y^2+z^2}}=\frac{\sqrt{\pi}}{2\,b}\,\exp\left(\frac{x^2}{4\,b^2}\right){\rm erfc}\left(\frac{|x|}{2\,b}\right).
\end{equation}
This interaction is finite at $x=0$, where it has a cusp, behaves as $1/|x|$ for large $x$ and it is convex for $x\ge 0$. Other popular quasi-1D interactions are the soft Coulomb and the regularised Coulomb,
\begin{align}
    v_{ee}^{\rm soft}(x) & =\frac{1}{\sqrt{x^2+a^2}}, \label{eq:veesoft}\\
    v_{ee}^{\rm reg}(x) & =\frac{1}{|x|+a}.\label{eq:veereg}
\end{align}
Notice, however, that the 1D maps of Seidl \cite{Sei-PRA-07} are exact only for interactions (costs) that are convex for $x\ge 0$ \cite{ColDepDim-CJM-15}. This means that when using $v_{ee}^{\rm soft}(x)$, which is concave for $x\in [0,a/\sqrt{2}]$, the Seidl maps are not guaranteed to yield the true minimizer, as illustrated, for example, in Fig.~2 of Ref.~\cite{GroKooGieSeiCohMorGor-JCTC-17}.

Numerical realizations of the 1D Seidl maps are reported in Refs.~\cite{MalGor-PRL-12,MalMirCreReiGor-PRB-13,MenLin-PRB-13,MalMirGieWagGor-PCCP-14, GroKooGieSeiCohMorGor-JCTC-17,GroMusSeiGor-JPCM-20}. The implementation of the maps directly follows from Sec~\ref{sec:1D}: given a density $\rho(x)$ on a grid, the cumulant function $F_\rho(x)$ is evaluated on the same grid, and its inverse $F_\rho^{-1}(x)$ is simply obtained by swapping the columns. The grid can be restored by using a spline interpolation for $F_\rho^{-1}(x)$, and the maps are readily obtained. Numerical issues can appear in regions where the density is close to zero, with $F_\rho^{-1}(x)$ raising extremely steeply. An alternative method to obtain the 1D maps without the need to construct $F_\rho^{-1}(x)$ is discussed in Ref.~\cite{GroMusSeiGor-JPCM-20}.

\subsubsection{Spherically symmetric densities}\label{sec:SpherNumSCE}
For spherically symmetric densities the radial SGS maps \eqref{SGSans1}--\eqref{SGSans2} conjectured in \cite{Sei-PRA-99,SeiGorSav-PRA-07} have been implemented in Refs.~\cite{SeiGorSav-PRA-07,GiaVucGor-JCTC-18} for the 3D case using numerical densities for atoms from He to Ne, and in Ref.~\cite{MenMalGor-PRB-14} for the 2D case, where the SCE functional has been combined self-consistently with Kohn-Sham DFT to describe electrons confined in a parabolic potential at low density. 

The construction of the radial maps is implemented as in the 1D case. However, the computational complexity is now higher due to the evaluation of the reduced radial cost of Eq.~\eqref{eq:red_rad}, which requires an angular minimization for given radial distances. For the two-dimensional case treated in Ref.~\cite{MenMalGor-PRB-14}, where the number of relative angles to 
minimize was equal to $N-1$, the procedure has been the following. For an initial non-degenerate radial configuration and given initial starting angles, the quasi-Newton Broyden-Fletcher-Goldfarb-Shanno (BFGS) algorithm was used to find the closest local minimum. Then the radial position of the ``first'' electron was changed in small discrete steps, the radial positions of the remaining electrons were computed using the SGS maps, and the angles were optimized using the BFGS algorithm, with starting angles taken from the previous step. This procedure rests on the assumption that the optimal angles change continuously with the radial configuration. The starting angles for the initial radial configuration can be chosen by using simulated annealing as a global optimization strategy. 
It should be stressed that the angular minimization does not need to be performed for the whole set $N_{\rm grid}$ of radial grid points. In fact, the $N$ radial distances are periodic, as each circular shell $r\in[a_i, a_{i+1}]$ (with $a_i=F_\rho^{-1}(i)$, $i\in \N$), corresponds to the same physical situation,\cite{SeiGorSav-PRA-07} simply describing a permutation of the set of distances occurring in the first shell $r\in [0,a_1]$. Thus, by keeping track of the minimizing angles, and by readapting the grid in every circular shell, it is possible to do the angular minimization only $N_{\rm grid}/N$ times rather than $N_{\rm grid}$ times. 

\subsection{Methods based on linear programming}

Direct discretization of the SIL variational principle \eqref{VP.SIL} yields a linear program, which is numerically tractable when $N=2$. 

\subsubsection{The $N=2$ case}\label{sec:numN2} 

\begin{figure} 
\includegraphics[width=0.4\textwidth]{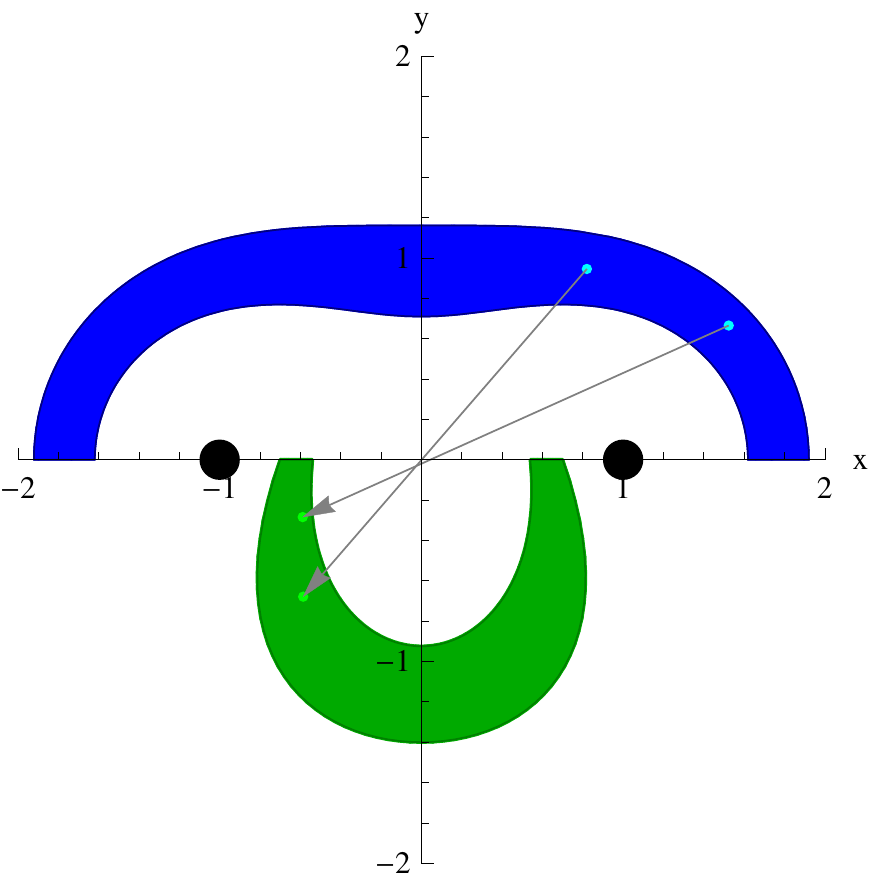} %\hspace*{1cm} \includegraphics[width=0.4\textwidth]{h2_rhox.pdf}
\caption{Co-motion function for the H$_2$ molecule \cite{CheFriMen-JCTC-14}.  The blue region -- corresponding to the points in a half plane adjacent to the molecular axis with density between $0.04$ and $0.08$ -- is mapped to the green region in the opposite half plane. The black dots indicate the positions of the nuclei.}
\label{fig:H2map}
\end{figure}

For two-electron systems in 3D with general density, Chen, Friesecke and Mendl \cite{CheFriMen-JCTC-14} have implemented a method to directly solve the SIL variational principle via linear programming and extract the co-motion function and the SCE potential from the SIL solution. They used this approach to compute the co-motion function and the KS-SCE binding curve of the H$_2$ molecule (see Figures \ref{fig:H2map} and \ref{F:H2binding}).

One truncates $\R^3$ to a bounded domain, discretizes it into $\ell$ finite regions $e_1,...,e_\ell$, and represents each element by a point $\bfa_\ell$ located at its barycenter. The single-particle density becomes a vector in $\R^\ell$ with components $\rho_\ell=\int_{e_\ell}\rho(\bfr)\, d\bfr$. The two-particle density $\pi$ is represented by a matrix $\gamma=(\gamma_{ij})\in\R^{\ell\times\ell}$ with $\gamma_{ij}=\int_{e_i}\int_{e_j}d\pi(\bfr_1,\bfr_2)$, and the interaction $V_{ee}(\bfr_1,\bfr_2)$ becomes a matrix $(c_{ij})\in\R^{\ell\times\ell}$ with $c_{ij}=\tfrac{1}{|\bfa_i-\bfa_j|}$. The SIL problem \eqref{VP.SIL} then becomes 
\begin{eqnarray}
  & \min\limits_{\gamma\in\R^{\ell\times\ell}} & \sum_{1\le i,j\le \ell} c_{ij} \gamma_{ij} \label{eq:LP} \\
  & \mbox{s/to} & \sum_{j=1}^\ell \gamma_{ij} = \frac{\rho_i}{2}, \; i=1,...,\ell, 
  \;\;\; \sum_{i=1}^\ell \gamma_{ij} = \frac{\rho_j}{2}, \; j=1,...,\ell, \nonumber \\[3mm]
  & & \gamma_{ij}\ge 0. \nonumber
\end{eqnarray}
This is a standard linear programming problem of the form $\min_x f^T x$ subject to $Ax=b$, $x_k\ge 0$, where $x$ is the vector containing the entries of $\gamma$. The solution can be obtained with a standard linear programming software (in \cite{CheFriMen-JCTC-14}, the authors used MOSEK). For a uniform discretization of the density, the number of degrees of freedom in the linear program would still be huge; instead an adaptive mesh was used in which all elements contain roughly the same amount of density, that is to say the mesh is much finer in the high-density region near the nuclei. (For automated generation of such a mesh, the finite element package PHK was used. See the chapter by Dai and Zhou for more information about this package.) The solution to \eqref{eq:LP} entails an approximation to the co-motion function $f$ at the barycenters $\{\bfa_i\}_{i=1}^\ell$, namely the barycenter of the image of $\bfa_i$ under the transport plan $X$:   
\begin{equation} \label{eq.bary}
    f^{(\ell)}(\bfa_i) = \sum_{j=1}^\ell \frac{\gamma_{ij}}{\rho_i/2}\, \bfa_j \; (i=1,...,\ell),
\end{equation}
where $\gamma_{ij}$ can be regarded as the mass transported from $\bfa_i$ to $\bfa_j$ and the normalization factor  $\rho_i/2$ guarantees that the barycentric weights sum to $1$. Since, for $N=2$, the optimal $N$-point density $\pi$ for the continuum problem is unique and of SCE form (see \eqref{eq.psiSCE2} and Theorem \ref{T:SCE}), if the discretization is sufficiently fine, i.e. $\ell$ is large enough, $f^{(\ell)}$ is a good approximation to $f$. The resulting co-motion function for the H$_2$ molecule is depicted in Figure \ref{fig:H2map}. 

\subsubsection{The $N>2$ case and the curse of dimension}
\label{sec:curse}

Since the above method uses a real-space discretization of the SIL variational principle whose unknown is the $N$-particle density on $\R^{3N}$, it is limited in practice to $N=2$, to keep the number of computational degrees of freedom manageable. Indeed, for general $N$ the $N$-particle density $\Pi$ must be represented by an order-$N$ tensor $(\gamma_{i_1...i_N})\in\R^{\ell \times ... \times \ell}$ with entries $\gamma_{i_1...i_N}=\int_{e_{i_1}\times ... \times e_{i_N}} d\pi(\bfr_1,...,\bfr_N)$. Since $\Pi$ can be assumed to be symmetric (see Remark \ref{R:symmetrization}), $\gamma$ can be assumed to be symmetric under permutation of indices and eq. \eqref{eq:LP} becomes
\begin{eqnarray}
  & \min\limits_{(\gamma_{i_1...i_N})\in\R^{\ell\times ... \times \ell}\, \mbox{\scriptsize symmetric}} & \sum_{1\le i_1,...,i_N\le \ell} V_{ee}(a_{i_1},...,a_{i_N})\, \gamma_{i_1...i_N}  \label{LPN} \\
  & \mbox{s/to} & \sum_{i_2,...,i_N=1}^\ell \gamma_{i_1i_2...i_\ell} = \frac{\rho_{i_1}}{N}, \; i_1=1,...,\ell,  \nonumber \\[4mm]
  & & \gamma_{i_1...i_N}\ge 0. \nonumber
\end{eqnarray}
This is still a linear program, but in $\ell^N$ (or, using symmetry, $\binom{N+\ell-1}{\ell+1}$) variables.

\subsection{Methods based on the dual formulation}
\label{sec:MenLin}

Mendl and Lin \cite{MenLin-PRB-13} have implemented a method for solving the dual formulation of the SCE functional, eq. \eqref{eq.SCEdual}, \eqref{eq.dualconstr2}, and applied it to the Beryllium atom, a four-electron quantum wire in 1D, and a model trimer in 3D. In the 3D case, they parametrized the (unknown) Kantorovich potential by a pseudocharge, 
$$
  v(\bfr) = \int \frac{m(\bfr')}{|\bfr-\bfr'|} \, d\bfr',
$$
with $m$ given by a small number of Gaussians and satisfying $\int m = N-1$ to account for the asymptotic behaviour $v(\bfr)\sim (N-1)/|\bfr|$ for large $|\bfr|$ (see \eqref{eq.vSCEasy}). They showed that the constrained maximization in \eqref{eq.SCEdual}, \eqref{eq.dualconstr2} is equivalent to a nested pair of unconstrained optimizations,
\begin{eqnarray} \label{eq:MeLi1}
  & &  V_{\rm ee}^{\rm SCE}[\rho] = \sup_v \Bigl(\int v(\bfr)\rho(\bfr)d\bfr + g[v]\Bigr) \; \mbox{ with} \\
  & & g[v] = \min_{(\bfr_1,...,\bfr_N)\in\R^{dN}} \Bigl( V_{ee}(\bfr_1,..,\bfr_N) - \sum_{i=1}^N v(\bfr_i)\Bigr). \label{eq:MeLi2}
\end{eqnarray}
The inner optimization for given $v$ was implemented by a quasi-Newton method and the outer optimization via a gradient-free simplex algorithm. For the Beryllium atom, using just two Gaussians for $m$ resulted in a relative error of the SCE energy of only 1.6\% compared to the SCE energy obtained via the SGS co-motion functions for radially symmetric densities \cite{SeiGorSav-PRA-07} as described in section \ref{sec:SpherNumSCE}. Also, the obtained SCE potential was in good agreement with the one based on the radial co-motion functions. 

As the authors point out, this approach is in practice limited to small systems, because the inner optimization is high-dimensional, nonlinear, and highly degenerate for the optimal $v$ (recall that the set of minimizers is typically $d$-dimensional), and the outer optimization \eqref{eq:MeLi1} is nonlinear and nonsmooth, and hence unsuitable for numerical optimization over a large number of degrees of freedom.  

\subsection{Multi-marginal Sinkhorn algorithm}\label{sec:sinkhorn}

In optimal transport, a standard computational method \cite{PeyCut-FTML-19} is to pass to the entropic regularization (in our case, problem \eqref{eq:EntrFunctional2}) and solve the latter via the Sinkhorn algorithm. This is a simple and robust algorithm which goes back to Sinkhorn in the context of estimating Markov transition matrices \cite{Sin-TAMS-64}; it was introduced into two-marginal optimal transport in \cite{Cut-ANIPS-13} and generalized to several marginals in \cite{BenCarCutNenPey-arXiv-15}. The multi-marginal Sinkhorn algorithm with Coulomb cost was implemented by 
Benamou, Carlier and Nenna \cite{BenCarNen-SMCISE-16} (see also \cite{Nen-PhD-16}) to compute the SCE energy and potential for the He and Li atoms. 

The multi-marginal Sinkhorn algorithm goes as follows; we state it here in the continuous setting. One starts from the exact form \eqref{eq:defai} of the optimizer. One now allows the $N$ entropic weight functions $a_j(\bfr_j)=a^\tau(\bfr_j)$ in this form to be different (so as to be able to update them one by one). One updates them iteratively so as to enforce the $j$-th marginal constraint, \eqref{eq:entr3} for $j$: 
\begin{equation} \label{eq:Sin1}
   a_j(\bfr_j) \int_{\R^{d(N-1)}} \prod_{i\neq j} a_i(\bfr_i) e^{-V_{\rm ee}(\bfr_1,...,\bfr_N)/\tau} \prod_{i\neq j} d\bfr_i \overset{!}{=} \rho(\bfr_j)/N. 
\end{equation}
Solving for $a_j$ yields an explicit formula for $a_j$ in terms of the other $a_i$. Thus a single updating cycle consists of the $N$ steps
\begin{equation} \label{eq:Sin2}
   a_j^{\rm new}(\bfr_j) = \frac{\rho(\bfr_j)/N}{\int_{\R^{d(N-1)}} \prod_{i< j} a_i^{\rm new}(\bfr_i) \prod_{i>j} a_i^{\rm old}(\bfr_i) e^{-V_{\rm ee}(\bfr_1,...,\bfr_N)/\tau} \prod_{i\neq j} d\bfr_i}, \;\;\; j=1,...,N.
\end{equation}
One then repeats the cycle until convergence. 

Convergence of the Sinkhorn algorithm is rigorously guaranteed under mild conditions on the interaction potential and the density (e.g., bounded potentials and $\rho\log\rho \in L^1$ are sufficient); see \cite{Sin-TAMS-64} for the discretized $N=2$ case, \cite{Rus-TAS-95} for the general  $N=2$ case, and \cite{DMaGer-JSC-20} for $N\ge 2$. The (linear) rate of convergence for the Sinkhorn algorithm was obtained in \cite{CheGeoPav-SIAMAM-16,FraLor-LAA-89} in the $N=2$ case, and in \cite{Car21} for the multi-marginal Sinkhorn algorithm. For a two-electron example in dimension one computed with the Sinkhorn algorithm see Figure \ref{fig:gammacoul}.

In \cite{BenCarNen-SMCISE-16}, Benamou, Carlier and Nenna demonstrated that for the He atom (and the choice $\tau=0.02$) the algorithm yields an accurate approximation to the SCE energy and the SCE potential compared to the (in this case rigorously justified) SGS map based solution; the relative error of the potential in the $L^\infty$ norm was only 0.4\%. Moreover, for the Li atom the numerical Sinkhorn solution exhibited very good qualitative agreement with the SGS solution.  

Some regularization is essential for the Sinkhorn approach. As $\tau$ approaches zero -- so that the entropic regularization $V^{\tau}_{\rm ee}[\rho]$ from  \eqref{eq:EntrFunctional2} approaches the exact SCE functional \eqref{eq.SIL2} -- the convergence speed of the algorithm also goes to zero (see e.g. \cite{PeyCut-FTML-19, Fri-Book-22}), and numerical instabilities can appear associated with the extremely small order $e^{-1/\tau}$ of the integrand (see e.g. \cite{BenCarCutNenPey-arXiv-15}). 

The idea of regularization underlying the algorithm fits well into our DFT context as the optimal $N$-point density is smeared out anyway off the strongly interacting limit. However, a significant limitation from the point of view of DFT is the high-dimensionality of the integral in \eqref{eq:Sin1}, \eqref{eq:Sin2}. For a discretization of the one-body density by $\ell$ gridpoint values, the cost of a single integral evaluation for fixed $\bfr_j$ is O($\ell^{N-1}$), limiting the method to small $N$.

\subsection{Towards large $N$} 

Very recently, some promising methods have been proposed which should, at least in principle, be suitable for tackling the case of large $N$. These have been demonstrated to show good performance on one-dimensional test examples where the Seidl solution from section \ref{sec:1D} is available for comparison.
%: for the semidefinite convex relaxation method by Khoo and Ying with N=8; for the mesh-free Lagrangian method by Alfonsi, Coyaud, and Ehrlacher with N=5 marginals; and by the genetic column generation method of Friesecke, Schulz, and V\"ogler for up to N=30 marginals. 
At the time of writing, it has yet to be demonstrated that any of these methods is capable of accurately computing the SCE energy for large $N$ in three dimensions. 

\subsubsection{Semidefinite convex relaxation}
\label{sec:relax}

Starting point of this method, introduced by Khoo and Ying \cite{KhoYin-SIAMJSC-19}, is the fact that the SIL problem \ref{VP.SIL} can, due to the fact that $V_{ee}$ is a two-body potential \eqref{eq:2body}, be reformulated as a minimization over $N$-representable 2-point probability measures: 
\begin{equation} \label{VP.2body}
   V_{ee}^{SCE}[\rho] = \min_{\substack{\Gamma\in {\mathcal P}(\R^d\times\R^d) \\ \Gamma \, N-\mbox{\scriptsize representable}, \, \Gamma\mapsto\rho}} 
   \binom{N}{2} \int_{\R^d\times\R^d} w_{ee}(\bfr-\bfr')\,  d\Gamma(\bfr,\bfr').
\end{equation}
Here a two-point probability measure on $\R^d\times\R^d$ is called {\it N-representable} if it is the 2-marginal of a symmetric $N$-point probability measure on $\R^{dN}$. This two-body formulation of the SCE functional was introduced in \cite{FriMenPasCotKlu-JCP-13}, and is a direct adaptation of the well known two-body reduced density matrix formulation \cite{CoYu-book-00} of the Rayleigh-Ritz variational principle \eqref{GSE} to the strongly correlated limit of DFT. 

After discretization as described in section \ref{sec:numN2}, the two-point marginal becomes a matrix $\Gamma=(\Gamma_{ij})\in\R^{\ell\times\ell}$, and $N$-representability means that $\Gamma$ is obtained from some symmetric tensor $(\gamma_{i_1...i_\ell})\in\R^{\ell\times ... \times\ell}$ with nonnegative entries which sum to one by $\Gamma_{i_1i_2}=\sum_{i_3,...,i_N}\gamma_{i_1i_2i_3...i_N}$. 

The extreme points of the set of discrete $N$-representable 2-marginals have been determined explicitly \cite{FriVog-SIAMJMA-18} (see \cite{KhoYin-SIAMJSC-19, CarFriVoe-21} for generalizations to $3$-marginals respectively general $k$-marginals). 

\begin{theorem} \label{T:extrpts} {\rm \cite{FriVog-SIAMJMA-18}} The set of extreme points of the set ${\mathcal R}_2$ of discrete $N$-representable 2-marginals is
\begin{equation} 
    {\mathcal R}_2^{\rm ext} = \left\{ \frac{N}{N-1}\lambda\lambda^T  - \frac{1}{N-1} diag(\lambda) \, : \, \lambda\in\R^\ell, \, \lambda_i\ge 0 \, \forall i, \, \mathbf{1}^T\lambda=1, \, \lambda_i\in  \{0,\tfrac{1}{N},\tfrac{2}{N},...\} \right\}.
\end{equation}
In particular, ${\mathcal R}_2$ is the convex hull of ${\mathcal R}_2^{\rm ext}$. 
\end{theorem}

Here $\mathbf{1}$ denotes the vector in $\R^\ell$ with all components equal to $1$. The discretized problem is then 
\begin{eqnarray} 
  & \min\limits_\Gamma & \sum_{1\le i,j\le \ell} c_{ij} \Gamma_{ij} \label{VP.2body2} \\
  & \mbox{s/to} & \Gamma\in{\mathcal R}_2, \, \Gamma \mathbf{1} = \frac{\rho}{N}. \nonumber
\end{eqnarray}
Khoo and Ying \cite{KhoYin-SIAMJSC-19} introduced the following convex relaxation of this problem in which ${\mathcal R}_2$ is replaced by a slighty larger but simpler set:
\begin{eqnarray} 
  & \min\limits_\Gamma & \sum_{1\le i,j\le \ell} c_{ij} \Gamma_{ij} \label{VP.2body2approx} \\
  & \mbox{s/to} & \Gamma\in \widetilde{{\mathcal R}_2} =
  \left\{\frac{N}{N-1}\Lambda - \frac{1}{N-1}diag(\Lambda \mathbf{1}) \, : \, \Lambda_{ij}\ge 0 \forall i,j,\, \Lambda\ge 0, \mathbf{1}^T\Lambda\mathbf{1}=1\right\}, \, \Gamma \mathbf{1} = \frac{\rho}{N}. \nonumber
\end{eqnarray}
Here $\Lambda\ge 0$ means matrix positivity of $\Lambda$. 

It is clear that $\widetilde{{\mathcal R}_2}\supset {\mathcal R}_2$, since  $\widetilde{{\mathcal R}_2}$ is convex and -- by inspection -- contains the set of extreme points of ${\mathcal R}_2$ given in Theorem \ref{T:extrpts}. A theoretical argument in support of the approximation \eqref{VP.2body2approx} is:
\begin{theorem} {\rm \cite{KhoYin-SIAMJSC-19}} The extreme points of the true set ${\mathcal R}_2$ of discrete $N$-representable 2-marginals are still extreme points of $\widetilde{{\mathcal R}_2}$. 
\end{theorem}
Intuitively this means that, at least near the extreme points of the exact set ${\mathcal R}_2$ of $N$-representable 2-marginals, the relaxation is very tight. 

Viewed as a minimization over $\Lambda$, \eqref{VP.2body2approx} is a semidefinite program (SDP), i.e. a problem of minimizing a linear cost subject to finitely many linear equalities or inequalities and a matrix positivity constraint. It has been implemented in \cite{KhoYin-SIAMJSC-19} using a uniform grid and the large-scale SDP solver SDPNAL+. For 1D problems with $N=8$, up to $\ell=1600$ gridpoints, and different one-body densities, the solutions reported in \cite{KhoYin-SIAMJSC-19} are in excellent qualitative agreement with the pair density of the exact Seidl solution. The relative energy error compared to the unapproximated discrete problem \eqref{VP.2body} is estimated to be of the order of $10^{-2}$ to $10^{-4}$, depending on the choice of one-body density. Also, \eqref{VP.2body2approx} is solved for $6$ electrons in 2D with a Gaussian density on a $10\times 10$ grid. 

Khoo and Ying \cite{KhoYin-SIAMJSC-19} also give a dual formulation of the SDP \eqref{VP.2body2approx} which yields an approximation to the Kantorovich potential. For 1D test problems with 8 electrons and 200 gridpoints, a relative accuracy of $10^{-2}$ to $10^{-3}$ in the $L^2$ norm is reported compared to the exact potential obtained from the Seidl solution and eq.~\eqref{eq.gradpot2}. 

\subsubsection{Langevin dynamics with moment constraints}
\label{sec:Lagr}

This approach was proposed by Alfonsi, Coyaud, and Ehrlacher, and Lombardi \cite{AlfCoyEhrLom-21, AlfCoyEhr-21}. The idea is to only discretize the density constraint, but not the $N$-point density, and then use a stochastic particle method to simulate the many-electron density. One performs a Galerkin (or ``moment'') discretization of the marginal  constraint \eqref{eq.margs} by requiring only a fixed number $M$ of integral constraints, of the form \begin{equation} \label{eq:momcon}
    \int_{\R^{Nd}}  \varphi_m(\bfr_i) \, d\gamma(\bfr_1,...,\bfr_N) = \int_{\R^d} \varphi_m \, d\mu \; \forall i=1,...,N, \, \forall m=1,...,M,
\end{equation}
where $\mu=\rho/N$ is the prescribed single-particle density and $\varphi_1,...,\varphi_M$ are suitable single-particle basis functions on $\R^d$. Moreover since the marginal constraint has been relaxed, one introduces a mild additional constraint on the class of admissible $N$-electron densities $\gamma$ to prevent mass from escaping to infinity, 
\begin{equation} \label{eq:noescape}
    \int_{\R^{dN}} \sum_{i=1}^N \theta(|\bfr_i|) \, d\gamma(\bfr_1,...,\bfr_N) \; \le \; A
\end{equation}
for some nonnegative increasing function $\theta \, : \, [0,\infty)\to [0,\infty)$ with $\theta(r)\to\infty \, (r\to\infty)$ and some constant $A>0$. The SIL problem \eqref{VP.SIL} is now approximated by:
\begin{equation} \label{VP.mom}
    {\rm Minimize } \int_{\R^{Nd}} V_{ee} \, d\gamma \mbox{ over }\gamma\in{\mathcal P}(\R^{Nd})\mbox{ subject to } \eqref{eq:momcon}, \, \eqref{eq:noescape}.
\end{equation}
 Under suitable assumptions on the basis functions, and for $A$ chosen sufficiently large, the minimum value of \eqref{VP.mom} can be shown to converge to the SCE energy $\VSCE[\rho]$ as the number $M$ of basis functions tends to infinity \cite{AlfCoyEhrLom-21}. The key property of \eqref{VP.mom} opening the door to numerical methods is the following. 
\begin{theorem} {\rm \cite{AlfCoyEhrLom-21}} Assume $\mu\in{\mathcal P}(\R^d)$, and suppose that the basis functions  $\varphi_1,..,\varphi_M\ : \, \R^d\to\R$ are continuous, belong to $L^1(d\mu)$, and satisfy the growth bound $|\varphi_m(\bfr)|\le const(1 + \theta(|\bfr|))^s$ for some $s\in(0,1)$. Assume that $V_{ee}\, : \, \R^{dN}\to\R\cup\{+\infty\}$ is nonnegative and $\int V_{ee}d\gamma$ is finite for some $\gamma$ satisfying \eqref{eq:momcon}, and that $A$ is sufficiently large. Then there exists a
%n $A$-independent 
minimizer of \eqref{VP.mom} of the
form $d\gamma(\bfr_1,...,\bfr_N) = \sum_{\nu=1}^K \alpha_\nu S_N \delta(\bfr_1-\bfa_1^{(\nu)}) ... \delta(\bfr_N - \bfa_N^{(\nu)})$ for some $K\le M+2$, some coefficients $\alpha_\nu\ge 0$, and some $\bfa_{i}^{(\nu)}\in\R^d$.
\end{theorem}
Thus a sparse ansatz for the many-electron density consisting of $K\le M+2$ symmetrized Dirac measures (where $M$ is the number of constraints discretizing the marginal condition) is sufficient. This result generalizes Theorem \ref{T:FV} from discrete problems to to semi-discrete problems with continous state space and discretized marginal constraint. 

In order to numerically solve \eqref{VP.mom}, in \cite{AlfCoyEhr-21} a stochastic particle method  in continuous state space has been implemented. More precisely, the authors use constrained overdamped Langevin 
dynamics in the potential $V_{ee}$, which is a natural stochastic evolution equation for minimizing $V_{ee}$, applied to weighted sums of $K$ symmetrized Dirac measures moving on the constraint manifold \eqref{eq:momcon}. For 5 electrons in a one-dimensional interval and the regularized Coulomb interaction \eqref{eq:veereg} with $a=0.1$, up to $M=40$ basis functions taken to be Legendre polynomials, and superpositions of up to 
$K=10~000$ symmetrized Dirac measures, the method achieves good agreement with the Seidl solution described in section \ref{sec:1D}.
The implementation uses an iterative method to maintain the constraints (which are nonlinear in the particle positions), as well as judicious choices of the time steps, temperature profile, and numbers of symmetrized Diracs to balance accuracy and computational efficiency.
%Computationally challenging aspects are the maintaining of the constraints which - due to their nonlinearity with respect to particle positions - requires iterative methods, and judicious choices of the time steps, temperature profile, and numbers of symmetrized Diracs to balance accuracy and computational efficiency. 

An  attractive feature of this method besides its feasibility for large numbers of electrons is the fact that space is not discretized. In \cite{AlfCoyEhr-21} simulations are reported for 100 electrons in three dimensions subject to 52 marginal constraints, again using superpositions of 10~000 symmetrized Dirac measures. At the time of writing, it remains an interesting open question to assess, in such situations, the accuracy of the model \eqref{VP.mom} and its numerical solutions.
%and understand some of their interesting features which are reported.

%
%
\subsubsection{Genetic column generation}
\label{sec:gencol}

This method was proposed recently by Friesecke, Schulz, and V\"ogler \cite{FriSchVoe-21}. It directly solves the discretized SIL problem \eqref{LPN}, by combining the sparse but exact quasi-SCE or quasi-Monge ansatz (see Theorem \ref{T:FV}), the method of column generation from discrete optimization, and basic ideas from machine learning. 

The idea is to alternate between solving the SIL problem on a small but otherwise unconstrained subset of the many-electron configuration space, and updating the subset based on the (primal and dual) SIL solution. Recall that 
after discretization, the many-electron density becomes a density $\gamma$ on $X^N$, where $X=\{\bfa_1,...,\bfa_\ell\}$ is a set of discretization points (e.g., a grid) for the single-electron configuration space $\R^d$. One now  starts from the  quasi-SCE or quasi-Monge ansatz in the form \eqref{eq:sparse} which suffices to solve the discrete SIL problem \eqref{LPN} exactly (see Theorem \ref{T:FV}), but -- for computational reasons -- allows a slightly larger number of delta functions:
\begin{equation} \label{eq:GCG1}
 \gamma(\bfr_1,...,\bfr_N) = \sum_{\nu=1}^{\ell'} \alpha_\nu S_N \delta(\bfr_1-\bfr_1^{(\nu)}) ... \delta(\bfr_N-\bfr_N^{(\nu)}), \;\;\; \ell\le\ell'\le \beta \ell.
\end{equation}
Here the $\bfr^{(\nu)}=(\bfr_1^{(\nu)},...,\bfr_N^{(\nu)})$ are arbitrary $N$-point configurations in $X^N$ and $\beta>1$ is a hyperparameter (taken to be 5 in \cite{FriSchVoe-21})  which limits the number of $N$-point configurations to $O(\ell)$ instead of the naively required $O(\ell^N)$. To achieve a unique correspondence between symmetrized Diracs and $N$-point configurations one restricts the  $\bfr^{(\nu)}$ to the sector $X^N_{sym}=\{(\bfa_{i_1},...,\bfa_{i_N})\in X^N \, : \, i_1\le ... \le i_N\}$, making the expansion coefficients $\alpha_\nu$ in \eqref{eq:GCG1} unique. 

The ansatz \eqref{eq:GCG1} involves two sets of degrees of freedom, the subset $\Omega=\{\bfr^{(1)},...,\bfr^{(\ell')}\}$ of the many-electron configuration space and the coefficient vector $(\alpha_1,...,\alpha_{\ell'})$, which are updated alternatingly. For fixed $\Omega$, the coefficient vector is governed by the SIL problem \eqref{LPN} restricted to the ansatz \eqref{eq:GCG1}, which reads, using that $S_N\delta(\bfr_1-\bfr_1^{(\nu)})...\delta(\bfr_N-\bfr_N^{(\nu)})$ has single-particle density $\rho^{(\nu)}(\bfr)=\sum_{i=1}^N \delta(\bfr-\bfr_i^{(\nu)})$, 
\begin{equation} \label{eq:GCG2}
   \min\limits_{\alpha\in\R^{\ell'}}  \sum_{\nu=1}^{\ell'} \alpha_\nu V_{ee}(\bfr^{\nu)}) \;\; 
   \mbox{s/to} \;\;   \sum_{\nu=1}^{\ell'} \alpha_\nu\rho^{(\nu)}(\bfa_i) = \frac{\rho_i}{N}, \;\; i=1,...,\ell, \;\;\; \alpha_\nu \ge 0.
%  \label{GCG2} \\
%  & \mbox{s/to} & \sum_{\nu=1}^{\ell'} \alpha_\nu\rho^{(\nu)}(a_i) = \frac{\rho_i}{N}, \;\; i=1,...,\ell, \;\;\; \alpha_\nu \ge 0. \nonumber 
\end{equation}
This is just a small linear program with an $\ell\times O(\ell)$ constraint matrix.  Updating the set $\Omega$ is done in a simple but subtle manner, as standard methods would encur the curse of dimension (see below). One also uses the dual problem
\begin{eqnarray} \label{eq:GCG3} 
  & \max\limits_{u \, : \, X\to \R} & \sum_{i=1}^\ell u(\bfa_i) \rho_i \;\; \mbox{s/to} \;\; u(\bfr_1^{(\nu)})+...+u(\bfr_N^{(\nu)}) \le V_{ee}(\bfr^{(\nu)}) \; \forall \nu=1,...,\ell',
%  & \mbox{s/to} & u(\bfr_1^{(\nu)})+...+u(\bfr_N^{(\nu)}) \le V_{ee}(\bfr^{(\nu)}) \; \forall \nu=1,...,\ell', \nonumber
\end{eqnarray}
whose solution $u$ is an approximation to the Kantorovich potential. 

An updating cycle in the genetic column generation ({\tt GenCol}) method  goes as follows: 

1. Given a set $\Omega\subset X_{sym}^N$ of $N$-particle configurations, update the primal solution $\alpha$ and the dual solution $u$ by solving \eqref{eq:GCG2}, \eqref{eq:GCG3}. 

2. Given the updates $\alpha^{new}$ and $u^{new}$, update $\Omega$ by the following genetic learning method: %Let $\Omega^{active}$ consist of those configurations $\bfr^{(\nu)}\in\Omega$ such that $\alpha_\nu$ and 
\vspace*{-8mm}

\begin{eqnarray} \label{eq:GCG4}
    & & \mbox{pick a random ``parent'' configuration } \bfr^{(\nu)}\in\Omega \mbox{ satisfying }\alpha^{new}_\nu > 0 \\[-1mm]
    & & \mbox{create a random ``child'' }\bfr^*\in X_{sym}^N \mbox{ by moving one electron position to a nearest neighbour } \nonumber \\[-1mm]
    & & \mbox{repeat these steps until }u^{new}(\bfr_1^*)+...+u^{new}(\bfr_N^*) > V_{ee}(\bfr^*) \mbox{ and set }\Omega^{new}=\Omega\cup\{\bfr^*\}. \nonumber 
\end{eqnarray}
Steps 1. and 2. are iterated until convergence, with the oldest configurations which do not contribute to the current optimal plan (i.e. satisfy $\alpha^{new}_\nu = 0$) being deleted from $\Omega$ whenever its size $\ell'$ exceeds the maximum allowed size $\beta\ell$.

The simple but powerful genetic learning aspect of the search rule in \eqref{eq:GCG4} is that only ``successful'' $N$-electron configurations in $\Omega$ (i.e. ones that contribute to the current optimal plan \eqref{eq:GCG1} with a nonzero coefficient $\alpha_\nu$) are allowed to bear offspring. Numerical observations and theoretical considerations show that this is essential for overcoming the curse of dimension. An unbiased random search of new configurations, or the updating step in the classical column generation method of solving the so-called pricing problem\footnote{which consists in our case in finding a configuration $\bfr^*$ which maximizes the difference $u^{new}(\bfr_1^*)+...+u^{new}(\bfr_N^*)-V_{ee}(\bfr^*)$}, would merely turn the curse of dimension with respect to the size of the state space 
%($X^N_{sym}$, which contains combinatorially many configurations)
into a curse of dimension with respect to the number of search steps. 

The rationale behind the acceptance criterion in \eqref{eq:GCG4} is that any new configuration $\bfr^*$ satisfying it represents a constraint of the full dual problem (eq.~\eqref{eq:GCG3} with the $\bfr^{(\nu)}$ being replaced by all configurations in $X_{sym}^N$) which the current dual solution $u^{new}$ violates. Adding this configuration to the set $\Omega$ ``cuts off'' $u^{new}$ from the optimization domain of the dual problem, yielding a new dual solution and an energy decrease. For a rigorous justification see \cite{FriSchVoe-21}. 
%Except in degenerate cases, this also leads to a new primal solution and the energy decrease is strict. 

\begin{figure}[http!]
\hspace*{0.1\textwidth}
\includegraphics[width=0.33\textwidth ]{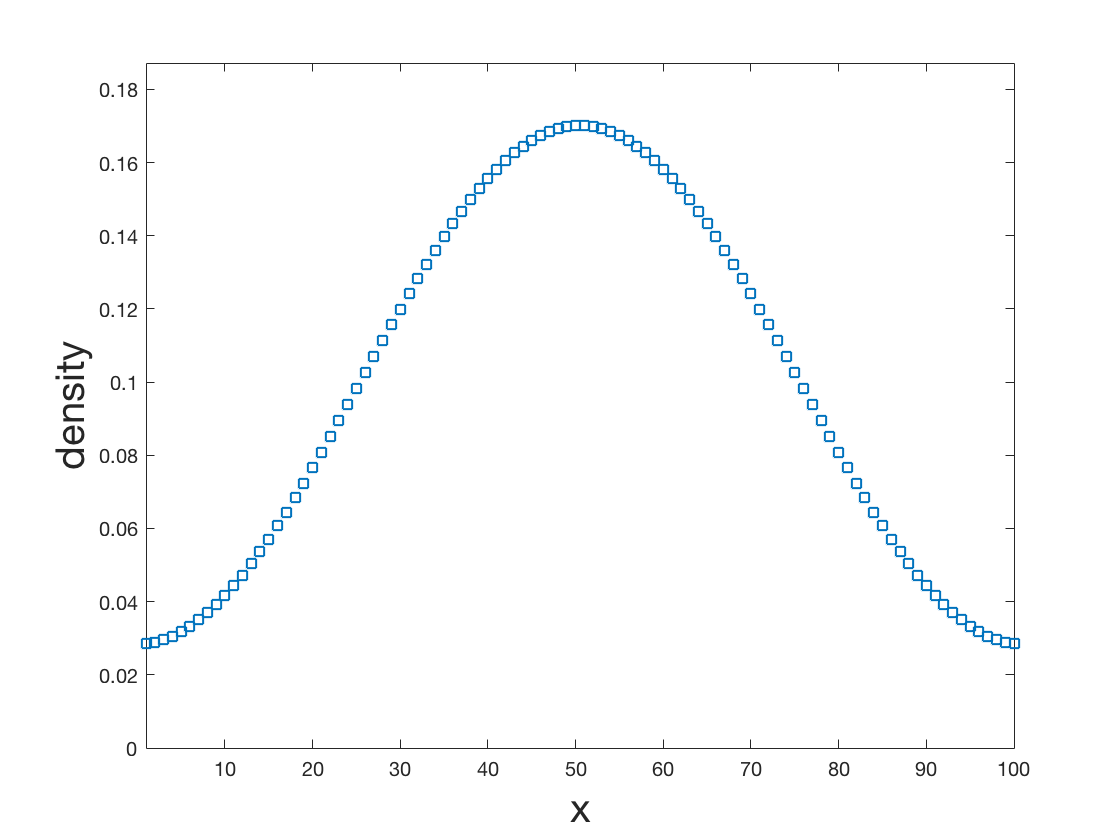} \\[1mm]
\includegraphics[width=0.12\textwidth ]{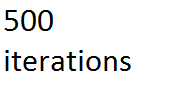}
\hspace*{-0.6cm}
\includegraphics[width=0.41\textwidth ]{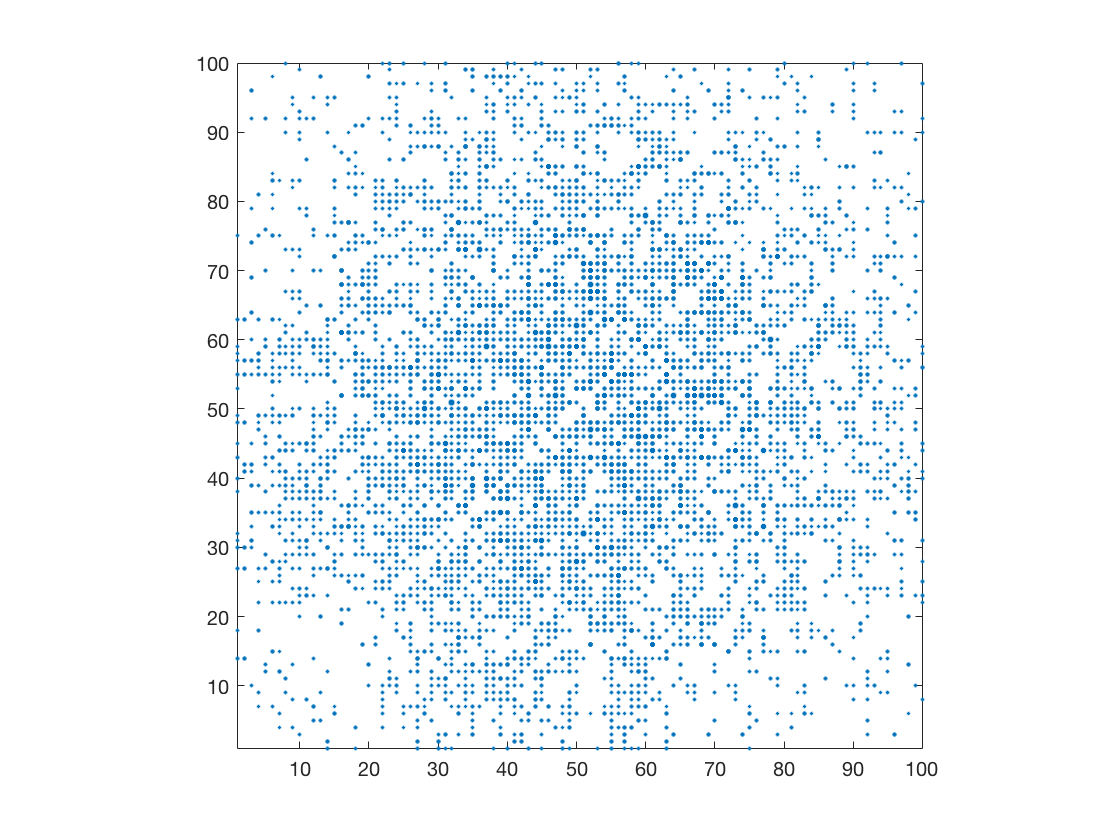} 
\includegraphics[width=0.32\textwidth ]{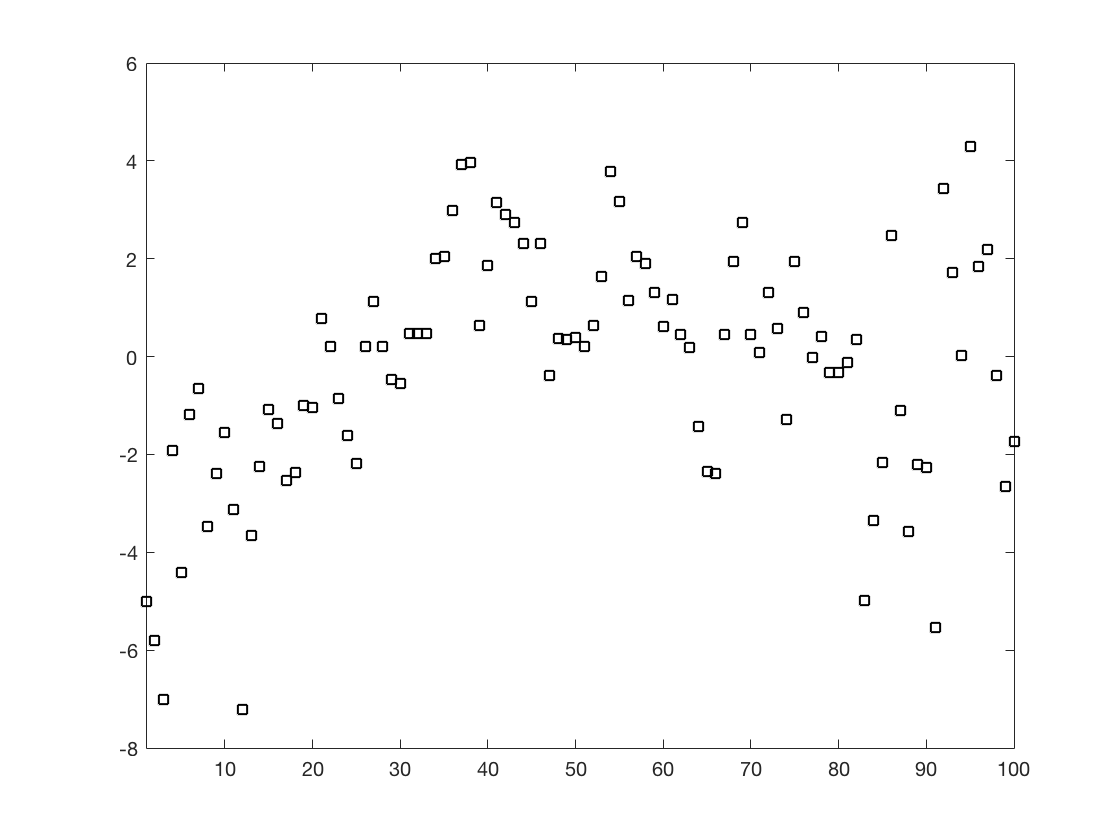} \\
\includegraphics[width=0.12\textwidth ]{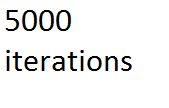}
\hspace*{-0.6cm}
\includegraphics[width=0.41\textwidth ]{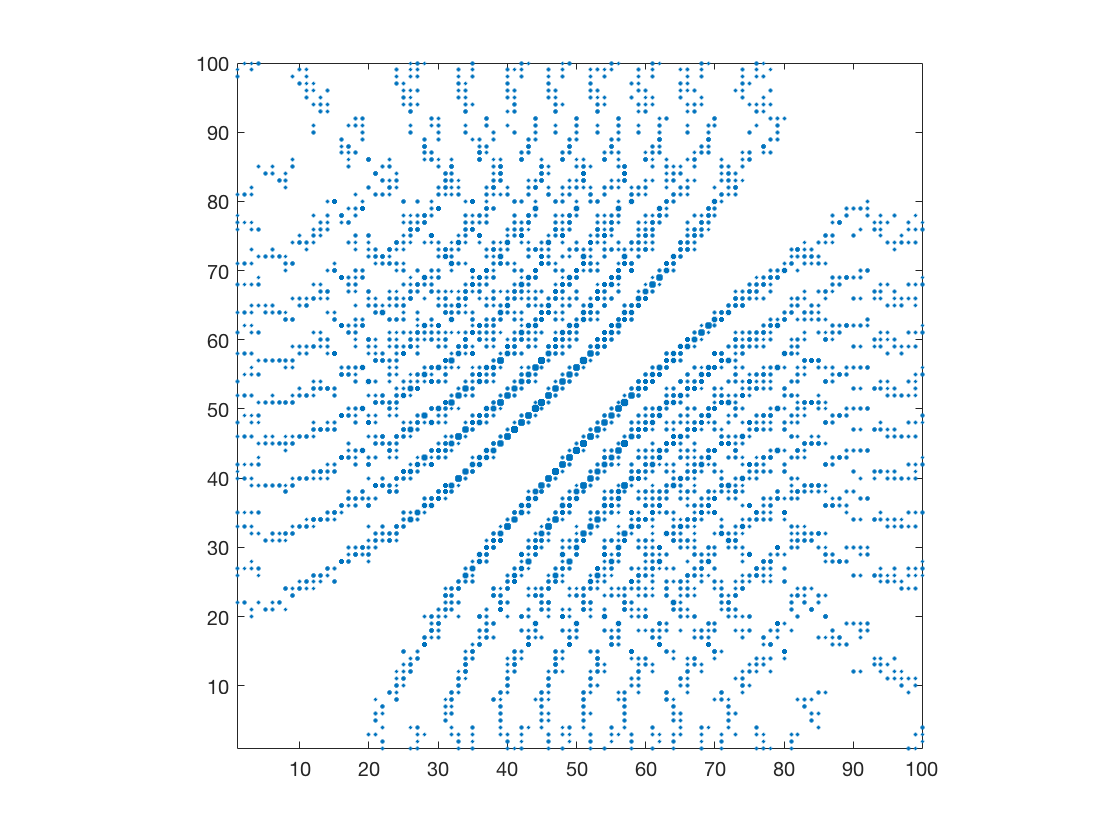} 
\includegraphics[width=0.32\textwidth ]{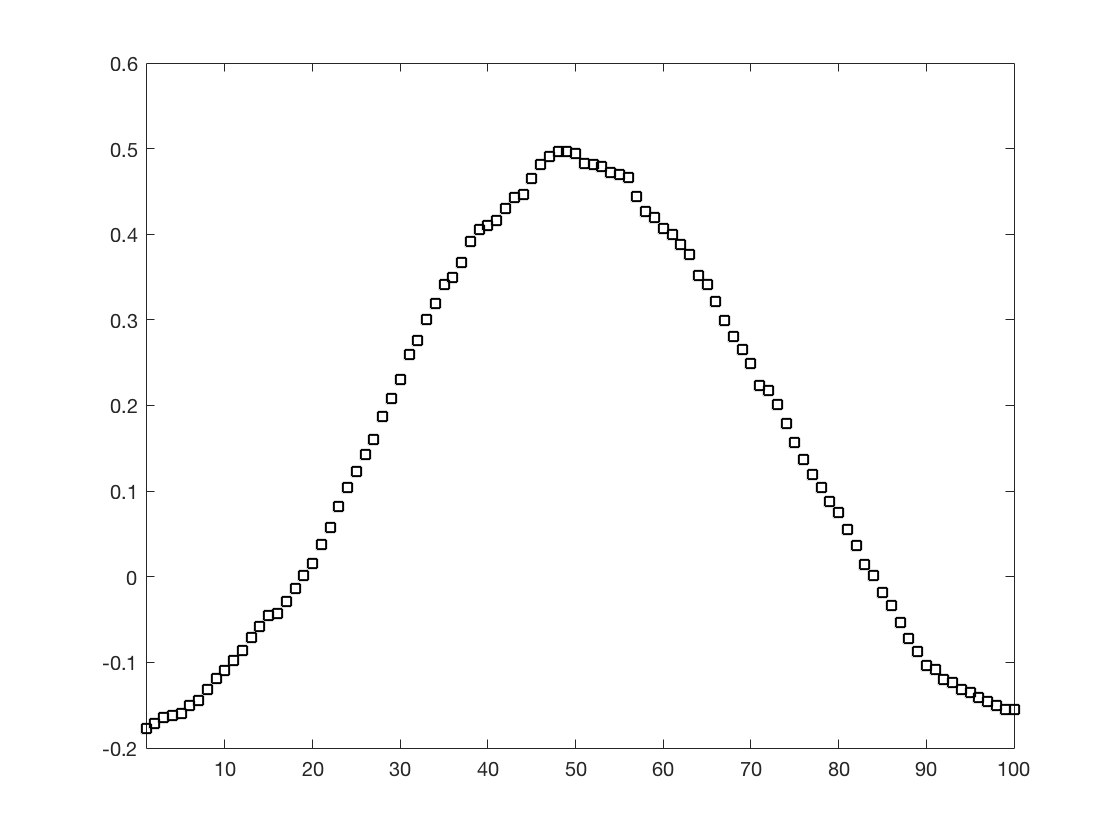} \\
%\parbox[c]{0.15\textwidth}
%{6000 iterations}
%\includegraphics[width=0.42\textwidth ]{N_10_ell_100_iter_6000_i} 
%\includegraphics[width=0.38\textwidth ]{N_10_ell_100_iter_6000_di} \\
\includegraphics[width=0.12\textwidth ]{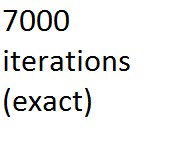}
\hspace*{-0.6cm}
\includegraphics[width=0.41\textwidth ]{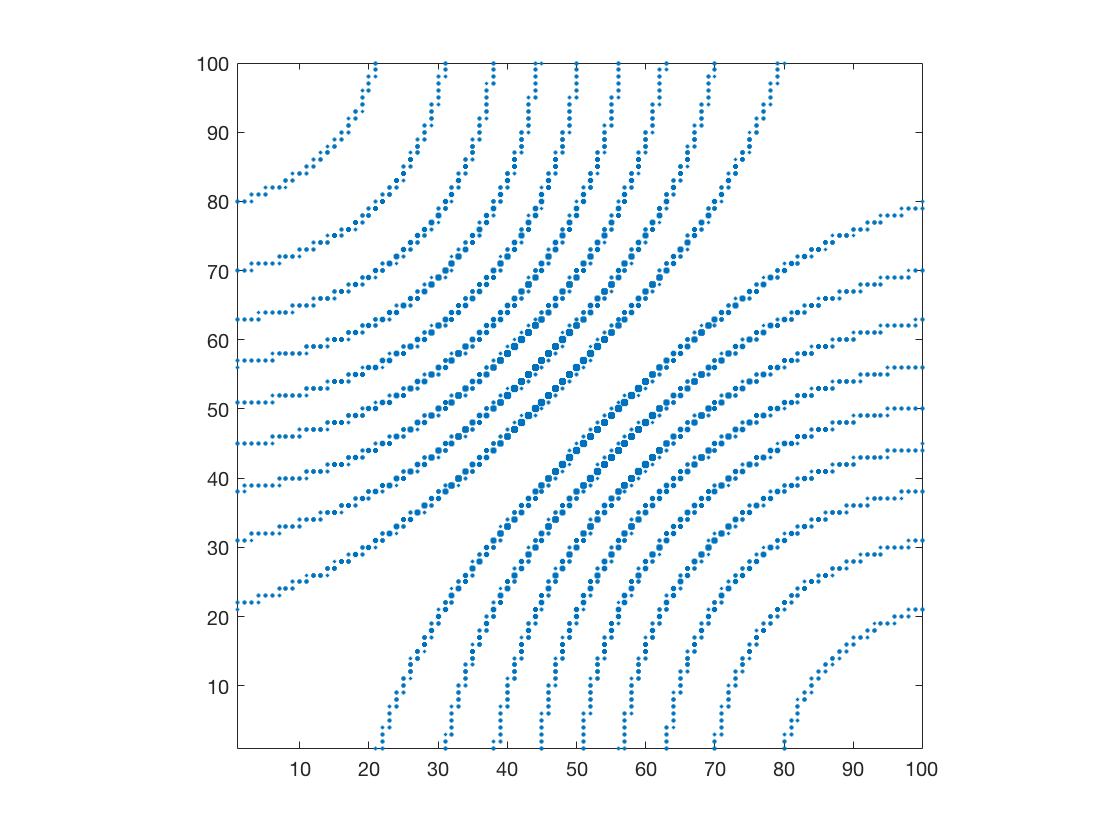} 
\includegraphics[width=0.32\textwidth ]{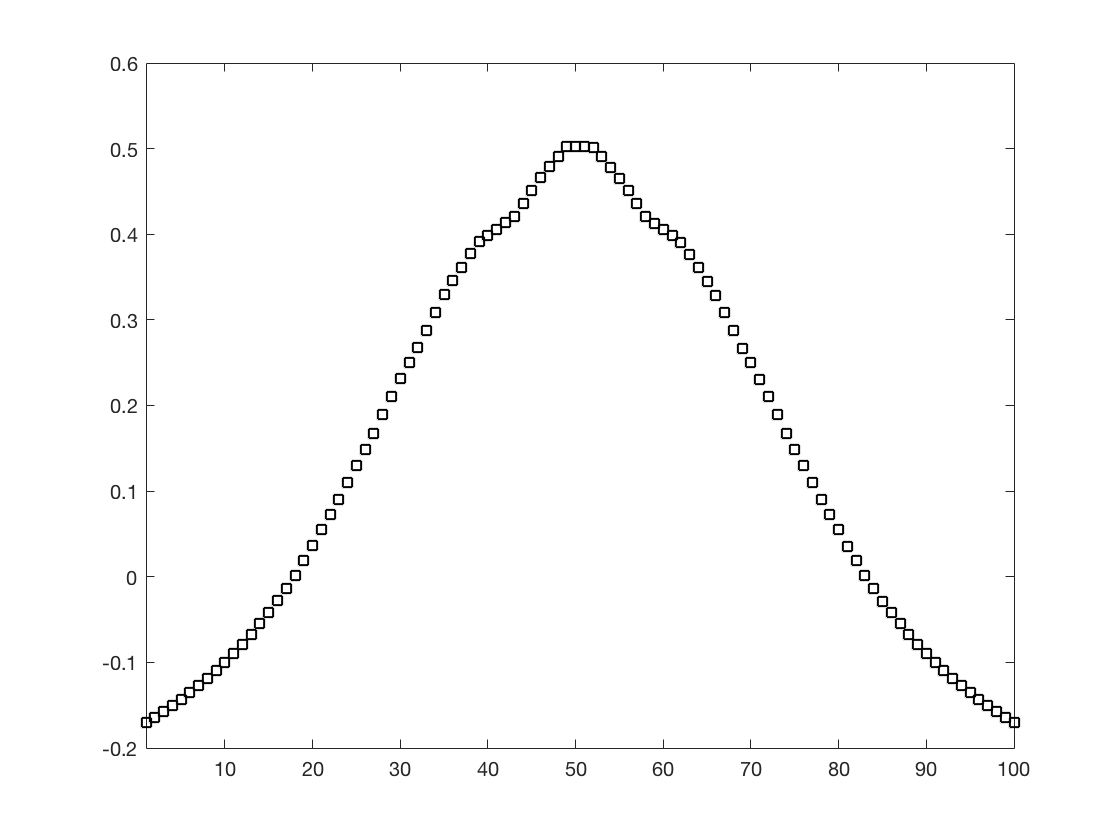}

\caption{Solution to the SIL problem \eqref{VP.SIL} for $10$ electrons in 1D with the {\tt GenCol} algorithm \cite{FriSchVoe-21}. {\it Top:} prescribed density.  {\it Left:} Evolution under {\tt GenCol} of the $N$-electron density from a random initial state, visualized via its two-point marginal (pair density). Gridpoints with nonzero values (i.e.,  ``successful'' configurations) are shown in blue, with larger markers indicating higher values. {\it Right:} Evolution of the Kantorovich potential. The final $N$-point density recovers Seidl's SCE state for the discretized problem with machine precision.}
\label{F:1Di}
\end{figure}

Figure \ref{F:1Di}, taken from \cite{FriSchVoe-21},  shows the solution of the SIL problem \eqref{VP.SIL} computed by the {\tt GenCol} algorithm for 10 electrons in a 1D interval discretized by 100 gridpoints. In this example, the grid spacing is normalized to $1$, the density is taken to be  $\rho(x)=const (0.2 + \sin^2(\tfrac{x}{\ell+1}))$, and the interaction is the soft Coulomb potential \eqref{eq:veesoft} with $a=0.1$. With the initial set of many-electron configurations chosen randomly, the algorithm always found the exact Seidl solution (see section \ref{sec:1D}) of the discretized problem to machine precision using less than 7000 iterations and less than 5 samples per iteration. This  means that only a tiny fraction of the configuration space was accessed. The energy decreased steadily at an exponential rate. 

Tests reported in \cite{FriSchVoe-21} on larger 1D  systems with up to $N=30$ electrons on $120$ grid points (corresponding to a space of $N$-point densities of dimension  $\ell^N \approx 2.4\times 10^{62}$) show only a slow polynomial growth in $N$ of the number of iterations required to find the exact solution to machine precision, with the average number of samples needed per iteration to satisfy the acceptance criterion remaining approximately constant. 

Apart from its simplicity and efficiency in high dimensions, attractive features of the genetic column generation method are that after discretization no further approximations are made (and the discrete SIL problem is solved accurately), and that the method also provides the Kantorovich potential for use within Kohn-Sham DFT.  

Tests for accurately discretized three-dimensional densities
%, which require significantly larger values of $\ell$, 
are not yet available at the time of writing.

%The multi-marginal Kantorovich plan (or $N$-point density), visualized via its two-point marginal (or pair density), is seen to concentrate on the graphs of $N-1=9$ maps, thereby accurately reproducing the known behaviour of the continuous problem as predicted by Seidl \cite{Se99} and rigorously proved in \cite{CDD15}.

\subsection{Approximations}
As explained, there are not at the moment efficient algorithms to solve the SCE problem in an exact or very accurate way for the general three-dimensional case. In the usual spirit of DFT, several approximations for the functional $\VSCE[\rho]$ have been proposed and used in combination with Kohn-Sham DFT.  We review the approximations in this section, and their use within Kohn-Sham DFT in section~\ref{sec:KSSCE}.
\subsubsection{Gradient Expansion: Point-charge-plus-continuum model (PC)}\label{sec:PC}
The first gradient expansion approximation (GEA) for the indirect energy functional $W_\infty[\rho]=\VSCE[\rho]-U[\rho]$ has been proposed by Seidl, Perdew and Kurth \cite{SeiPerKur-PRA-00}, and it is called point-charge-plus continuum (PC) model,
\begin{equation}
    W_\infty^{\rm PC}[\rho]=\int d\bfr\;\left[A\,\rho({\bf r})^{4/3}
   \;+\;B\,\frac{|\nabla\rho({\bf r})|^2}{\rho({\bf r})^{4/3}}\right],
\label{eq:WinfPC}
\end{equation}
where $A=-\frac9{10}(\frac{4\pi}3)^{1/3}$     
and $B=\frac3{350}(\frac3{4\pi})^{1/3}$. The model is built from the physical interpretation of  $W_\infty[\rho]$ as the electrostatic energy of a system of perfectly correlated electrons with density $\rho$ inside a classical background with the same charge density $\rho$ of opposite sign \cite{SeiPerKur-PRA-00}. Notice that the electrons are not allowed to relax in this fictitious external potential, as they are kept in the SCE state with the prescribed density. Only when the density is uniform the energy of the SCE state is the same as the one we would obtain by letting the electrons relax in the positive background external potential \cite{LewLieSei-PRB-19}. 
The idea of the PC model is that when the density is slowly varying the energy should be well approximated by making each electron be surrounded by a PC cell (given by the combined effect of the background and the remaining electrons) that neutralises its charge and it is such that the electron plus its cell have zero dipole moment \cite{SeiPerKur-PRA-00}.

The PC approximation works rather well: for example, for the atomic densities from He to Ne, the values $W_\infty^{\rm PC}[\rho]$ agree within 1\% with the values obtained by using the radial co-motion functions (maps) described in sec.~\ref{sec:RadSymm}, as shown in Table I of Ref.~\cite{SeiGorSav-PRA-07}. This is quite remarkable as, usually, gradient expansions for the exchange-correlation functionals fail in providing accurate quantitative results.
 
\subsubsection{Generalized gradient approximations: the modified PC model}
Although quantitatively accurate for the SIL energy, the main drawback of the PC model is that its functional derivative,
\begin{equation}
    v_{xc}^{\rm PC}(\bfr)=\frac{\delta W_\infty^{\rm PC}[\rho]}{\delta \rho(\bfr)},
\end{equation}
diverges to $-\infty$ in the tail of atomic and molecular densities \cite{FabSmiGiaDaaDelGraGor-JCTC-19}, making a self-consistent Kohn-Sham calculation not possible. Moreover, the PC model fails for quasi-2D and quasi-1D systems \cite{Con-PRB-19} .

To overcome these problems, Constantin \cite{Con-PRB-19} has proposed a generalised gradient approximation (GGA)  for $W_\infty[\rho]$, called modified PC model (mPC), which reads
\begin{equation}
    W_\infty^{\rm mPC}[\rho]=A\int d\bfr \rho(\bfr)^{4/3}\, \frac{1+a\, s(\bfr)^2}{1+(a+0.14)\,s(\bfr)^2},\qquad s(\bfr)=\frac{|\nabla\rho(\bfr)|}{2(3\pi)^{1/3}\rho(\bfr)^{4/3}},
\end{equation}
where $A$ has the same value as in the original PC model, and $a=2$. This approximation is less accurate for the SIL of atomic densities with respect to the original PC model (with errors around 9-10\%), but has the advantage of a well behaved  functional derivative, and of achieving a physical description of the crossover from three to two dimensions.

\subsubsection{Approximations with some non-locality: the non-local radius (NLR) and the shell model}
The PC and mPC are semilocal approximations, while, as we have seen, the exact SIL physics has an extreme non-local dependence on the density. Approximations that retain some (albeit limited) non-locality are the non-local radius (NLR) \cite{WagGor-PRA-14} and the shell models \cite{BahZhoErn-JCP-16}. Both approximations use as key ingredient the spherically averaged density $\tilde{\rho}(\bfr,u)$ around a given position $\bfr$, obtained by integrating out the angular dependence of ${\bf u}$,
\begin{equation}
   \tilde{\rho}(\bfr,u)=\int \rho(\bfr + {\bf u})\frac{d \hat{{\bf u}}}{4\pi},
\end{equation}
and, in analogy with the SCE structure for spherical densities conjectured in Ref.~\cite{SeiGorSav-PRA-07} and illustrated in Sec.~\ref{sec:RadSymm}, its cumulant
\begin{equation}
    N_e(\bfr,u)=\int_0^u 4\pi\,x^2\,\tilde{\rho}(\bfr,x)\,dx.
\end{equation}
In the NLR model \cite{WagGor-PRA-14} the functional $W_\infty[\rho]$ is approximated as
\begin{equation}
    W_\infty^{\rm NLR}[\rho]=-\int d\bfr\, \rho(\bfr)\int_0^{R(\bfr)}2\pi\,\tilde{\rho}(\bfr,u)\,u\,du
    %-\frac{1}{2}\int d\bfr \rho(\bfr)\int_{\Omega(\bfr)}d\bfr' \frac{\rho(\bfr')}{|\bfr -\bfr'|},
\end{equation}
  where the radius $R(\bfr)$ is defined by the condition that the underlying exchange-correlation hole be normalised:
  \begin{equation}
      N_e(\bfr,R(\bfr))=1.
  \end{equation}
  This simple approximation is less accurate than  the PC and mPC models for the case of the uniform electron gas, giving a too high energy. For non-uniform densities, the NLR has the advantage, with respect to the PC and mPC models, of being exact for one-electron systems. For atomic densities, NLR makes errors, with respect to the SCE results of Ref.~\cite{SeiGorSav-PRA-07}, of the order of 8-9\% \cite{WagGor-PRA-14}.
  
  The shell model \cite{BahZhoErn-JCP-16} substantially improves the NLR approximation, by making it exact for a uniform density, and reducing its error with respect to the SCE results for atomic densities by almost a factor of 10. While the NLR model approximates the exchange-correlation hole with a sphere depleting one electron from the spherically averaged density, the shell model adds a single positive oscillation, and reads
  \begin{equation}
      W_\infty^{\rm shell}[\rho]=\int d\bfr\, \rho(\bfr)\,2\pi\,\left(-\int_0^{u_s(\bfr)}\tilde{\rho}(\bfr,u)\,u\,du+\int_{u_s(\bfr)}^{u_c(\bfr)}\tilde{\rho}(\bfr,u)\,u\,du\right),
  \end{equation}
  where for all $\bfr$ we have $u_s=0.849488\,u_c$, which is the condition needed to make the model exact for a uniform density. The value of $u_c(\bfr)$ is then obtained again by the normalization condition,
  \begin{equation}
      2\, N_e(\bfr,0.849488\,u_c(\bfr))-N_e(\bfr,u_c(\bfr))=1.
  \end{equation}

\section{Kohn-Sham combined with the strong-interaction limit}\label{sec:KSSCE}

\subsection{Kohn-Sham with the SCE functional (KS SCE)}

The Kohn-Sham scheme with the SCE functional (KS SCE) was first proposed and implemented in \cite{MalGor-PRL-12}, and corresponds to a crude, but well defined approximation for the HK functional,
\begin{equation}
    F_{\rm KS SCE}[\rho]=T_s[\rho]+\VSCE[\rho],
\end{equation}
in which we replace the minimum of the sum of kinetic energy and electron-electron repulsion at fixed density, with the sum of the two minima. As such, the KS SCE will always provide a lower bound for the HK functional. When implemented self-consistently, the KS SCE scheme yields the usual KS equations with the Hartree-exchange-correlation potential given by the SCE or Kantorovich potential (written below for simplicity for a closed-shell system),
\begin{equation}
    -\frac{1}{2}\nabla^2\phi_i(\bfr)+\left(v_{\rm SCE}(\bfr, [\rho])+v_{ne}(\bfr)\right)\phi_i({\bfr})=\epsilon_i\,\phi_i(\bfr),\qquad \rho(\bfr)=2\sum_{i=1}^{N/2} |\phi_i(\bfr)|^2, \label{eq:selfconsistentKSSCE}
\end{equation}
where the SCE potential is equal to
\begin{equation}
    v_{\rm SCE}(\bfr, [\rho])=u(\bfr,[\rho])+C[\rho],
\end{equation}
with $u(\bfr,[\rho])$ the maximizer in Eq.~\eqref{eq.SCEdual}, and the constant $C[\rho]$ a shift that ensures $v_{\rm SCE}(\bfr,[\rho])$ tends to $0$ as $|\bfr|\to\infty$, see eq.~\eqref{vsce}.  This shift is the same, in the $\lambda\to\infty$ limit of the density-fixed adiabatic connection, as the one introduced by Levy and Zahariev \cite{LevZah-PRL-14,VucLevGor-JCP-17}. If we want to compute the ground-state density and the ground state energy only, one could better work with $u$ instead of $v_{\rm SCE}$,  as with the former the energy becomes simply \cite{LevZah-PRL-14,CheFriMen-JCTC-14,VucLevGor-JCP-17} the sum of the occupied orbital energies, $E_0=2\sum_{i=1}^{N/2}\epsilon_i$. 
The shift is needed if we want to estimate the ionisation potential $I=E_0^{N-1}-E_0^N$ from the highest occupied molecular orbital energy (HOMO), as $I=-\epsilon_{N/2}$ holds only when the exchange-correlation potential goes to zero far from the barycentre of nuclear charge \cite{LevPerSah-PRA-84,AlmBar-PRB-85}. For further
discussion of this point see the chapter by Toulouse in this volume.

\subsubsection{1D case}
The self-consistent KS SCE equations have been solved for 1D systems with the interaction $v_{ee}^{\rm wire}(x)$ of Eq.~\eqref{eq:veewire} when the external potential is harmonic, $v_{ne}(x)=\frac{1}{2}\omega^2\,x^2$, \cite{MalGor-PRL-12,MalMirCreReiGor-PRB-13,GroMusSeiGor-JPCM-20}, and with the soft Coulomb interaction $v_{ee}^{\rm soft}(x)$ of Eq.~\eqref{eq:veesoft} for model 1D atoms and molecules with 'nuclei' that attract the electrons with the same soft Coulomb potential
\cite{MalMirGieWagGor-PCCP-14}. At each KS iteration, the 1D co-motion functions \cite{Sei-PRA-07} were computed numerically as explained in Sec.~\ref{sec:1DnumSCE}, and the potential $v_{\rm SCE}(x,[\rho])$ was obtained by simply integrating the force equation
\begin{equation}
    v'_{\rm SCE}(x,[\rho])=\sum_{i=2}^N w'(|x-f_i(x)|){\rm sgn}(x-f_i(x)),
\end{equation}
with boundary condition $v_{\rm SCE}(x\to \pm \infty,[\rho])=0$, and
where $w(x)$ is the chosen 1D interaction (wire or soft Coulomb, see Sec.~\ref{sec:1DnumSCE}). In addition, at low density the highest occupied KS SCE eigenvalue gives a very accurate ionization energy of the system  \cite{MalMirCreReiGor-PRB-13}. 
\begin{figure}
\centering
%\subfloat[$\tau=0.1$][]
   {\includegraphics[width=.9\textwidth]{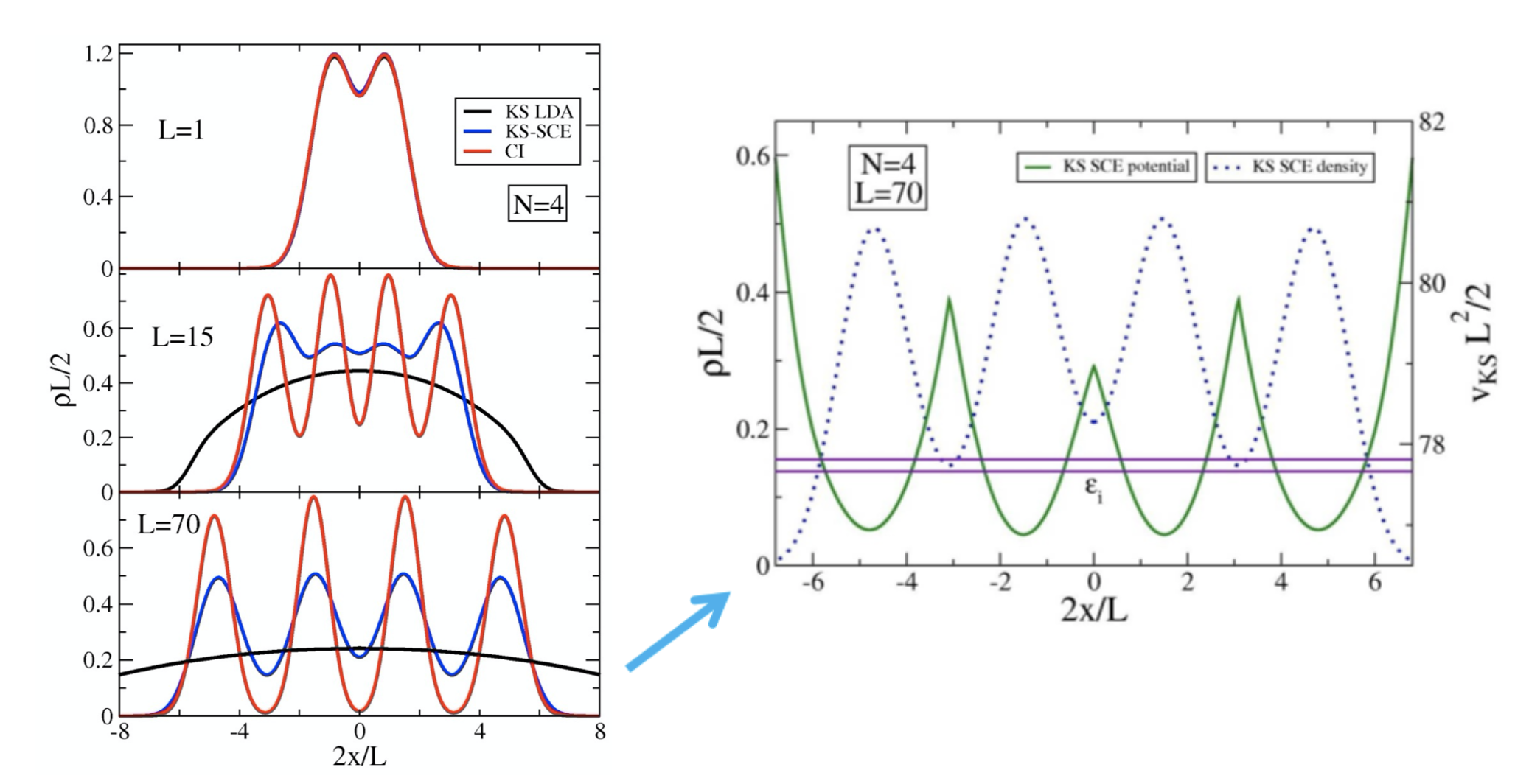}} 
\caption{Left: self-consistent KS SCE densities for $N=4$ electrons interacting with $v_{ee}^{\rm wire}(x)$ of Eq.~\eqref{eq:veewire} when the external potential is harmonic, $v_{ne}(x)=\frac{1}{2}\omega^2\,x^2$, compared with very accurate configuration interaction results (CI) and with KS LDA. Right: the total KS potential at self consistency, $v_{\rm KS}=v_{ne}+v_{\rm SCE}$, for the most correlated case. The horizontal lines are the two eigenvalues of the occupied KS SCE orbitals. Results are in scaled units, where $L=2\,\omega^{-1/2}$, and are taken from Ref.~\cite{MalMirCreReiGor-PRB-13}. }
\label{fig:KSSCEwire}
\end{figure}

{\it Harmonic external potential} -- In Fig.~\ref{fig:KSSCEwire} we show the self-consistent KS SCE densities for $N=4$ electrons interacting with $v_{ee}^{\rm wire}(x)$ of Eq.~\eqref{eq:veewire} when the external potential is harmonic, using scaled units in terms of  $L=2\,\omega^{-1/2}$, compared with accurate many-body results from configuration interaction (CI) and with KS within the local density approximation (LDA), provided for this interaction in Ref.~\cite{CasSorSen-PRB-06}. We see that, as the system is driven to low density by reducing the strength of the harmonic confinement (large $L$), the exact many-body solution undergoes a so called ``$2 k_F\to 4 k_F$'' transition, in which the number of peaks in the density is doubled. At high density, in fact, the number of peaks is dictated by the number of occupied orbitals, $N/2$ for a closed shell system. At low density, we have an incipient Wigner molecular structure, in which the electrons are well separated. Notice that with the Coulomb interaction this Wigner molecular phase exhibits different properties than the simpler case of very short-range interactions, in which the physics can be captured by making the system spin-polarised (i.e. by occupying $N$ orbitals instead of $N/2$). This is clearly illustrated in Ref.~\cite{WanLiCheXiaRonPol-PRB-12}. 

It is well-known that the local and semilocal approximations to the XC functional, as well as exact exchange, are not able to capture this ``$2 k_F\to 4 k_F$'' transition \cite{VieCap-JCTC-10,Vie-PRB-12} without introducing artificial symmetry breaking. This is also clearly shown by the KS LDA results of Fig.~\ref{fig:KSSCEwire}, which become very close in this limit to the Thomas-Fermi result (minus the external potential in the classically allowed region) predicting a too delocalized density. The KS SCE self-consistent density, although not quantitatively very accurate, has the correct qualitative behavior, with two peaks at high density and four at low density, and with the correct extension. The KS SCE HOMO energy is also very close to the exact many-body ionisation potential \cite{MalGor-PRL-12,MalMirCreReiGor-PRB-13}. In the right panel of Fig.~\ref{fig:KSSCEwire} we show the total KS potential at self consistency, $v_{\rm KS}=v_{ne}+v_{\rm SCE}$, for the most correlated case. The horizontal lines are the two occupied KS SCE eigenvalues. We see that the SCE functional is able to self-consistently build barriers that create classically forbidden regions inside the harmonic trap. Classically forbidden regions for the KS orbitals created by the Hartree-exchange-correlation potential seem to play a crucial role to describe strong correlation within KS DFT \cite{BuiBaeSni-PRA-89,HelTokRub-JCP-09,YinBroLopVarGorLor-PRB-16}.

{\it Model 1D Chemistry with soft Coulomb potential} -- In Ref.~\cite{MalMirGieWagGor-PCCP-14} the KS SCE method has been tested for model 
chemical systems in 1D, consisting of ``nuclei'' and electrons attracting each other with the soft-Coulomb potential (for the use of these 1D models to test DFT approximations see also Refs.~\cite{HelEtAl-PRA-11,WagStoBurWhi-PCCP-12}). While in the harmonic external potential we can drive the system to low density where the SCE becomes a very good approximation to the exact KS exchange-correlation functional, chemical systems (bound by the Coulomb external potential) are never in this regime. For this reason, KS SCE does not in general yield accurate results, with total energies that are way too low. An exception seems to be the good agreement between the eigenvalue of the highest occupied KS SCE orbital and the many-body chemical potential, as shown in Table 2 of Ref.~\cite{MalMirGieWagGor-PCCP-14}.

\subsubsection{2D case}
The circularly-symmetric 2D case of electrons interacting with the $1/r$ repulsion in the harmonic external potential has been studied with KS SCE in Ref.~\cite{MenMalGor-PRB-14}, using the SGS radial co-motion functions and the reduced radial cost of Eq.~\eqref{eq:c1reduced} implemented as described in Sec.~\ref{sec:SpherNumSCE}. As in 1D, the aim is to model electrons strongly confined in one direction, found, e.g., at the interface of semiconductor etherostructures. For this reason, the interaction remains the same as the 3D Coulomb one. 

As discussed in Sec.~\ref{sec:RadSymm}, the SGS state defined by \eqref{SGSans1}-\eqref{SGSans2} are not guaranteed to yield the absolute minimum for the electron-electron interaction in a given radial density $\rho(r)$. Nonetheless, it can be proven \cite{SeiDiMGerNenGieGor-arxiv-17} that, for a spherically-symmetric density, if we reduce the admissible class of maps ${\cal T}_\rho$ in the SCE functional \eqref{SCE3'} to a class ${\cal T}^{{\rm SGS}}_\rho\subset {\cal T}_\rho$ of maps given by the SGS ansatz defined in equations \eqref{SGSans1}-\eqref{SGSans2}
as an approximation for $\VSCE[\rho]$,
%of Sec.~\ref{conj.Seidl} (using the notation of Remark~\ref{rem:notationSGS}),
%\begin{equation}
%    \rho(r)\to {\bf f}_i^{\rm SGS}({\bf r})\to V_{ee}^{\rm SGS}[\rho]=\frac{1}{2}\int \rho(r)\sum_{i=2}^N \frac{1}{|{\bf r}-{\bf f}_i^{\rm SGS}({\bf r})|} d\bfr,
%\end{equation}
even when the SGS maps are not optimal the functional derivative of this approximate $\VSCE[\rho]$ with respect to $\rho(r)$ still satisfies the force equation (written using the notation of equation~\eqref{eq.gradpot2}),
\begin{equation}
    \nabla v_{\rm SGS}(\bfr)=-\sum_{i=2}^N\frac{\bfr-{\bf f}_i^{\rm SGS}(\bfr)}{|\bfr-{\bf f}_i^{\rm SGS}(\bfr)|^3},
\end{equation}
which we can integrate to obtain a potential $v_{\rm SGS}(r)$. In other words, the SGS maps provide a well defined approximation to the exact SCE functional, with a functional derivative easy to evaluate, which, in turn, can be used in the KS equations. \\
\begin{figure}
\centering
%\subfloat[$\tau=0.1$][]
   {\includegraphics[width=.4\textwidth]{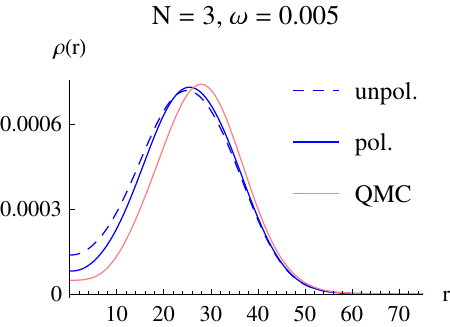}} 
   {\includegraphics[width=.4\textwidth]{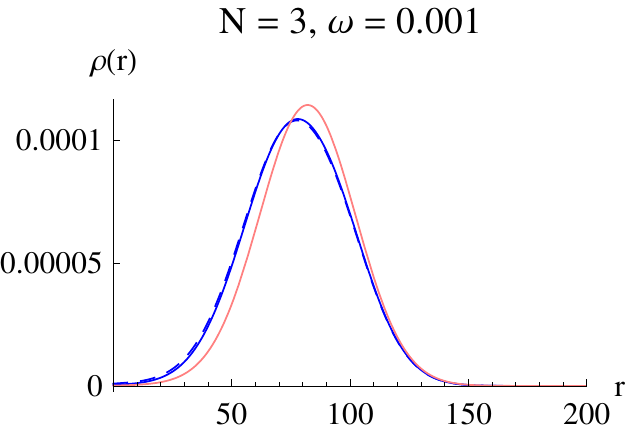}} 
\caption{Self-consistent radial KS SCE densities $\rho(r)$ for $N=3$ electrons in 2D, with external potential $v_{ne}(r)=\frac{1}{2}\omega^2\,r^2$ \cite{MenMalGor-PRB-14}, compared with accurate Quantum Monte Carlo (QMC) results \cite{GhoGucUmrUllBar-NP-06,GucGhoUmrBar-PRB-08}. The KS SCE densities are shown for both the unpolarized (2 orbitals, of which only the lowest is doubly occupied) and spin polarized (3 different singly occupied KS orbitals) case. }
\label{fig:KSSCE2Dradial}
\end{figure}

In Fig.~\ref{fig:KSSCE2Dradial} we show the resulting KS SCE self-consistent radial density for $N=3$ electrons for two low-density cases, compared with accurate Quantum Monte Carlo (QMC) results from Refs.~\cite{GhoGucUmrUllBar-NP-06,GucGhoUmrBar-PRB-08}. The KS SCE calculations have been done for both the unpolarized case (2 KS orbitals, of which only the lowest is doubly occupied) and the spin-polarized case (3 different singly occupied KS orbitals). We see that the KS SCE densities are very close to the QMC ones, predicting the right shell structure with one peak. Total energies are in agreement with QMC within $\sim 4-6\%$ \cite{MenMalGor-PRB-14}. Notice that at such low densities it is very hard to even obtain converged results using KS with the local-spin density (LSD) approximation. We thus see that even if the SGS maps are not optimal for these densities (see \cite{SeiDiMGerNenGieGor-arxiv-17}), they yield very good results when used in the self-consistent KS equations at low density. However, we have to mention that QMC predicts that at such small $\omega$'s the ground state is spin-polarized, while in KS SCE the unpolarised case always yields the lowest energy, due to the lack of any spin dependence in the SCE functional.\\
\begin{figure}
\centering
%\subfloat[$\tau=0.1$][]
   {\includegraphics[width=.3\textwidth]{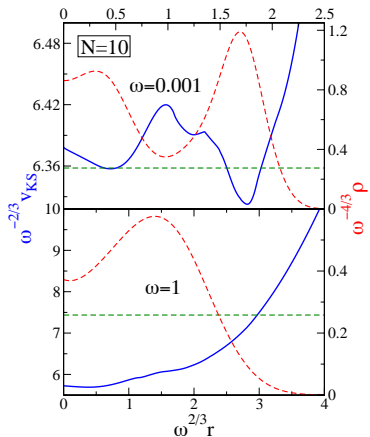}} 
\caption{Self-consistent KS SCE radial potential $v_{\rm KS}(r)=v_{\rm SCE}(r)+\frac{1}{2}\omega^2 r^2$ (blue solid line) and radial densities (red dashed line) for a strongly and weakly correlated case (top and bottom, respectively) of a 2D system composed of $N=10$ electrons inside a circularly symmetric harmonic trap. The green dashed horizontal lines correspond to the energies of the highest occupied KS orbital. Notice the presence of classically forbidden regions inside the trap in the strongly correlated case ($\omega = 0.001$). }
\label{fig:KSSCE2Dpot}
\end{figure}
Figure~\ref{fig:KSSCE2Dpot} shows the self-consistent KS SCE total potential and density for $N=10$ electrons (spin unpolarised) \cite{MenMalGor-PRB-14}. The green dashed curve is the energy of the highest occupied KS orbital. We clearly see, as in the 1D case of Fig.~\ref{fig:KSSCEwire}, that when the system is driven to low-density (small $\omega$ case), KS SCE is able to self-consistently create classically forbidden regions inside the trap.
\subsubsection{3D case}\label{sec:kssce3dcase}
KS SCE has been tested on the anions of the He isoelectronic series  \cite{MirUmrMorGor-JCP-14}  %\textcolor{blue}{on the atoms He, Li, and Be \cite{FriMenPasCotKlu-JCP-13}},
and on the dissociation curve of the 
H$_2$ molecule \cite{CheFriMen-JCTC-14}.

{\it Anions of the He isoelectronic series} -- In this case, i.e., $N=2$ electrons with $v_{ne}(r)=-\frac{Z}{r}$, where $Z$ is lowered until the system cannot bind anymore two particles, the co-motion function and the SCE potential are simply built following the original work of Seidl \cite{Sei-PRA-07} (see Example~\ref{ex:SCEN2Seidl}), which is a special case of the SGS maps. While very accurate wavefunction results predict \cite{FreHuxMor-PRA-84} that one electron is lost by the system at a critical nuclear charge $Z_{\rm crit}^{\rm exact}\approx 0.911$, KS SCE binds two electrons down to  $Z_{\rm crit}^{\rm KSSCE}\approx 0.7307$ \cite{MirUmrMorGor-JCP-14}. This is because in the KS SCE case the two electrons can get much closer to the nucleus by perfectly avoiding each other, without raising too much the kinetic energy, which is only treated within KS. 

%\textcolor{blue}{{\it The atoms He, Li, Be} -- KS SCE correctly reproduces the shell structure of these atoms, with the radial densities for Li and Be exhibiting a second (valence) peak \cite{FriMenPasCotKlu-JCP-13}. In this work, following \cite{MenLin-PRB-13} the SCE potential within the self-consistent iteration loop was obtained via the dual formulation as described in section \ref{sec:MenLin}.} 

\begin{figure}
  \includegraphics[width=.38\textwidth]{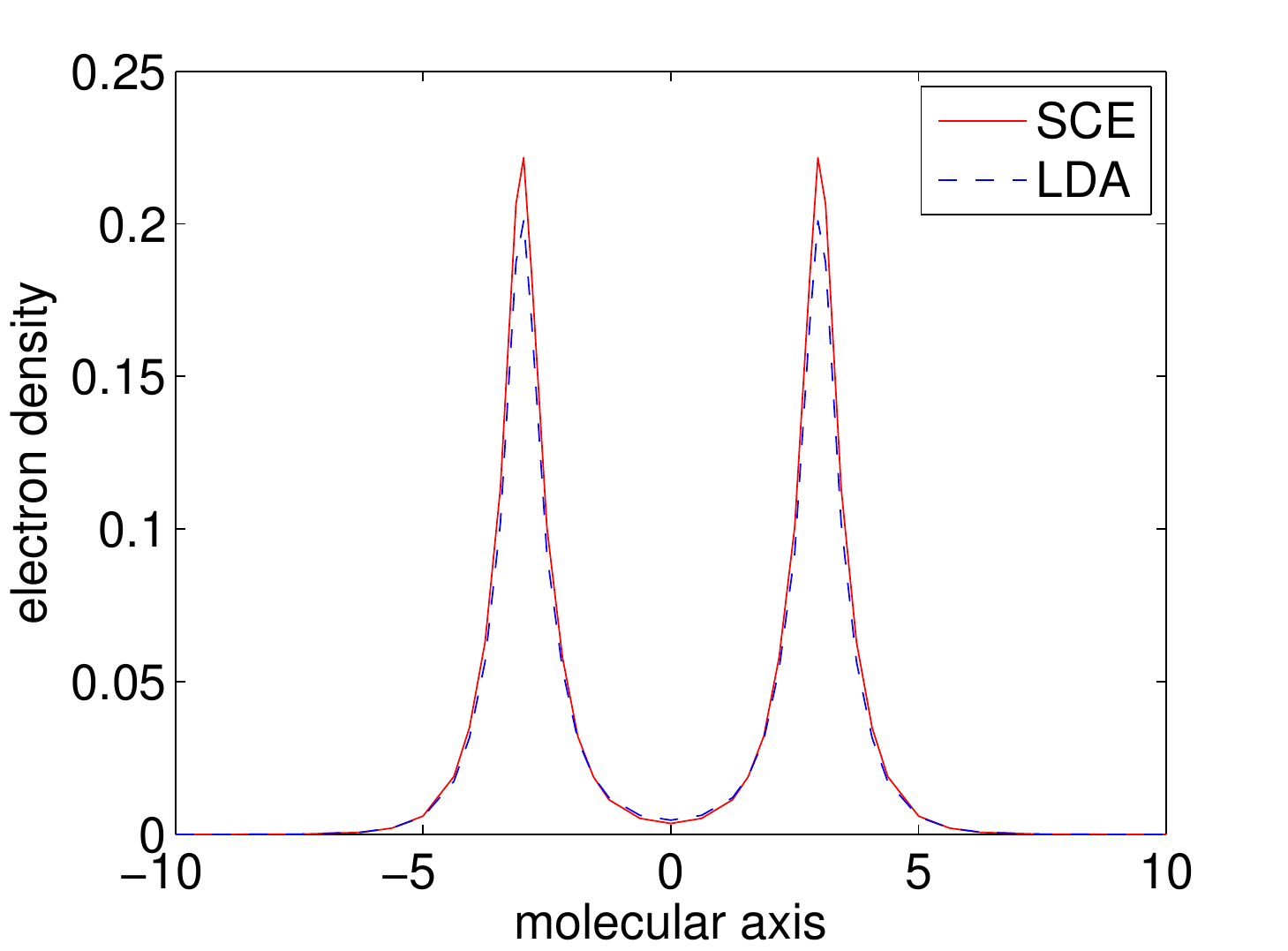} \vspace*{2mm}
  \includegraphics[width=.45\textwidth]{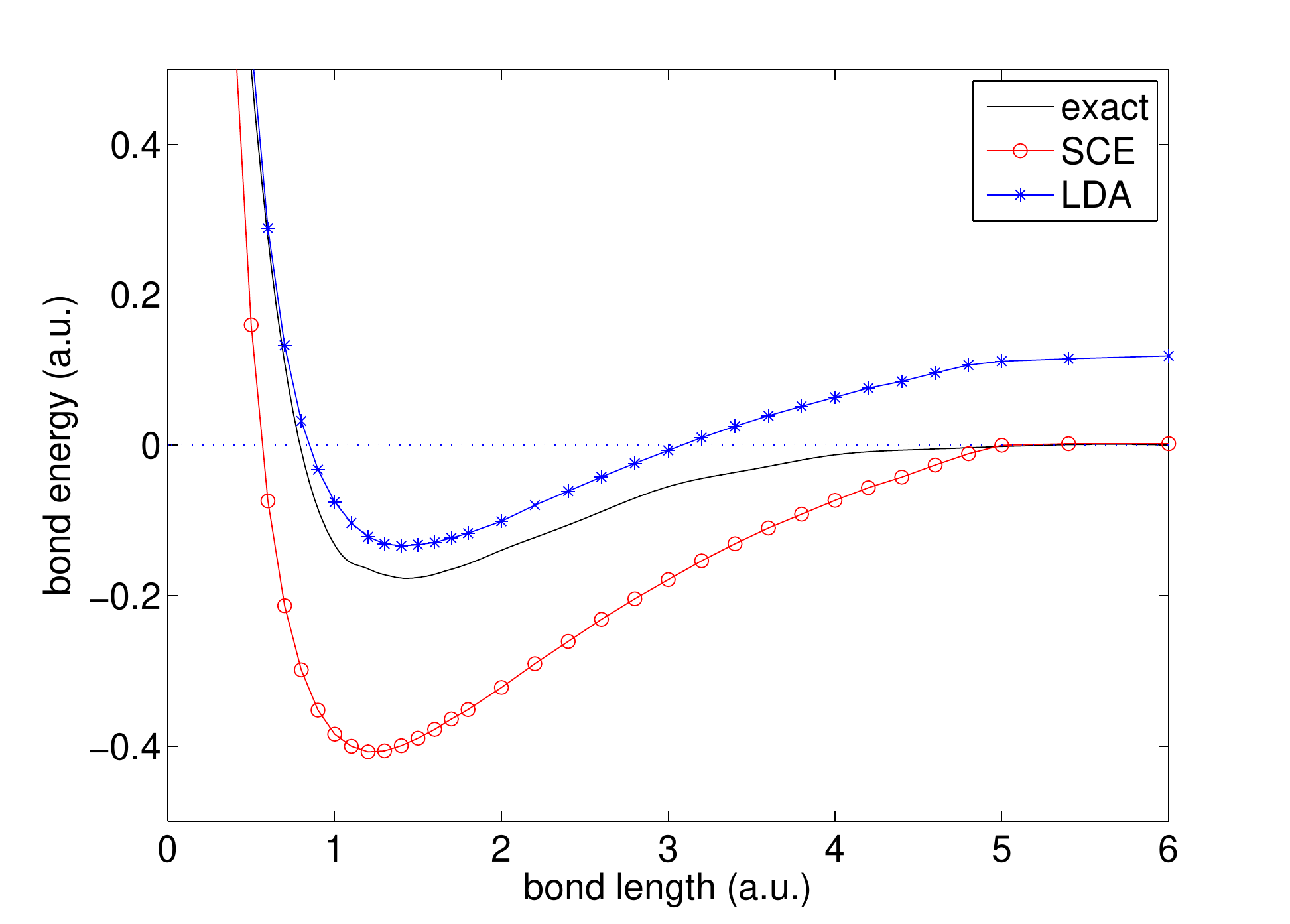}
\caption{{\it Right:} Dissociation curve of H$_2$ in KS SCE \cite{CheFriMen-JCTC-14}, that is, energy of H$_2$ minus twice the energy of the isolated H atom.  For comparison, the KS LDA curve computed on the same mesh and the exact curve from Ref.~\cite{KolRoo-60} are also shown. Note that KS SCE, unlike the local density approximation, dissociates H$_2$ correctly. {\it Left:} Corresponding self-consistent KS SCE density and KS LDA density  near dissociation.}
\label{F:H2binding}
\end{figure}

{\it The H$_2$ molecule} -- The dissociation curve of the H$_2$ molecule has been computed within KS SCE in Ref.~\cite{CheFriMen-JCTC-14}.  %\paola{Gero can you write here some details on that? Do you have a figure?} \gero{Done.} 
The result is shown in Figure \ref{F:H2binding}. 
To compute the self-consistent density and energy, an  
accurate adaptive three-dimensional finite element discretization was used and the SIL problem was solved using linear programming, as described in section \ref{sec:numN2}. The co-motion function for H$_2$ was then obtained from the SIL density via eq.~\eqref{eq.bary}, and the SCE potential via  the force equation \eqref{eq.gradpot2} and \eqref{vsce}. 

\begin{figure}
\centering
%\subfloat[$\tau=0.1$][]
   {\includegraphics[width=.3\textwidth]{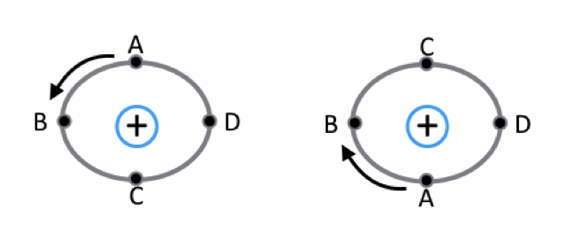}} 
\caption{When the distance $R$ between the two atoms of the H$_2$ molecule gets very large, the optimal map describes the physics of perfectly coupled dipoles. The figure shows four pairs of electronic positions $\{\bfr,{\bf f}(\bfr)\}$, labelled with the same letter A,B,C,D, with respect to the two positive nuclei.}
\label{fig:H2mapAsymp}
\end{figure}

Not surprisingly, KS SCE predicts a binding energy that is way too low. A remarkable feature, though, is the ability of KS SCE to dissociate correctly the H$_2$ molecule, i.e., the molecular energy tends to twice the energy of the isolated H atom as the internuclear distance $R$ becomes very large (see \cite{CheFriMen-JCTC-14} for a rigorous proof). Local and semilocal approximations to the XC functionals are unable to do that, and exact exchange (or Hartree-Fock) perform even worse, unless we allow spin-symmetry breaking. Indeed, the extremely stretched H$_2$ molecule is often regarded as a severe test for XC functionals to check whether they are able to describe strong (or ``static'') correlation \cite{CohMorYan-SCI-08}.\\
Although the SCE functional yields the exact energy when $R\to\infty$, we see that at large but finite $R$ the KS SCE curve immediately start to deviate from the exact one. We can understand this error by making the following simple analysis. 
%Consider two H atoms at a very large distance $R$, and approximate the total density as the sum of the two isolated densities, $\rho_R(\bfr)\approx \rho_0^A(\bfr)+\rho_0^B(\bfr)$, where $\rho_0^{A/B}=\rho_{\rm H}(|\bfr \pm {\bf R}/2|)$, and $ \rho_{\rm H}(r)=\frac{1}{\pi}e^{-r}$ is the H atom ground state density. 
With the internuclear vector ${\bf R}$ directed along the $x$-axis, we can expand the electron-electron interaction at large $R$, which, without considering one-body terms and neglecting higher orders in $R^{-1}$ yields
\begin{equation}\label{eq:largeR}
    \frac{1}{\sqrt{(x_1-x_2-R)^2+(y_1-y_2)^2+(z_1-z_2)^2}}\to \frac{2(x_1-x_2)^2-(y_1-y_2)^2-(z_1-z_2)^2}{R^3},
\end{equation}
where the origins of $\bfr_1$ and $\bfr_2$ are placed on their respective nuclei.
The SCE functional for large $R$ then corresponds to the minimization of this interaction at fixed one-body density (hence, the neglect of one-body terms that will not affect the minimizer). The SCE problem in this limit reduces then to the attractive harmonic cost  \cite{GanSwi-CPAM-98} in the bond ($x$) direction and to the repulsive harmonic cost \cite{DMaGerNen-TOOAS-17} in the two directions perpendicular to the bond axis. %Both problems have been analyzed in Sec.~\ref{sec:harmonic}. 
The optimal map will then approach for large $R$ the solution
\begin{equation}
    f_x(\bfr)=x,\qquad f_y(\bfr)=-y,\qquad f_z(\bfr)=-z, 
\end{equation}
which corresponds to perfectly coupled dipoles (see fig.~\ref{fig:H2mapAsymp}). Such maps will give a finite (negative) expectation value for the r.h.s. of eq.~\eqref{eq:largeR} even when the total density of the molecule is given by the sum of two spherical atomic densities, yielding an interaction energy that is too attractive, decaying as $\sim R^{-3}$ instead of the exact $\sim R^{-6}$. What is missing in the KS SCE approach is the raising in kinetic energy associated with the perfectly correlated dipoles of fig.~\ref{fig:H2mapAsymp}. A strategy to include the raise in kinetic energy in this asymptotic large-$R$ regime is described in Ref.~\cite{KooGor-JPCL-19}.

\subsection{Interaction strength interpolation (ISI) functionals}
Another  way to use the SIL in KS DFT is the interaction strength interpolation (ISI) construction, originally proposed by Seidl, Perdew and Levy (SPL) \cite{SeiPerLev-PRA-99}. ISI is essentially the extension to non uniform densities of Wigner's original idea \cite{Wig-PR-34} of approximating the energy  of the uniform electron gas by interpolating between its high- and low-density asymptotics, which, by scaling, correspond to the weak- and strong-interaction limits, respectively.

The starting point is to use the Hellmann-Feynman theorem to write the exchange-correlation energy as an integral over the coupling-strength parameter $\lambda$ of \eqref{eq.LLalpha}\footnote{For further
discussion see also the chapter by Toulouse in this volume.} 
\begin{equation}
    E_{xc}[\rho]=\int_0^1 W_\lambda[\rho],
\end{equation}
where 
\begin{equation}\label{eq:Wlambda}
    W_\lambda[\rho]=V_{ee}[\psi_\lambda[\rho]]-U[\rho],
\end{equation}
with $\psi_\lambda[\rho]$ the minimizer in ~\eqref{eq.LLalpha}. The idea is then to construct approximations for the adiabatic connection integrand $W_\lambda[\rho]$ by interpolating between the $\lambda\to 0$ asymptotic expansion,
\begin{equation}
    W_{\lambda\to 0}\to E_x+2\,\lambda\,E_c^{\rm GL2}+\dots,
\end{equation}
with $E_x$ the exchange energy and $E_c^{\rm GL2}$ the second-order G\"orling-Levy perturbation theory correlation energy \cite{GorLev-PRB-93}, and the large-$\lambda$ limit provided by the SIL, and possibly by the conjectured next leading term of Eq.~\eqref{eq:ZPOfunc},
\begin{equation}
    W_{\lambda\to \infty}\to W_\infty[\rho]+\frac{F^{\rm ZPE}[\rho]}{2\sqrt{\lambda}}+\dots.
\end{equation}
For example, SPL \cite{SeiPerLev-PRA-99} proposed the following simple form to interpolate between the two limits, without using the term with $F^{\rm ZPE}[\rho]$:
\begin{equation}\label{spl_eq}
W_\lambda^\SPL  = W_\infty  +\frac{E_x -W_\infty }{\sqrt{1+2\lambda \chi }}\ ,
\end{equation}
with
\begin{equation}
\chi = \frac{2E_c^{\rm GL2}}{W_\infty-E_x}\ .
\end{equation}
The SPL XC functional then reads
\begin{equation}\label{eq:ExcSPL}
E_{xc}^\SPL = \left(E_x-W_\infty\right)\left[\frac{\sqrt{1+2\chi}-1-\chi}{\chi}\right] + E_x .
\end{equation}
Several other interpolating functions that may or may not include $F^{\rm ZPE}[\rho]$ have been proposed in the literature, \cite{Ern-CPL-96,SeiPerKur-PRL-00,SeiPerKur-PRA-00,GorVigSei-JCTC-09,LiuBur-PRA-09} and are reported, for example, in the appendix of Ref.~\cite{KooGor-TCA-18}.

\subsubsection{Global Interpolations}
Interpolations such as the one of  eq.~\eqref{spl_eq} have been implemented and tested on several chemical systems by using for the $\lambda\to\infty$ limit the PC model of Sec.~\ref{sec:PC} (and its extension \cite{SeiPerKur-PRA-00} to $F^{\rm ZPE}[\rho]$ when needed). In a practical calculation, KS orbitals with a given approximate semilocal or hybrid functional are used to compute the density $\rho$, the exchange energy $E_x$, and the second-order energy $E_c^{\rm GL2}$, which are then fed into formulas such as \eqref{eq:ExcSPL} to obtain improved energies. The result is thus dependent on the chosen starting approximate functional used to generate the KS orbitals. 

A basic problem of these {\em global} (in the sense that they are done on quantities that have been already integrated over all space) interpolations is the violation of size-consistency, i.e., if we take two different systems $A$ and $B$ that do not interact with each other, it is easy to verify from \eqref{eq:ExcSPL} that, in general,
\begin{equation}\label{eq:nosizecons}
    E_{xc}^\SPL(A+B)\neq E_{xc}^\SPL(A)+E_{xc}^\SPL(B),
\end{equation}
an issue shared by all the other interpolation formulas proposed in the literature \cite{CohMorYan-JCP-07}. Notice that size-consistency of approximate electronic-structure methods is a very delicate issue when $A$ and/or $B$ have a degenerate ground state  \cite{GorSav-JPCS-08,Sav-CP-09}. Here we stress that even when degeneracy is not present, the fact that the input ingredients ($E_x$, $E_c^{\rm GL2}$ and $W_\infty$) enter in a non-linear way in the ISI formulas introduces anyway a size-consistency error. However, this error can be easily corrected \cite{VucGorDelFab-JPCL-18}. In fact, the reason why in chemistry size-consistency is crucial is that we are interested in interaction energies rather than total energies. All what we need to do is then to set the limit of a molecular dissociation curve (when $A$ and $B$ are infinitely far apart) at the value given by the left-hand side of eq.~\eqref{eq:nosizecons} rather than the one given by the right-hand side. Notice that both sides of this equation can be evaluated at exactly the same computational cost, as all what is needed is the input ingredients of the fragments $A$ and $B$ \cite{VucGorDelFab-JPCL-18}. With this size-consistency correction it is possible to extract meaningful interaction energies from the ISI functionals \cite{VucGorDelFab-JPCL-18}. 

The ISI functionals have been tested on several chemical data sets and systems \cite{FabGorSeiDel-JCTC-16,GiaGorDelFab-JCP-18,VucGorDelFab-JPCL-18}. They have been found to work reasonably well for interaction energies (especially of non-covalent systems) when Hartree-Fock orbitals (rather than KS ones) are used as input. This observation has triggered  the study of the strong-interaction limit in Hartree-Fock theory \cite{SeiGiaVucFabGor-JCP-2018,DaaGroVucMusKooSeiGieGor-JCP-20}, which, in turn, has lead to new interpolation schemes in this framework able to give very accurate results for a large variety of non-covalent interaction energies, ranging from small to medium-large systems \cite{DaaFabDelGorVuc-JPCL-21}.

If one wants to overcome the dependence on the input orbitals, one should evaluate the energy using the ISI functionals within a fully self-consistent KS scheme. For this, their functional derivative with respect to the density is needed, which is challenging due to the presence of second-order perturbation theory. Nonetheless, first attempts in the computation of the ISI functional derivatives have been reported in Refs.~\cite{FabSmiGiaDaaDelGraGor-JCTC-19,SmiCon-JCTC-20}, and self-consistent calculations are likely to appear soon.

\subsubsection{Local Interpolations}
Another possibility is to build the interpolations {\em locally}, in each point of space, by defining an energy density $w_\lambda(\bfr;[\rho])$ for the coupling-constant integrand $W_\lambda[\rho]$ of eq.~\eqref{eq:Wlambda}, writing
$E_{xc}[\rho]$ as
\begin{equation}
\label{eq:energydensity}
 E_{xc}[\rho] = \int d \mathbf{r} \, \rho(\mathbf{r})\int_0^1  w_\lambda(\mathbf{r}; [\rho])\,\text{d} \lambda .
\end{equation}
Energy densities are obviously not uniquely defined, and the only important requirement here is to use local quantities defined in the same way at weak and strong coupling. Some different choices for energy densities in the $\lambda$-interpolation context have been analyzed in \cite{VucLevGor-JCP-17}, where it has been found that the  electrostatic potential of the exchange-correlation hole \footnote{For further
discussion of this point see the chapter by Toulouse in this volume.} $h^\lambda_{xc}(\mathbf{r}_1, \mathbf{r}_2)$ seems to be the most suitable ,
\begin{equation}
 w_\lambda(\mathbf{r}) = \frac{1}{2} \int \frac{h^\lambda_{xc}(\mathbf{r}, \mathbf{r}_2)}{|\mathbf{r}-\mathbf{r}_2|} d \mathbf{r}_2,
 \label{eq:energydef}
\end{equation}
 where $h^\lambda_{xc}(\mathbf{r}_1, \mathbf{r}_2)$ is defined in terms of the pair-density $P_2^\lambda(\mathbf{r}_1, \mathbf{r}_2)$ and the density $\rho$,
\begin{equation}
 h^\lambda_{xc}(\mathbf{r}_1, \mathbf{r}_2) = \frac{P_2^\lambda(\mathbf{r}_1, \mathbf{r}_2)}{\rho(\mathbf{r}_1)} - \rho(\mathbf{r}_2),
\end{equation}
 with $P_2^\lambda$ obtained from $\Psi_\lambda [\rho]$,
\begin{equation}
 P_2^\lambda(\mathbf{r}, \mathbf{r}') = N(N-1) \sum_{\sigma, \sigma', \sigma_3, \dots, \sigma_N} \int | \Psi_\lambda(\mathbf{r}\sigma, \mathbf{r}',\sigma', \mathbf{r}_3,\sigma_3, \dots, r_N, \sigma_N)|^2 d \mathbf{r}_3 \dots d \mathbf{r}_N.
 \label{pddef}
\end{equation} 
Local interpolations within this definition have been analysed and tested in Refs.~\cite{VucIroSavTeaGor-JCTC-16,KooGor-TCA-18} on small systems, with mixed results.

\section*{Acknowledgment}
AG acknowledges partial support of his research by the Canada Research Chairs Program and Natural Sciences and Engineering Research Council of Canada. This work started when AG was at the Vrije Universiteit Amsterdam and has received funding from the European Union's Horizon 2020 research and innovation programme under the Marie Sk\l odowska-Curie grant agreement No. [795942].

\section{Appendix: Kantorovich duality} \label{App:Kantdual}
The dual construction of the SCE functional and potential (see Theorem \ref{T:SCE-dual})   relies on Kantorovich duality. 
In this Appendix we give a precise mathematical statement of Kantorovich duality for multi-marginal optimal transport, and show how it implies Theorem \ref{T:SCE-dual} (1). 

Recall the general Kantorovich optimal transport problem introduced in section \ref{sec:OT}: for given marginal measures $\mu_1,...,.\mu_N$ defined on closed subsets $X_1,...,X_N$ of $\R^d$, 
minimize a cost functional 
$$
   {\cal C}[\gamma] = \int_{X_1\times ... \times X_N} c(\bfr_1,...,\bfr_N) \, d\Pi(\bfr_1,...,\bfr_N)
$$
over probability measures $\Pi$ on the product space  $X_1\times ... \times X_N$ subject to the marginal constraints 
$$
   \int_{X_1\times ... \times X_{i-1} \times A_i \times X_{i+1}\times ... \times X_N} d\Pi = \int_{A_i} 
   d\mu_i\mbox{ for all measurable sets }A_i\subseteq X_i \mbox{ and all }i\in \{1,...,N\}.
$$
Here $c \, : \, X_1\times ... \times X_N\to\R\cup\{+\infty\}$ is a given measurable cost function. 

This problem is related to a certain dual variational problem: maximize the functional
$$
      J[u_1,...,u_N] = \sum_{i=1}^N \int_{X_i} u_i d\mu_i
$$
over potentials $u_i \, : \, X_i\to\R$ ($i=1,...,N$) which must satisfy the pointwise constraint
\begin{equation} \label{eq.dualconstr}
      \sum_{i=1}^N u_i(\bfr_i) \le c(\bfr_1,...,\bfr_N) \;\; \forall (\bfr_1,...,\bfr_N)\in X_1\times ... \times X_N.
\end{equation}

The following nontrivial statement, taken from the recent textbook \cite{Fri-Book-22}, summarizes what is known in $\R^d$, and is general enough to cover the Coulomb cost. 

\begin{theorem} \label{T:KantDual} {\rm (Kantorovich duality)} For given probability measures $\mu_1,...,\mu_N$ defined on closed subsets $X_1,...,X_N$ of $\R^d$, provided the cost function $c \, : \, X=X_1\times ..\times X_N\to\R\cup\{ +\infty \}$ is bounded from below and lower semi-continuous and the optimal cost is finite, 
\begin{equation} \label{eq.duality}
   \inf_{\substack{\Pi\in {\cal P}(X) \\ \gamma\mapsto\mu_1,...,\mu_N}} \int_X c \, d\Pi = \sup_{(u_1,...,u_N)\in{\cal A}(c)} \sum_{i=1}^N \int_{X_i} u_i \, d\mu_i,
\end{equation}
where ${\cal A}(c)$ is any of the following increasingly general sets of admissible potentials: \\
{\rm (1)} ${\cal A}(c)=\{(u_1,...,u_N) \, : \, u_i\in C_0(X_i) \, \forall i, \, \mbox{\eqref{eq.dualconstr} holds } \forall \bfr\in X\}$ \\
{\rm (2)} ${\cal A}(c)$ as in {\rm (1)}, with $C_b(X_i)$ in place of $C_0(X_i)$ \\
{\rm (3)} ${\cal A}(c)$ as in {\rm (1)}, with $B(X_i)=\{u :  X_i\to\R\, : \, u \mbox{ bounded measurable}\}$ in place of $C_0(X_i)$ \\
{\rm (4)} ${\cal A}(c)=\{(u_1,...,u_N) \, : \, u_i \in L^1(X_i;d\mu_i)\, \forall i, \, \mbox{\eqref{eq.dualconstr} holds for } \mu_1\!\otimes\!\cdots\!\otimes\!\mu_N\mbox{-a.e. }\bfr\in X\}$.
\end{theorem}

Here we have used the standard notation $C_b(X_i)$ for the space of bounded continuous functions on $X_i$, and $C_0(X_i)$ for the space of decaying continuous functions on $X_i$ (i.e. those $u$ which in addition satisfy $u(\bfr_i)\to 0$ if $|\bfr_i|\to\infty$).

In the special case of two marginals defined on compact sets, cost functions $c$ which are a metric (such as $|\bfr_1-\bfr_2|$), and the choice (2) for the potentials, this fundamental result was discovered by Kantorovich  \cite{Kan-DAN-42}. A  great many variants and modifications have subsequently appeared in the mathematics literature. Some of them replace the $X_i$ by abstract spaces; many are worked out only for two marginals; almost all of them differ in the precise assumptions on the cost function and the class of admissible potentials. For instance, \cite{RacRus-BOOK-98} (Theorems 2.1.4(b) and 2.1.1) and \cite{Kel-ZWG-84} cover bounded continuous cost functions and the class (3) for $N$ marginals; \cite{Vil-BOOK-03} (Theorem 1.3) covers bounded-below lower semi-continuous cost functions and the class (4) for two marginals. Strictly speaking, none of the versions published prior to the discovery of the {\it optimal transport/SCE theory} connection applied directly to the multi-marginal Coulomb case, even though the underlying ideas essentially did. For the proof of Theorem \ref{T:KantDual} we refer the reader to \cite{Fri-Book-22}. 

%(for early exceptions see \cite{Kel-ZWG-84} and \cite{RacRus-BOOK-98}) even though the analysis could be extended to $N$ marginals;  

{\bf Technical remark.} From a functional analysis point of view, the natural class of admissible potentials in \eqref{eq.duality} is the smallest one, (1). This choice reflects the duality between potentials $u_i$ and measures $\mu_i$ in the integral $\int\! u_i d\mu_i$; note that the linear hull of the space ${\cal P}(\R^d)$ of probability measures, that is, the space ${\cal M}(\R^d)$ of signed measures, is the dual of $C_0(\R^d)$. Enlarging this class from (1) to (2)--(4) has the virtue that the supremum of the dual problem is attained for increasingly general cost functions $c$.
\\[2mm]
{\bf Proof of Theorem \ref{T:SCE-dual}~(1) using Theorem  \ref{T:KantDual}.} Applying the Kantorovich duality theorem with $X_i$, $\mu_i$, and $c$ as in Example \ref{ex:coul}, and making the choice (3) for the class of admissible potentials, one obtains 
\begin{equation} \label{eq.preSCEdual}
     \inf_{\substack{\Pi\in {\cal P}(\R^{Nd}) \\ \Pi\mapsto\rho}} \int_{\R^{Nd}} V_{ee} \, d\Pi = \sup_{\substack{(u_1,...,u_N) \, : \, u_i\in B(\R^d) \, \forall i, \\ u_1(\bfr_1)+...+u_N(\bfr_N) \le V_{ee}(\bfr_1,...,\bfr_N) \, \forall (\bfr_1,...,\bfr_N)}} \sum_{i=1}^N \int_{\R^d} u_i(\bfr_i) \frac{\rho(\bfr_i)}{N} \, d\bfr_i.
\end{equation}
The left hand side is the enlarged-search definition \eqref{eq.SIL2} of the SCE functional $V_{ee}^{\rm SCE}[\rho]$ (which, by Theorem \ref{T:basic}, is equivalent to the original definition  \eqref{eq.SIL-WF}). The right hand side can be simplified. For any collection of potentials $(u_1,...,u_N)$, the sum of the integrals on the right hand side is preserved under the replacement $(u_1,...,u_N)\mapsto (\bar{u},...,\bar{u})$, where $\bar{u}$ denotes the average $(u_1+...+u_N)/N$; moreover the constraint in \eqref{eq.preSCEdual} is also preserved, thanks to the symmetry of $V_{ee}$. Thus the right hand side of \eqref{eq.preSCEdual} 
stays unaltered if the supremization is restricted to $N$ equal potentials, $u_1=...=u_N=u$. But in this case the right hand side reduces to that of \eqref{eq.SCEdual}, establishing Theorem \ref{T:SCE-dual} (1).

\begin{small}
%\bibliographystyle{siam}
%\bibliography{refs}
%\bibliography{bib_clean}

\begin{thebibliography}{100}

\bibitem{AlfCoyEhr-21}
{\sc A.~Alfonsi, R.~Coyaud, and V.~Ehrlacher}, {\em Constrained overdamped
  langevin dynamics for symmetric multimarginal optimal transportation}, arXiv
  preprint: arXiv:2102.03091,  (2021).

\bibitem{AlfCoyEhrLom-21}
{\sc A.~Alfonsi, R.~Coyaud, V.~Ehrlacher, and D.~Lombardi}, {\em Approximation
  of optimal transport problems with marginal moments constraints}, Math.
  Comp., 90 (2021), pp.~689--737.

\bibitem{AlmBar-PRB-85}
{\sc C.-O. Almbladh and U.~von Barth}, {\em Exact results for the charge and
  spin densities, exchange-correlation and density-functional eigenvalues},
  Phys. Rev. B, 31 (1985), pp.~3232--3244.

\bibitem{Ambrosio-03}
{\sc L.~Ambrosio}, {\em Lecture notes on optimal transport problems}, in
  Mathematical {A}spects of {E}volving {I}nterfaces, vol.~1812, Springer
  Lecture Notes in Mathematics, pp.~1--52.

\bibitem{BahZhoErn-JCP-16}
{\sc H.~Bahmann, Y.~Zhou, and M.~Ernzerhof}, {\em The shell model for the
  exchange-correlation hole in the strong-correlation limit}, J. Chem. Phys.,
  145 (2016), p.~124104.

\bibitem{BakStoMilWagBurSte-PRB-15}
{\sc T.~E. Baker, E.~M. Stoudenmire, L.~O. Wagner, K.~Burke, and S.~R. White},
  {\em One-dimensional mimicking of electronic structure: The case for
  exponentials}, Phys. Rev. B, 91 (2015), p.~235141.
\newblock Err. \textbf{93} 119912 (2016).

\bibitem{BenCarCutNenPey-arXiv-15}
{\sc J.-D. Benamou, G.~Carlier, M.~Cuturi, L.~Nenna, and G.~Peyr\'e},
  arXiv:1412.5154,  (2015).

\bibitem{BenCarNen-SMCISE-16}
{\sc J.-D. Benamou, G.~Carlier, and L.~Nenna}, {\em A numerical method to solve
  multi-marginal optimal transport problems with coulomb cost},  (2016),
  pp.~577--601.

\bibitem{BinDep-JEP-17}
{\sc U.~Bindini and L.~De~Pascale}, {\em Optimal transport with {C}oulomb cost
  and the semiclassical limit of density functional theory}, J. \'{E}c.
  polytech. Math., 4 (2017), pp.~909--934.

\bibitem{BinDepKau-arxiv-20}
{\sc U.~Bindini, L.~De~Pascale, and A.~Kausamo}, {\em On {S}eidl-type maps for
  multi-marginal optimal transport with {C}oulomb cost}, arXiv preprint
  arXiv:2011.05063,  (2020).

\bibitem{BuiBaeSni-PRA-89}
{\sc M.~A. Buijse, E.~J. Baerends, and J.~G. Snijders}, {\em Analysis of
  correlation in terms of exact local potentials: Applications to two-electron
  systems}, Phys. Rev. A, 40 (1989), pp.~4190--4202.

\bibitem{ButChaDeP-AMO-18}
{\sc G.~Buttazzo, T.~Champion, and L.~De~Pascale}, {\em Continuity and
  estimates for multimarginal optimal transportation problems with singular
  costs}, Applied Mathematics \& Optimization, 78 (2018), pp.~185--200.

\bibitem{ButDepGor-PRA-12}
{\sc G.~Buttazzo, L.~De~Pascale, and P.~Gori-Giorgi}, {\em Optimal-transport
  formulation of electronic density-functional theory}, Phys. Rev. A, 85
  (2012), p.~062502.

\bibitem{Car21}
{\sc G.~Carlier}, {\em On the linear convergence of the multi-marginal
  {S}inkhorn algorithm}, HAL Id: hal-03176512,  (2021).

\bibitem{CarFriVoe-21}
{\sc G.~Carlier, G.~Friesecke, and D.~V\"ogler}, {\em Convex geometry of finite
  exchangeable laws and de finetti style representation with universal
  correlated corrections}, arXiv preprint: arXiv:2106.09101,  (2021).

\bibitem{Car-SIMA-19}
{\sc G.~Carlier and M.~Laborde}, SIAM Journal on Mathematical Analysis, 52
  (2020), pp.~709--717.

\bibitem{CasSorSen-PRB-06}
{\sc M.~Casula, S.~Sorella, and G.~Senatore}, {\em Ground state properties of
  the one-dimensional coulomb gas using the lattice regularized diffusion monte
  carlo method}, Phys. Rev. B, {74} (2006), p.~245427.

\bibitem{ChaCha-JSP-84}
{\sc J.~Chayes and L.~Chayes}, {\em On the validity of the inverse conjecture
  in classical density functional theory}, Journal of statistical physics, 36
  (1984), pp.~471--488.

\bibitem{ChaChaLie-CMP-84}
{\sc J.~Chayes, L.~Chayes, and E.~H. Lieb}, {\em The inverse problem in
  classical statistical mechanics}, Communications in Mathematical Physics, 93
  (1984), pp.~57--121.

\bibitem{CheFri-MMS-15}
{\sc H.~Chen and G.~Friesecke}, {\em Pair densities in density functional
  theory}, Multiscale Modeling \& Simulation, 13 (2015), pp.~1259--1289.

\bibitem{CheFriMen-JCTC-14}
{\sc H.~Chen, G.~Friesecke, and C.~B. Mendl}, {\em Numerical methods for a
  {K}ohn-{S}ham density functional model based on optimal transport}, J. Chem.
  Theory Comput, 10 (2014), pp.~4360--4368.

\bibitem{CheGeoPav-SIAMAM-16}
{\sc Y.~Chen, T.~Georgiou, and M.~Pavon}, {\em Entropic and displacement
  interpolation: a computational approach using the {H}ilbert metric}, SIAM
  Journal on Applied Mathematics, 76 (2016), pp.~2375--2396.

\bibitem{CohMorYan-SCI-08}
{\sc A.~Cohen, P.~Mori-S{\'a}nchez, and W.~Yang}, {\em Insights into current
  limitations of density functional theory}, Science, 321 (2008), pp.~792--794.

\bibitem{CohMorYan-JCP-07}
{\sc A.~J. Cohen, P.~Mori-S{\'a}nchez, and W.~Yang}, J. Chem. Phys., 127
  (2007), p.~034101.

\bibitem{CoYu-book-00}
{\sc A.~J. Coleman and V.~I. Yukalov}, {\em Reduced Density Matrices}, Lecture
  Notes in Chemistry Vol. 72, Springer, 2000.

\bibitem{ColDepDim-CJM-15}
{\sc M.~Colombo, L.~De~Pascale, and S.~Di~Marino}, {\em Multimarginal optimal
  transport maps for one-dimensional repulsive costs}, Canad. J. Math, 67
  (2015), pp.~350--368.

\bibitem{ColDiM-INC-13}
{\sc M.~Colombo and S.~{Di Marino}}, {\em Equality between monge and
  kantorovich multimarginal problems with coulomb cost}, in Annali di
  Matematica Pura ad Applicata, Springer, Berlin Heidelberg, 2013, pp.~1--14.

\bibitem{ColDMaStra-arxiv-21}
{\sc M.~Colombo, S.~Di~Marino, and F.~Stra}, {\em First order expansion in the
  semiclassical limit of the levy-lieb functional}, arXiv preprint
  arXiv:2106.06282,  (2021).

\bibitem{ColStr-M3AS-15}
{\sc M.~Colombo and F.~Stra}, {\em Counterexamples in multimarginal optimal
  transport with coulomb cost and spherically symmetric data}, Mathematical
  Models and Methods in Applied Sciences, 26 (2016), pp.~1025--1049.

\bibitem{Con-PRB-19}
{\sc L.~A. Constantin}, {\em Correlation energy functionals from adiabatic
  connection formalism}, Phys. Rev. B, 99 (2019), p.~085117.

\bibitem{CotFriKlu-CPAM-13}
{\sc C.~Cotar, G.~Friesecke, and C.~Kl\"uppelberg}, {\em Density functional
  theory and optimal transportation with coulomb cost}, Comm. Pure Appl. Math.,
  66 (2013), pp.~548--99.

\bibitem{CotFriKlu-ARMA-18}
{\sc C.~Cotar, G.~Friesecke, and C.~Kl{\"u}ppelberg}, {\em Smoothing of
  transport plans with fixed marginals and rigorous semiclassical limit of the
  hohenberg--kohn functional}, Arch. Ration. Mech. An., 228 (2018),
  pp.~891--922.

\bibitem{Cut-ANIPS-13}
{\sc M.~Cuturi}, {\em Sinkhorn distances: Lightspeed computation of optimal
  transport}, in Advances in neural information processing systems, 2013,
  pp.~2292--2300.

\bibitem{PeyCut-FTML-19}
{\sc M.~Cuturi and G.~Peyr{\'e}}, {\em Computational optimal transport},
  vol.~11, Now Publishers, Inc., 2019.

\bibitem{DaaFabDelGorVuc-JPCL-21}
{\sc T.~J. Daas, E.~Fabiano, F.~Della~Sala, P.~Gori-Giorgi, and S.~Vuckovic},
  {\em Noncovalent interactions from models for the m{\o}ller--plesset
  adiabatic connection}, The journal of physical chemistry letters, 12 (2021),
  pp.~4867--4875.

\bibitem{DaaGroVucMusKooSeiGieGor-JCP-20}
{\sc T.~J. Daas, J.~Grossi, S.~Vuckovic, Z.~H. Musslimani, D.~P. Kooi,
  M.~Seidl, K.~J. Giesbertz, and P.~Gori-Giorgi}, {\em Large coupling-strength
  expansion of the m{\o}ller--plesset adiabatic connection: From paradigmatic
  cases to variational expressions for the leading terms}, The Journal of
  chemical physics, 153 (2020), p.~214112.

\bibitem{Dep-ESAIMMMNA-15}
{\sc L.~De~Pascale}, {\em Optimal transport with coulomb cost. approximation
  and duality}, ESAIM: Math. Model. Numer. Anal., 49 (2015), pp.~1643--1657.

\bibitem{del2020statistical}
{\sc E.~del Barrio and J.-M. Loubes}, {\em The statistical effect of entropic
  regularization in optimal transportation},  (2020).

\bibitem{DMaGer-JSC-20}
{\sc S.~Di~Marino and A.~Gerolin}, {\em An {O}ptimal {T}ransport approach for
  the {S}chr{\"o}dinger bridge problem and convergence of {S}inkhorn
  algorithm}, Journal of Scientific Computing, 85 (2020).

\bibitem{DMaGer-arXiv-2020}
\leavevmode\vrule height 2pt depth -1.6pt width 23pt, {\em Optimal transport
  losses and {S}inkhorn algorithm with general convex regularization}, arXiv
  preprint arXiv:2007.00976,  (2020).

\bibitem{DMaGerNen-TOOAS-17}
{\sc S.~Di~Marino, A.~Gerolin, and L.~Nenna}, {\em Optimal transport for
  repulsive costs}, Topological Optimization and Optimal Transport In the
  Applied Sciences,  (2017).

\bibitem{Ern-CPL-96}
{\sc M.~Ernzerhof}, {\em Construction of the adiabatic connection}, Chem. Phys.
  Lett., {263} (1996), p.~499.

\bibitem{FabGorSeiDel-JCTC-16}
{\sc E.~Fabiano, P.~Gori-Giorgi, M.~Seidl, and F.~Della~Sala}, {\em
  Interaction-strength interpolation method for main-group chemistry:
  Benchmarking, limitations, and perspectives}, J. Chem. Theory. Comput., 12
  (2016), pp.~4885--4896.

\bibitem{FabSmiGiaDaaDelGraGor-JCTC-19}
{\sc E.~Fabiano, S.~Smiga, S.~Giarrusso, T.~J. Daas, F.~Della~Sala,
  I.~Grabowski, and P.~Gori-Giorgi}, {\em Investigation of the
  exchange-correlation potentials of functionals based on the adiabatic
  connection interpolation}, Journal of chemical theory and computation, 15
  (2019), pp.~1006--1015.

\bibitem{FraLor-LAA-89}
{\sc J.~Franklin and J.~Lorenz}, {\em On the scaling of multidimensional
  matrices}, Linear Algebra and its applications, 114 (1989), pp.~717--735.

\bibitem{FreHuxMor-PRA-84}
{\sc D.~E. Freund, B.~D. Huxtable, and J.~D. Morgan}, {\em Variational
  calculations on the helium isoelectronic sequence}, Phys. Rev. A, {29}
  (1984), pp.~980--982.

\bibitem{Fri-ARMA-03}
{\sc G.~Friesecke}, {\em The multiconfiguration equations for atoms and
  molecules: charge quantization and existence of solutions}, Archive for
  Rational Mechanics and Analysis, 169 (2003), pp.~35--71.

\bibitem{Fri-SIAMJMA-19}
\leavevmode\vrule height 2pt depth -1.6pt width 23pt, {\em A simple
  counterexample to the {M}onge ansatz in multi-marginal optimal transport,
  convex geometry of the set of {K}antorovich plans, and the
  {F}renkel-{K}ontorova model}, SIAM J. Math. Analysis, 51 (2019),
  pp.~4332--4355.

\bibitem{Fri-Book-22}
\leavevmode\vrule height 2pt depth -1.6pt width 23pt, {\em Lectures on optimal
  transport}, SIAM, 2022, to appear.

\bibitem{FriMenPasCotKlu-JCP-13}
{\sc G.~Friesecke, C.~B. Mendl, B.~Pass, C.~Cotar, and C.~Kl\"uppelberg}, {\em
  N-density representability and the optimal transport limit of the
  {H}ohenberg-{K}ohn functional}, J. Chem. Phys., 139 (2013), p.~164109.

\bibitem{FriSchVoe-21}
{\sc G.~Friesecke, A.~S. Schulz, and D.~V\"ogler}, {\em Genetic column
  generation: Fast computation of high-dimensional multi-marginal optimal
  transport problems}, to appear in SIAM J. Sci. Comp., arXiv preprint:
  arXiv:2103.12624,  (2021).

\bibitem{FriVog-SIAMJMA-18}
{\sc G.~Friesecke and D.~V\"ogler}, {\em Breaking the curse of dimension in
  multi-marginal kantorovich optimal transport on finite state spaces}, SIAM J.
  Math. Analysis, 50 (2018), pp.~3996--4019.

\bibitem{Gal-CEPR-10}
{\sc A.~Galichon and B.~Salani{\'e}}, {\em Matching with trade-offs: Revealed
  preferences over competing characteristics}, CEPR Discussion Paper No.
  DP7858,  (2010).

\bibitem{Gangbo-McCann-96}
{\sc W.~Gangbo and R.~J. McCann}, {\em The geometry of optimal transportation},
  Acta Math., 177 (1906), pp.~113--161.

\bibitem{GanSwi-CPAM-98}
{\sc W.~Gangbo and A.~Swiech}, {\em Optimal maps for the multidimensional
  monge-kantorovich problem}, Commun. Pure Appl. Math., 51 (1998), p.~23.

\bibitem{GerThesis}
{\sc A.~Gerolin}, {\em Multi-marginal optimal transport and potential
  optimization problems for Schr{\"o}dinger operators}, PhD thesis,
  Universit{\`a} degli studi di Pisa, 2016.

\bibitem{GerGroGor-JCTC-20}
{\sc A.~Gerolin, J.~Grossi, and P.~Gori-Giorgi}, {\em Kinetic correlation
  functionals from the entropic regularisation of the strictly-correlated
  electrons problem}, Journal of Chemical Theory and Computation, 16 (2019),
  pp.~488--498.

\bibitem{GerKauRaj-SIAMJMA-19}
{\sc A.~Gerolin, A.~Kausamo, and T.~Rajala}, {\em Non-existence of optimal
  transport maps for the multi-marginal repulsive harmonic cost}, SIAM Journal
  on Mathematical Analysis, 51 (2019).

\bibitem{GerKauRaj-CVPDE-20}
\leavevmode\vrule height 2pt depth -1.6pt width 23pt, {\em Multi-marginal
  {E}ntropy-{T}ransport with repulsive cost}, Calc. Var. PDEs, 59 (2020).

\bibitem{GhoGucUmrUllBar-NP-06}
{\sc A.~Ghosal, A.~D. Guclu, C.~J. Umrigar, D.~Ullmo, and H.~U. Baranger},
  Nature Phys., 2 (2006), p.~336.

\bibitem{GiaGorDelFab-JCP-18}
{\sc S.~Giarrusso, P.~Gori-Giorgi, F.~Della~Sala, and E.~Fabiano}, {\em
  Assessment of interaction-strength interpolation formulas for gold and silver
  clusters}, J. Chem. Phys., 148 (2018), p.~134106.

\bibitem{GiaVucGor-JCTC-18}
{\sc S.~Giarrusso, S.~Vuckovic, and P.~Gori-Giorgi}, {\em Response potential in
  the strong-interaction limit of dft: Analysis and comparison with the
  coupling-constant average}, J. Chem. Theory Comput., 14 (2018),
  pp.~4151--4167.

\bibitem{GiuVig-BOOK-05}
{\sc G.~F. Giuliani and G.~Vignale}, {\em Quantum Theory of the Electron
  Liquid}, Cambridge University Press, New York, 2005.

\bibitem{GorSav-JPCS-08}
{\sc P.~Gori-Giorgi and A.~Savin}, J. Phys.: Conf. Ser., {117} (2008),
  p.~012017.

\bibitem{GorSei-PCCP-10}
{\sc P.~Gori-Giorgi and M.~Seidl}, {\em Density functional theory for
  strongly-interacting electrons: perspectives for physics and chemistry},
  Phys. Chem. Chem. Phys, 12 (2010), pp.~14405--14419.

\bibitem{GorSeiVig-PRL-09}
{\sc P.~Gori-Giorgi, M.~Seidl, and G.~Vignale}, {\em Density-functional theory
  for strongly interacting electrons}, Phys. Rev. Lett., {103} (2009),
  p.~166402.

\bibitem{GorVigSei-JCTC-09}
{\sc P.~Gori-Giorgi, G.~Vignale, and M.~Seidl}, {\em Electronic zero-point
  oscillations in the strong-interaction limit of density functional theory},
  J. Chem. Theory Comput., 5 (2009), pp.~743--753.

\bibitem{GorLev-PRB-93}
{\sc A.~G\"{o}rling and M.~Levy}, Phys. Rev. B, 47 (1993), p.~13105.

\bibitem{GozChr-MPRF-10}
{\sc N.~Gozlan and C.~L{\'e}onard}, {\em Transport inequalities. a survey},
  Markov Processes and Related Fields, 16 (2010), pp.~635--736.

\bibitem{GroKooGieSeiCohMorGor-JCTC-17}
{\sc J.~Grossi, D.~P. Kooi, K.~J.~H. Giesbertz, M.~Seidl, A.~J. Cohen,
  P.~Mori-S{\'a}nchez, and P.~Gori-Giorgi}, {\em Fermionic statistics in the
  strongly correlated limit of density functional theory}, J. Chem. Theory
  Comput., 13 (2017), pp.~6089--6100.

\bibitem{GroMusSeiGor-JPCM-20}
{\sc J.~Grossi, Z.~Musslimani, M.~Seidl, and P.~Gori-Giorgi}, {\em Kohn-sham
  equations with functionals from the strictly-correlated regime: Investigation
  with a spectral renormalization method.}, Journal of Physics: Condensed
  Matter,  (2020).

\bibitem{GucGhoUmrBar-PRB-08}
{\sc A.~D. Guclu, A.~Ghosal, C.~J. Umrigar, and H.~U. Baranger}, Phys. Rev. B,
  77 (2008), p.~041301.

\bibitem{HelEtAl-PRA-11}
{\sc N.~Helbig, J.~I. Fuks, M.~Casula, M.~J. Verstraete, M.~Marques,
  I.~Tokatly, and A.~Rubio}, {\em Density functional theory beyond the linear
  regime: Validating an adiabatic local density approximation}, Physical Review
  A, 83 (2011), p.~032503.

\bibitem{HelTokRub-JCP-09}
{\sc N.~Helbig, I.~V. Tokatly, and A.~Rubio}, {\em Exact {Kohn--Sham} potential
  of strongly correlated finite systems}, J. Chem. Phys., 131 (2009),
  p.~224105.

\bibitem{Hel-book-16}
{\sc T.~Helgaker, P.~J{\o}rgensen, and J.~Olsen}, {\em Density-functional
  Theory: A Convex Treatment}, Wiley Blackwell, 2016.

\bibitem{Hof-77}
{\sc M.~Hoffmann-Ostenhof and T.~Hoffmann-Ostenhof}, {\em "{S}chr\"odinger
  inequalities" and asymptotic behavior of the electron density of atoms and
  molecules}, Phys. Rev. A, 16 (1977), pp.~1782--1785.

\bibitem{HohKoh-PR-64}
{\sc P.~Hohenberg and W.~Kohn}, {\em Inhomogeneous electron gas}, Phys. Rev.,
  {136} (1964), p.~B 864.

\bibitem{Jan-NEURIPS-2020}
{\sc H.~Janati, B.~Muzellec, G.~Peyr\'{e}, and M.~Cuturi}, {\em Entropic
  optimal transport between unbalanced gaussian measures has a closed form}, in
  Advances in Neural Information Processing Systems, H.~Larochelle, M.~Ranzato,
  R.~Hadsell, M.~F. Balcan, and H.~Lin, eds., vol.~33, Curran Associates, Inc.,
  2020, pp.~10468--10479.

\bibitem{Kan-DAN-42}
{\sc L.~V. Kantorovich}, {\em On the transfer of masses}, Dokl. Akad. Nauk.
  SSSR., 37 (1942), p.~227.

\bibitem{Kel-ZWG-84}
{\sc H.~G. Kellerer}, {\em Duality theorems for marginal problems}, Zeitschrift
  f{\"u}r Wahrscheinlichkeitstheorie und verwandte Gebiete, 67 (1984),
  pp.~399--432.

\bibitem{KhoYin-SIAMJSC-19}
{\sc Y.~Khoo and L.~Ying}, {\em Convex relaxation approaches for strictly
  correlated density functional theory}, SIAM J. Sci. Comput., 41 (2019),
  pp.~B773--B795.

\bibitem{Knott-Smith-84}
{\sc M.~Knott and C.~S. Smith}, {\em On the optimal mapping of distributions},
  J. Optimization Theory and Appl., 43 (1984), pp.~39--49.

\bibitem{Koh-PRL-83}
{\sc W.~Kohn}, {\em v-representability and density functional theory}, Physical
  review letters, 51 (1983), p.~1596.

\bibitem{KS65}
{\sc W.~Kohn and L.~J. Sham}, {\em Self-consistent equations including exchange
  and correlation effects}, Phys. Rev., 140 (1965), pp.~A1133--A1138.

\bibitem{KolRoo-60}
{\sc W.~Kolos and C.~C.~J. Roothaan}, {\em Accurate electronic wave functions
  for the ${\mathrm{h}}_{2}$ molecule}, Rev. Mod. Phys., 32 (1960),
  pp.~219--232.

\bibitem{KooGor-TCA-18}
{\sc D.~P. Kooi and P.~Gori-Giorgi}, {\em Local and global interpolations along
  the adiabatic connection of dft: a study at different correlation regimes},
  Theoretical chemistry accounts, 137 (2018), pp.~1--12.

\bibitem{KooGor-JPCL-19}
\leavevmode\vrule height 2pt depth -1.6pt width 23pt, {\em A variational
  approach to london dispersion interactions without density distortion}, The
  journal of physical chemistry letters, 10 (2019), pp.~1537--1541.

\bibitem{Leo-DCDSA-14}
{\sc C.~L{\'e}onard}, {\em A survey of the schr{\"o}dinger problem and some of
  its connections with optimal transport}, Discrete Cont. Dyn.-A, 34 (2014),
  pp.~1533--1574.

\bibitem{Lev-PNAS-79}
{\sc M.~Levy}, {\em Universal variational functionals of electron densities,
  first-order density matrices, and natural spin-orbitals and solution of the
  v-representability problem}, Proc. Natl. Acad. Sci., 76 (1979),
  pp.~6062--6065.

\bibitem{Lev-PRA-82}
\leavevmode\vrule height 2pt depth -1.6pt width 23pt, {\em Electron densities
  in search of hamiltonians}, Phys. Rev. A, 26 (1982), pp.~1200--1208.

\bibitem{Lev-Per-85}
{\sc M.~Levy and J.~Perdew}, {\em Hellman-{F}eynman, virial, and scaling
  requisites for the exact universal density functionals. {S}hape of the
  correlation potential and diamagnetic susceptibility for atoms}, Phys. Rev.
  A, 32 (1985), pp.~2010--2021.

\bibitem{LevPerSah-PRA-84}
{\sc M.~Levy, J.~P. Perdew, and V.~Sahni}, {\em Exact differential equation for
  the density and ionization energy of a many-particle system}, Phys. Rev. A,
  30 (1984), pp.~2745--2748.

\bibitem{LevZah-PRL-14}
{\sc M.~Levy and F.~Zahariev}, Phys. Rev. Lett., 113 (2014), p.~113002.

\bibitem{Lew-CRM-18}
{\sc M.~Lewin}, {\em Semi-classical limit of the {L}evy--{L}ieb functional in
  {D}ensity {F}unctional {T}heory}, C. R. Math., 356 (2018), pp.~449--455.

\bibitem{LewLieSei-PRB-19}
{\sc M.~Lewin, E.~H. Lieb, and R.~Seiringer}, {\em Floating wigner crystal with
  no boundary charge fluctuations}, Physical Review B, 100 (2019), p.~035127.

\bibitem{Lie-IJQC-83}
{\sc E.~H. Lieb}, {\em Density functionals for {CouIomb} systems}, Int. J.
  Quantum. Chem., 24 (1983), pp.~243--277.

\bibitem{LiuBur-PRA-09}
{\sc Z.-F. Liu and K.~Burke}, {\em Adiabatic connection in the low-density
  limit}, Phys. Rev. A, 79 (2009), p.~064503.

\bibitem{LorMah20cont}
{\sc D.~Lorenz and H.~Mahler}, {\em Orlicz space regularization of continuous
  optimal transport problems}, arXiv preprint arXiv:2004.11574,  (2020).

\bibitem{MalGor-PRL-12}
{\sc F.~Malet and P.~Gori-Giorgi}, {\em Strong correlation in kohn-sham density
  functional theory}, Phys. Rev. Lett., 109 (2012), p.~246402.

\bibitem{MalMirCreReiGor-PRB-13}
{\sc F.~Malet, A.~Mirtschink, J.~C. Cremon, S.~M. Reimann, and P.~Gori-Giorgi},
  {\em Kohn-sham density functional theory for quantum wires in arbitrary
  correlation regimes}, Phys. Rev. B, 87 (2013), p.~115146.

\bibitem{MalMirGieWagGor-PCCP-14}
{\sc F.~Malet, A.~Mirtschink, K.~J.~H. Giesbertz, L.~O. Wagner, and
  P.~Gori-Giorgi}, {\em Exchange-correlation functionals from the strong
  interaction limit of dft: applications to model chemical systems}, Phys.
  Chem. Chem. Phys., 16 (2014), pp.~14551--14558.

\bibitem{Mal-InfoGeo-2021}
{\sc A.~Mallasto, A.~Gerolin, and H.~Q. Minh}, {\em Entropy-regularized
  2-wasserstein distance between gaussian measures}, Information Geometry,
  (2021), pp.~1--35.

\bibitem{MenLin-PRB-13}
{\sc C.~B. Mendl and L.~Lin}, {\em Kantorovich dual solution for strictly
  correlated electrons in atoms and molecules}, Phys. Rev. B, 87 (2013),
  p.~125106.

\bibitem{MenMalGor-PRB-14}
{\sc C.~B. Mendl, F.~Malet, and P.~Gori-Giorgi}, {\em Wigner localization in
  quantum dots from kohn-sham density functional theory without symmetry
  breaking}, Phys. Rev. B, 89 (2014), p.~125106.

\bibitem{MirUmrMorGor-JCP-14}
{\sc A.~Mirtschink, C.~J. Umrigar, J.~D. Morgan~III, and P.~Gori-Giorgi}, {\em
  Energy density functionals from the strong-coupling limit applied to the
  anions of the he isoelectronic series}, J. Chem. Phys., 140 (2014),
  p.~18A532.

\bibitem{MoaPas-ESAIMCOCV-17}
{\sc A.~Moameni and B.~Pass}, {\em Solutions to multi-marginal optimal
  transport problems concentrated on several graphs}, ESAIM: Control Optim.
  Calc. Var.,  (2017), pp.~551--567.

\bibitem{Mon-BOOK-1781}
{\sc G.~Monge}, {\em M\'emoire sur la th\'eorie des d\'eblais et des remblais},
  Histoire Acad. Sciences, Paris, 1781.

\bibitem{MorCoh-JPCL-17}
{\sc P.~Mori-S{\'a}nchez and A.~J. Cohen}, {\em Exact density functional
  obtained via the {L}evy constrained search}, The journal of physical
  chemistry letters, 9 (2018), pp.~4910--4914.

\bibitem{Nen-PhD-16}
{\sc L.~Nenna}, {\em Numerical methods for multi-marginal optimal
  transportation}, PhD thesis, 2016.

\bibitem{Pas-Thesis-11}
{\sc B.~Pass}, {\em Structural results on optimal transportation plans}, PhD
  thesis, University of Toronto, 2011.

\bibitem{Pas-CVPDE-12}
\leavevmode\vrule height 2pt depth -1.6pt width 23pt, {\em On the local
  structure of optimal measures in the multi-marginal optimal transportation
  problem}, Calculus of Variations and Partial Differential Equations, 43
  (2012), pp.~529--536.

\bibitem{Pas-NL-13}
{\sc B.~Pass}, Nonlinearity, {26} (2013), p.~2731.

\bibitem{Pas-IOP-13}
{\sc B.~Pass}, {\em Remarks on the semi-classical hohenberg--kohn functional},
  Nonlinearity, 26 (2013), p.~2731.

\bibitem{Pas-DCDS-14}
\leavevmode\vrule height 2pt depth -1.6pt width 23pt, {\em Multi-marginal
  optimal transport and multi-agent matching problems: uniqueness and structure
  of solutions}, Discrete Contin. Dyn. Syst., 34:1623-1639,  (2014).

\bibitem{Pas-ESAIM-15}
\leavevmode\vrule height 2pt depth -1.6pt width 23pt, {\em Multi-marginal
  optimal transport: theory and applications}, ESAIM: Mathematical Modelling
  and Numerical Analysis,  (2015).

\bibitem{Pra-AIHP-07}
{\sc A.~Pratelli}, {\em On the equality between monge's infimum and
  kantorovich's minimum in optimal mass transportation}, in Annales de
  l'Institut Henri Poincare (B) Probability and Statistics, vol.~43, Elsevier,
  2007, pp.~1--13.

\bibitem{RacRus-BOOK-98}
{\sc S.~Rachev and L.~R\"uschendorf}, {\em Mass transportation problems},
  Springer-Verlag, New York, 1998.

\bibitem{Rus-TAS-95}
{\sc L.~Ruschendorf}, {\em Convergence of the iterative proportional fitting
  procedure}, The Annals of Statistics, 23 (1995), pp.~1160--1174.

\bibitem{San-Book-15}
{\sc F.~Santambrogio}, {\em Optimal Transport for Applied Mathematicians},
  Progress in Nonlinear Differential Equations and Their Applications.,
  Birkh{\"a}user, 2015.

\bibitem{Sav-CP-09}
{\sc A.~Savin}, Chem. Phys., {356} (2009), p.~91.

\bibitem{Sch-VAWK-31}
{\sc E.~Schr{\"o}dinger}, {\em {\"U}ber die umkehrung der naturgesetze}, Verlag
  Akademie der wissenschaften in kommission bei Walter de Gruyter u. Company,
  1931.

\bibitem{Sei-PRA-99}
{\sc M.~Seidl}, {\em Strong-interaction limit of density-functional theory},
  Phys. Rev. A, 60 (1999), pp.~4387--4395.

\bibitem{Sei-PRA-07}
\leavevmode\vrule height 2pt depth -1.6pt width 23pt, Phys. Rev. A, {75}
  (2007), p.~062506.

\bibitem{SeiDiMGerNenGieGor-XX-XX}
{\sc M.~Seidl, S.~Di~Marino, A.~Gerolin, L.~Nenna, K.~J. Giesbertz, and
  P.~Gori-Giorgi}, {\em The strictly-correlated electron functional for
  spherically symmetric systems revisited ii: Sgs conjecture}, in preparation.

\bibitem{SeiDiMGerNenGieGor-arxiv-17}
\leavevmode\vrule height 2pt depth -1.6pt width 23pt, {\em The
  strictly-correlated electron functional for spherically symmetric systems
  revisited}, arXiv preprint arXiv:1702.05022,  (2017).

\bibitem{SeiDiMGerNenGieGor-PRA-17}
\leavevmode\vrule height 2pt depth -1.6pt width 23pt, {\em The
  strictly-correlated electron functional for spherically symmetric systems
  revisited}, arXiv preprint arXiv:1702.05022,  (2017).

\bibitem{SeiGiaVucFabGor-JCP-2018}
{\sc M.~Seidl, S.~Giarrusso, S.~Vuckovic, E.~Fabiano, and P.~Gori-Giorgi}, {\em
  Communication: Strong-interaction limit of an adiabatic connection in
  hartree-fock theory}, The Journal of Chemical Physics, 149 (2018), p.~241101.

\bibitem{SeiGorSav-PRA-07}
{\sc M.~Seidl, P.~Gori-Giorgi, and A.~Savin}, {\em Strictly correlated
  electrons in density-functional theory: A general formulation with
  applications to spherical densities}, Phys. Rev. A, 75 (2007), p.~042511/12.

\bibitem{SeiPerKur-PRA-00}
{\sc M.~Seidl, J.~P. Perdew, and S.~Kurth}, Phys. Rev. A, {62} (2000),
  p.~012502.

\bibitem{SeiPerKur-PRL-00}
{\sc M.~Seidl, J.~P. Perdew, and S.~Kurth}, {\em Simulation of all-order
  density-functional perturbation theory, using the second order and the
  strong-correlation limit}, Phys. Rev. Lett., 84 (2000), pp.~5070--5073.

\bibitem{SeiPerLev-PRA-99}
{\sc M.~Seidl, J.~P. Perdew, and M.~Levy}, {\em Strictly correlated electrons
  in density-functional theory}, Phys. Rev. A, 59 (1999), pp.~51--54.

\bibitem{Sin-TAMS-64}
{\sc R.~Sinkhorn}, {\em A relationship between arbitrary positive matrices and
  doubly stochastic matrices}, The annals of mathematical statistics, 35
  (1964), pp.~876--879.

\bibitem{SmiCon-JCTC-20}
{\sc S.~Smiga and L.~A. Constantin}, {\em Modified interaction-strength
  interpolation method as an important step toward self-consistent
  calculations}, Journal of chemical theory and computation, 16 (2020),
  pp.~4983--4992.

\bibitem{Lee-AQC-03}
{\sc R.~van Leeuwen}, {\em Density functional approach to the many-body
  problem: key concepts and exact functionals}, Adv. Quantum Chem., 43 (2003),
  pp.~24--94.

\bibitem{Vie-PRB-12}
{\sc D.~Vieira}, Phys. Rev. B, {86} (2012), p.~075132.

\bibitem{VieCap-JCTC-10}
{\sc D.~Vieira and K.~Capelle}, J. Chem. Theory Comput., {6} (2010), p.~3319.

\bibitem{Vil-BOOK-03}
{\sc C.~Villani}, {\em Topics in Optimal Transportation}, Grad. Stud. Math.
  {\bf 58}, Amer. Math. Soc., Providence, 2003.

\bibitem{VucGorDelFab-JPCL-18}
{\sc S.~Vuckovic, P.~Gori-Giorgi, F.~Della~Sala, and E.~Fabiano}, {\em
  Restoring size consistency of approximate functionals constructed from the
  adiabatic connection}, J. Phys. Chem. Lett., 9 (2018), pp.~3137--3142.

\bibitem{VucIroSavTeaGor-JCTC-16}
{\sc S.~Vuckovic, T.~J.~P. Irons, A.~Savin, A.~M. Teale, and P.~Gori-Giorgi},
  {\em Exchange--correlation functionals via local interpolation along the
  adiabatic connection}, J. Chem. Theory Comput., 12 (2016), pp.~2598--2610.

\bibitem{VucLevGor-JCP-17}
{\sc S.~Vuckovic, M.~Levy, and P.~Gori-Giorgi}, {\em Augmented potential,
  energy densities, and virial relations in the weak-and strong-interaction
  limits of dft}, J. Chem. Phys., 147 (2017), p.~214107.

\bibitem{WagGor-PRA-14}
{\sc L.~O. Wagner and P.~Gori-Giorgi}, {\em Electron avoidance: A nonlocal
  radius for strong correlation}, Phys. Rev. A, 90 (2014), p.~052512.

\bibitem{WagStoBurWhi-PCCP-12}
{\sc L.~O. Wagner, E.~M. Stoudenmire, K.~Burke, and S.~R. White}, {\em
  Reference electronic structure calculations in one dimension}, Phys. Chem.
  Chem. Phys., {14} (2012), p.~8581.

\bibitem{WanLiCheXiaRonPol-PRB-12}
{\sc J.-J. Wang, W.~Li, S.~Chen, G.~Xianlong, M.~Rontani, and M.~Polini}, {\em
  Absence of wigner molecules in one-dimensional few-fermion systems with
  short-range interactions}, Physical Review B, 86 (2012), p.~075110.

\bibitem{Wig-PR-34}
{\sc E.~P. Wigner}, Phys. Rev., {46} (1934), p.~1002.

\bibitem{YinBroLopVarGorLor-PRB-16}
{\sc Z.-J. Ying, V.~Brosco, G.~M. Lopez, D.~Varsano, P.~Gori-Giorgi, and
  J.~Lorenzana}, {\em Anomalous scaling and breakdown of conventional density
  functional theory methods for the description of mott phenomena and stretched
  bonds}, Phys. Rev. B, 94 (2016), p.~075154.

\bibitem{Zhi-TMMO-60}
{\sc G.~M. Zhislin}, {\em Discussion of the spectrum of schr{\"o}dinger
  operators for systems of many particles}, Trudy Moskovskogo matematiceskogo
  obscestva, 9 (1960), pp.~81--120.

\end{thebibliography}
%\begin{thebibliography}{FGGSDS16}

\end{small}

\end{document}